\RequirePackage{fix-cm}

\documentclass[11pt]{amsart}
\usepackage[foot]{amsaddr}

\usepackage{graphicx}
\usepackage{amsmath}
\usepackage{amssymb}
\usepackage[english]{babel}              
\usepackage{geometry}
\usepackage{mathrsfs}

\usepackage{color}

\usepackage{ulem}

\usepackage{array,multirow,makecell}
\setcellgapes{1pt}
\makegapedcells

\usepackage{natbib}

\usepackage[noprefix]{nomencl}

\makeatletter
\renewcommand{\email}[2][]{%
  \ifx\emails\@empty\relax\else{\g@addto@macro\emails{,\space}}\fi%
  \@ifnotempty{#1}{\g@addto@macro\emails{\textrm{(#1)}\space}}%
  \g@addto@macro\emails{#2}%
}
\makeatother

\newcommand{\nnb}{\nonumber}
\newcommand{\be}{\begin{equation}}
\newcommand{\ee}{\end{equation}}

\newcommand{\qtext}[1]{\quad \mbox{ #1 } \quad}
\newcommand{\comment}[1]{}
\newcommand{\der }[2]{\frac{d#1}{d#2}}

\newcommand{\dron}[2]  {\frac{\partial#1   }{\partial#2}}
\newcommand{\drons}[2] {\frac{\partial^2 #1}{\partial #2^2}}
\newcommand{\dronss}[3]{\frac{\partial^2 #1}{\partial #2 \partial #3}}

\newcommand{\norm}[1]{\vert\vert#1 \vert\vert}

\newcommand\Hb{{\overline{H} }}

\newcommand\br  {{\bf r}}
\newcommand\brp {{\bf r'}}

\newcommand{\md}[1]{{\text{\d{$#1$}}}}
\newcommand{\und}[1]{\md{#1}}
\DeclareTextSymbol{\degre}{OT1}{23}

\newcommand\lam   {\lambda }

\newcommand\Lam   {\Lambda }

\newcommand\Gam   {\Gamma  }

\newcommand{\tLam} {{\widetilde{\Lam}}}

\newcommand\xb    {{\overline{x}}}

\newcommand\pb    {{\overline{p}}}

\newcommand\brd   {\dot{\br }}

\newcommand\dGam  {\dot{\Gam}}

\newcommand{\RR}{{\mathbb R}}
\newcommand{\CC}{{\mathbb C}}
\newcommand{\TT}{{\mathbb T}}

\newcommand{\NN}{{\mathbb N}}

\newcommand\cM{{\mathcal M}}
\newcommand\cC{{\mathcal C}}
\newcommand\cG{{\mathcal G}}
\newcommand\cN{{\mathcal N}}
\newcommand\cF{{\mathcal F}}
\newcommand\cH{{\mathcal H}}

\newcommand\cO{{\mathcal O}}

\newcommand\sM{{\mathscr{M}}}
\newcommand\sL{{\mathscr{L}}}

\newcommand\eps{{\varepsilon}}
\newcommand\ode{{\cO(\eps^2)}}
\newcommand\oue{{\cO(\eps)}}

\newcolumntype{R}[1]{>{\raggedleft\arraybackslash }b{#1}}
\newcolumntype{L}[1]{>{\raggedright\arraybackslash }b{#1}}
\newcolumntype{C}[1]{>{\centering\arraybackslash }b{#1}}

\newcommand\GLQ{{\cG_{L_4}^{e_0}}}
\newcommand\GLC{{\cG_{L_5}^{e_0}}}
\newcommand\NLQ{{\cN_{L_4}^{u}}}
\newcommand\NLC{{\cN_{L_5}^{u}}}
\newcommand\GQS{{\cG_{QS}^{e_0}}}
\newcommand\GLT{{\cG_{L_3}^{e_0}}}

\newcommand\GdLT{{G_{L_3}}}
\newcommand\GdLQ{{G_{L_4}}}
\newcommand\GdLC{{G_{L_5}}}
\newcommand\GdQS{{G_{QS}}}

\newcommand\GdLTs{{G_{L_3,1}^{e'}}}
\newcommand\GdLQs{{G_{L_4,1}^{e'}}}
\newcommand\GdLCs{{G_{L_5,1}^{e'}}}
\newcommand\GdQSs{{G_{QS,1}^{e'}}}

\newcommand\GdLTu{{G_{L_3,2}^{e'}}}
\newcommand\GdLQu{{G_{L_4,2}^{e'}}}
\newcommand\GdLCu{{G_{L_5,2}^{e'}}}
\newcommand\GdQSu{{G_{QS,2}^{e'}}}

\newcommand\LQs{{\sL_{4}^{s}}}
\newcommand\LCs{{\sL_{5}^{s}}}
\newcommand\LQl{{\sL_{4}^{l}}}
\newcommand\LCl{{\sL_{5}^{l}}}
\newcommand\LT {{\sL_{3}}}

\newcommand\Mmoy{{\overline{\sM}}}
\newcommand\MmoyI{{\overline{\sM}_{e_0}}}
\newcommand\MmoyIQ{{\overline{\sM}_{e_0}/SO(2)}}

\newcommand\QSh{{QS$_{h}~$}}
\newcommand\QSb{{QS$_{b}~$}}

\begin{document}
\title[the Quasi-satellite motion revisited]{On the co-orbital motion in the \\ Planar Restricted Three-Body Problem:\\the Quasi-satellite motion revisited}

\author{Alexandre Pousse}\address[A.Pousse]{\textit{supported by the H2020-ERC project 677793 StableChaoticPlanetM}
 Dipartimento di Matematica ed Applicazioni ``R.Caccioppoli"\\ Universit\`a di Napoli ``Federico II", Monte Sant’Angelo  Via Cinthia I, 80126 Napoli, Italia}
               \email{Alexandre.Pousse@unina.it}
\author{Philippe Robutel}\author{Alain Vienne}
\address[A.Pousse, P.Robutel, A.Vienne]{              IMCCE, Observatoire de Paris, UPMC Univ. Paris 6, Univ. Lille 1, CNRS, \\
              77 Av. Denfert-Rochereau, 75014 Paris, France}              \email{Philippe.Robutel@obspm.fr}\email{Alain.Vienne@obspm.fr}
              \date{\today}



%
%
\maketitle

\begin{abstract}
In the framework of the planar and circular restricted three-body problem, we consider an asteroid that orbits the Sun in quasi-satellite motion with a planet.
A quasi-satellite trajectory is a heliocentric orbit in co-orbital resonance with the planet, characterized by a non zero eccentricity and a resonant angle that librates around zero.
Likewise, in the rotating frame with the planet it describes the same trajectory as the one of a retrograde satellite even though the planet acts as a perturbator.\\
In the last few years, the discoveries of asteroids in this type of motion made the term ``quasi-satellite"  more and more present in the literature. 
However, some authors rather use the term ``retrograde satellite" when referring to this kind of motion in the studies of the restricted problem in the rotating frame.\\
In this paper we intend to clarify the terminology to use, in order to bridge the gap between the perturbative co-orbital point of view and the more general approach in the rotating frame.
Through a numerical exploration of the co-orbital phase space, we describe the quasi-satellite domain and highlight that it is not reachable by low eccentricities by averaging process.
We will show that the quasi-satellite domain is effectively included in the domain of the retrograde satellites and neatly defined in terms of frequencies.\\
Eventually, we highlight a remarkable high eccentric quasi-satellite orbit corresponding to a frozen ellipse in the heliocentric frame.
We extend this result to the eccentric case (planet on an eccentric motion) and show that two families of frozen ellipses originate from this remarkable orbit.

 \bigskip
 

\end{abstract}

\newpage

\subsection*{Abbreviations}$~$\\
RF: Rotating frame with the planet\\
AP: Averaged problem\\
RAP: Reduced averaged problem\\
RS: Retrograde-satellite\\
TP: Tadpole\\
HS: Horseshoe\\
QS: Quasi-satellite\\
sRS : ``Satellized" retrograde satellite\\
\QSb : Binary quasi-satellite\\
\QSh : Heliocentric quasi-satellite

\subsection*{List of symbols}$~$\\
$L_1$, $L_2$, $L_3$: Circular Eulerian aligned configurations\\
$L_4$, $L_5$: Circular Lagranian equilateral configurations\\
$\LQl$, $\LCl$: In the RF, long period families that originate from $L_4$ and $L_5$.\\
$\LT$, $\LQs$, $\LCs$: In the RF, short period families that originate from $L_3$, $L_4$ and $L_5$.\\[0.1cm]
Family $f$: In the RF, one-parameter family of  simple-periodic symmetrical retrograde satellite orbits that extends from an infinitesimal neighbourhood of the planet to the collision with the Sun. For $\eps<0.0477$, it is stable but contains two particular orbits where the frequencies $\nu$ and $1-g$ are in $1:3$ resonance.  These two orbits decomposed the neighbourhood of the family $f$ in three domains: sRS,  \QSb and  \QSh. \\
$1$, $\nu$, $g$: Frequencies respectively associated with the fast variations (the mean longitudes $\lam$ and $\lam'$), the semi-fast component of the dynamics (oscillation of the resonant angle $\theta$) and the secular evolution of a trajectory (precession of the periaster argument $\omega$).\\
$\NLQ$, $\NLC$: In the RAP, the AP and the RF, families of $2\pi/\nu$-periodic orbits parametrized by $|u|\leq 0$ and that originates from $L_4$ and $L_5$. Moreover, they correspond to $\LQl$ and $\LCl$ in the RF.\\
$\GLT$, $\GLQ$, $\GLC$: In the RAP, families of fixed points parametrized by $e_0$ and that originate from $L_3$, $L_4$ and $L_5$.  In the AP and the RF, these fixed points correspond to periodic orbits of frequency respectively $g$ and $1-g$. Moreover, they correspond to $\LT$, $\LQs$ and $\LCs$ in the RF.\\
$\GQS$: In the RAP, family of fixed points parametrized by $e_0$. In the AP and the RF, these fixed points correspond to periodic orbits of frequency respectively $g$ and $1-g$. Moreover, this family corresponds to a part of the family $f$ that belongs to the \QSh domain.\\
$\GdLT$, $\GdLQ$, $\GdLC$, $\GdQS$: In the RAP, fixed points that belong to $\GLT$, $\GLQ$, $\GLC$ and $\GQS$ and  characterized by $g=0$. In the AP, sets of fixed points (also denoted as ``circles of fixed points") parametrized by $\omega(t=0)$. In the RF, sets of $2\pi$-periodic orbits parametrized by $\big(\lam'-\omega\big)_{t=0}$. \\
$\GdLTs$, $\GdLTu$, $\GdLQs$, $\GdLQu$, $\GdLCs$, $\GdLCu$, $\GdQSs$, $\GdQSu$: In the AP with $e'\geq 0$, families of fixed points that originate from the circles of fixed points 
$\GdLT$, $\GdLQ$, $\GdLC$, $\GdQS$ when $e'=0$.


\section{Introduction}

Following the discoveries, in 1899 and 1908, of the retrograde moons Phoebe and Pasiphea moving at great distances from their respective primaries Saturn and Jupiter, \cite{Ja1913} published the first study dedicated to the motion of the retrograde satellites (RS).  
Seeking to understand how a moon could still be satellized at this remote distance (close to the limit of the planet Hill's sphere), he highlighted in the Sun-Jupiter system that where ``\textit{[...] the solar forces would prohibited direct motion, [...] the solar and the Jovian forces would go hand in hand to maintain a retrograde satellite}".
Thus, by this remark the author was the first to confirm the existence and stability of remote retrograde satellite objects in the solar system.\\
Afterwards, the existence and stability of some retrograde satellite orbits far from the secondary body have also been established in the planar restricted three-body problem with two equal masses
\citep{St1933, Mo1935, He1965, He1965a}\footnote{
The two firsts are works of the Copenhagen group that extensively explored periodic orbit solutions in the planar restricted three-body problem with two equal masses. 
The two lasts are the first numerical explorations of all the solutions of the restricted three-body problem that recovered and completed the precedent works.
} and in the Earth-Moon system \citep{Br1968}.


In the framework of the Hill's approximation, \cite{He1969} extended Jackson's study and highlighted that there exists a one-parameter family of simple-periodic symmetrical retrograde satellite  orbits (denoted family $f$) that could exists beyond the Eulerian configurations $L_1$ and $L_2$.
This has been confirm in \cite{HeGu1970} in the restricted three-body problem.
The authors showed  in the rotating frame with the planet (RF), that the family $f$ extends from the retrograde satellite orbits in an infinitesimal neighbourhood of the secondary to the collision orbit with the primary. 
Besides, they pointed out that if $\eps$, the ratio  of the secondary mass over the sum of the system masses, is less than $0.0477$, the whole family is stable.
\citet{Be1974,Be1975,Be1976} extended these results by studying the stability of the neighbourhood of the family $f$ in the configuration space for $0\leq\eps\leq 1$.

\begin{figure}
\begin{center}
\small
\def\svgwidth{0.85\textwidth}
\begingroup%
  \makeatletter%
  \providecommand\color[2][]{%
    \errmessage{(Inkscape) Color is used for the text in Inkscape, but the package 'color.sty' is not loaded}%
    \renewcommand\color[2][]{}%
  }%
  \providecommand\transparent[1]{%
    \errmessage{(Inkscape) Transparency is used (non-zero) for the text in Inkscape, but the package 'transparent.sty' is not loaded}%
    \renewcommand\transparent[1]{}%
  }%
  \providecommand\rotatebox[2]{#2}%
  \ifx\svgwidth\undefined%
    \setlength{\unitlength}{640.8bp}%
    \ifx\svgscale\undefined%
      \relax%
    \else%
      \setlength{\unitlength}{\unitlength * \real{\svgscale}}%
    \fi%
  \else%
    \setlength{\unitlength}{\svgwidth}%
  \fi%
  \global\let\svgwidth\undefined%
  \global\let\svgscale\undefined%
  \makeatother%
  \begin{picture}(1,0.35081149)%
    \put(0,0){\includegraphics[width=\unitlength]{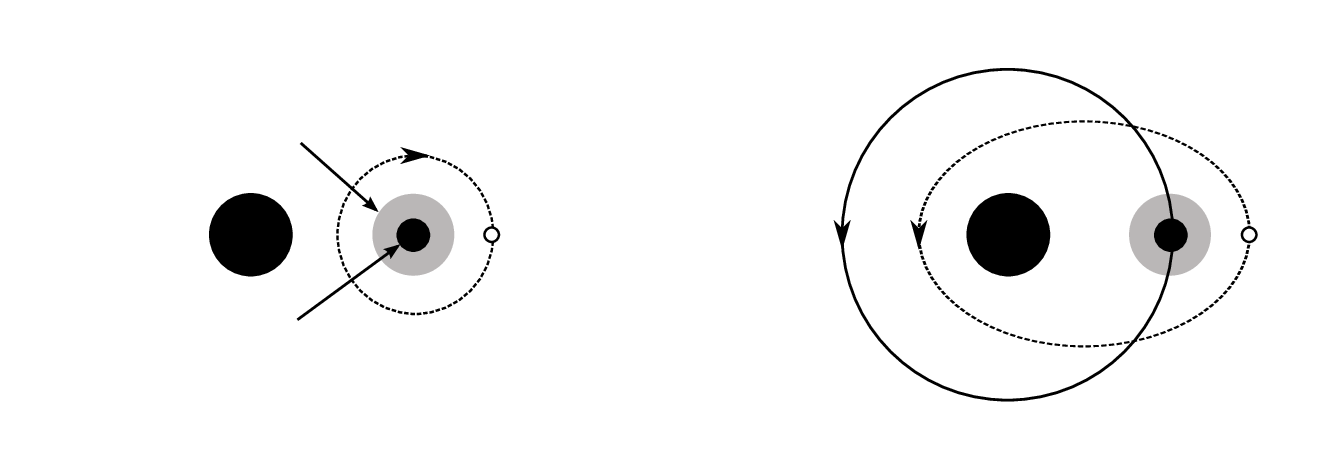}}%
    \put(0.09300874,0.16543423){\color[rgb]{0,0,0}\makebox(0,0)[lb]{\smash{Sun}}}%
    \put(0.18164794,0.09427318){\color[rgb]{0,0,0}\makebox(0,0)[lb]{\smash{Planet}}}%
    \put(0.38264669,0.16917955){\color[rgb]{0,0,0}\makebox(0,0)[lb]{\smash{Asteroid}}}%
    \put(0.05680413,0.31025321){\color[rgb]{0,0,0}\makebox(0,0)[lb]{\smash{Rotating frame with the planet (RF)}}}%
    \put(0.64481898,0.31025321){\color[rgb]{0,0,0}\makebox(0,0)[lb]{\smash{Heliocentric frame}}}%
    \put(0.05305868,0.0380934){\color[rgb]{0,0,0}\makebox(0,0)[lb]{\smash{a.}}}%
    \put(0.61985019,0.0380934){\color[rgb]{0,0,0}\makebox(0,0)[lb]{\smash{b.}}}%
    \put(0.15667915,0.24783123){\color[rgb]{0,0,0}\makebox(0,0)[lb]{\smash{Hill's sphere}}}%
  \end{picture}
\endgroup%
\caption{\small Asteroid on a quasi-satellite orbit (QS). In the rotating frame with the planet (RF) (a.), the trajectory is those of a retrograde satellite (RS) outside the planet Hill's sphere.
In the heliocentric frame (b.), the trajectory is represented by heliocentric osculating ellipses with a non zero eccentricity (in the circular case) and a resonant angle $\theta=\lam-\lam'$ that librates around zero.}
\label{DefQS}
\end{center}
\end{figure}
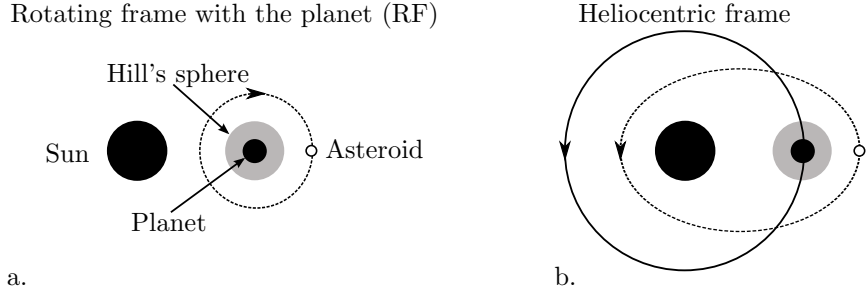

After these theoretical works, the study of the retrograde satellite orbits was addressed in a more practical point of view, with the project to inject a spacecraft in a circum-Phobos orbit.
Remark that as the Phobos Hill's sphere is too close to its surface, remote retrograde satellites  are particularly adapted trajectories.
Hence, at the end of the eighties, the terminology ``quasi-satellite"\footnote{
Let us still mention that the ``quasi-satellite" terminology has already been used in the paper of \cite{DaIp1972} but this was to describe the resonant behaviour of the near-Earth Object 1685 Toro and therefore was completely  disconnected to retrograde satellite motion.
} (QS) appeared in the USSR astrodynamicist community to define trajectories in the restricted three-body problem in rotating frame that correspond to retrograde satellite orbits outside the Hill's sphere of the secondary body (see Fig.\ref{DefQS}a).
The Phobos mission study led to the works of \cite{Ko1990} and \citet{LiVa1993,LiVa1994,LiVa1994a}.


At the end of the nineties, the quasi-satellite motion appeared in the celestial mechanics community in the view of asteroid trajectories in the solar system.\\
Let us suppose that a QS-type asteroid is far enough from the planet so that the influence of the Sun dominates its movement and therefore that the planet acts as a perturbator.
Then, its trajectory could be represented by heliocentric  osculating ellipses whose variations are governed by the influence of the planet. 
In this context, \cite{MiIn1997} remarked that the asteroid and the planet are in $1:1$ mean motion resonance and therefore that the quasi-satellite orbits correspond to a particular kind of configurations in the co-orbital resonance.
Unlike the tadpole (TP) orbits
that librate around the Lagrangian equilibria $L_4$ and $L_5$ or the horseshoes (HS)
 that encompass $L_3$, $L_4$ and $L_5$, the quasi-satellite orbits are characterized by a resonant angle $\theta = \lam - \lam'$ that librates around zero (where $\lam$ and $\lam'$ are the mean longitudes of the asteroid and the planet) and a non zero eccentricity if the planet gravitates on  a circle (see Fig.\ref{DefQS}b).
In their paper, these authors also introduced a first perturbative treatment to study the long term stability of quasi-satellites in the solar system.\\
At that time no natural object was known to orbit this configuration.
However, they suggested that, at least, the Earth and Venus could have quasi-satellite companions.
Following this work, \cite{WiInMi2000} also predicted, via a numerical investigation of the stability around the giant planets, that Uranus and Neptune could harbour QS-type asteroids whereas they did not found stable solutions for Jupiter and Saturn. \\ 
Subsequently, \cite{Na1999} and \cite{NaChMu1999} became the reference in term of co-orbital dynamics with close encounters.
Using Hill's approximation, these authors highlighted that in the spatial case, transitions between horseshoe and quasi-satellite trajectories could occurred. 
They exhibited  new kinds of compound trajectories denoted HS-QS, TP-QS or TP-QS-TP which means that there exists stable transitions exit between quasi-satellite, tadpole and horseshoe orbits. 
Later, \cite{NeThFe2002} recovered these new co-orbital structures in a global study of the co-orbital resonance phase space.
By developing a perturbative scheme using numerical averaging techniques, they showed how the tadpole, horseshoe, quasi-satellite and compound orbits vary with the asteroid eccentricity and inclination in the planar-circular, planar-eccentric and spatial-circular models.
Particularly, they showed that the higher the asteroid's eccentricity is, the larger the domain occupied by the quasi-satellite orbits in the  phase space is.\\
Eventually, the quasi-satellite long-term stability has been studied using perturbation theory in \cite{MiInWi2006} and \cite{SiNeAr2014}.
The first ones developed a practical algorithm to detect QS-type asteroids on temporary or perpetual regime, while the last ones established conditions of existence of quasi-satellite motion and also explore its different possible regimes. 


Following these theoretical works, many objects susceptible to be at least temporary quasi-satellites have been found in the solar system.
The first confirmed minor body was 2002 VE68 in co-orbital motion with Venus in \cite{MiBrWi2004}.
The Earth \citep{BrInCo2004, CoChMi2002, CoVeBr2004, DeDe2014, Wa2009, Wa2010} and Jupiter \citep{KiNa2007, WaKr2012} are the two planets with the largest number of documented QS-type objects.
Likewise, Saturn \citep{Ga2006}, Uranus \citep{Ga2006, DeDe2014}, Neptune \citep{DeDe2012} possess at least one of this type.\\
At last, let us mention that quasi-satellite motion could play a role in other celestial problems: according to \cite{Ko2005} and \citeyearpar{Ko2013}, planetesimals could be trapped in quasi-satellite motion around the protoplanet as well as interplanetary dust particles around Earth. Eventually, although no co-orbital exoplanet system has been found,  several studies on the planar planetary three-body problem
showed the existence and the stability of two co-orbital planets in quasi-satellite motion \citep{HaPsVo2009, HaVo2011, GiBeMi2010}.

	
During these last twenty years, even though the ``quasi-satellite" terminology becomes dominant in the literature, some studies use rather ``retrograde satellite" \citep{Na1999, NeThFe2002} in reference to the neighbourhood of the family $f$ in the restricted problem in rotating frame with the planet.
Hence, there exists an ambiguity in terms of terminology that is a consequence of the several approaches to describe these orbits, depending on the distance between the two co-orbitals.
One of our purposes is thus to clarify the terminology to use between ``quasi-satellite" and ``retrograde satellite".
Then, we chose to revisit the classical works on the family $f$ \citep{HeGu1970,Be1974} in the section \ref{sec:RF} and through a study on its frequencies, we show that the neighbourhood of the family is split in three different domains connected by an orbit; one corresponding to the ``satellized" retrograde satellite orbits while the two others to the quasi-satellites.
Among these two quasi-satellite domains, we identify one that is associated with asteroid trajectories in the solar system. 
This is on this last one that the paper is focussed.


An usual approach for these co-orbital trajectories in the restricted \citep{MiInWi2006,NeThFe2002,SiNeAr2014} and planetary \citep{RoPo2013} problems consists on  averaging the Hamiltonian over the fast angle of the system (the planet mean longitude)  to reduce the study of the problem to its semi-fast and secular components. 
This approach is generally denoted as the ``averaged problem" (AP).
However, as mentioned in \cite{RoPo2013} and \cite{RoNiPo2015}, this one has the important drawback to reflect poorly the dynamics close to the singularity associated with the collision with the planet.
Some quasi-satellite trajectories having close encounter with the planet, these are located close to the singularity in the averaged problem which implies that this approach would not be appropriate for them.
Thus, in order to estimate a validity limit of the averaged problem for the study of quasi-satellite motion, we also revisit the co-orbital resonance via the averaged problem.


Firstly, in the section \ref{sec:AP}, we develop the Hamiltonian formalism of the problem and introduce the averaged problem.
Subsequently, in the section \ref{sec:CC}, we focus on the circular case (i.e. planet on a circular orbit) that allows possible reduction.
We introduce the reduced averaged problem (RAP)
that seems to be the most adapted approach to understand the dynamics in the co-orbital resonance.
Focussing on quasi-satellite motion, we exhibit a family of fixed points in the reduced averaged problem  representing the family $f$ that allows us to estimates the validity limit of the averaged problem.

Next, to bridge a gap between the averaged problem and the works of \cite{HeGu1970} and \cite{Be1975}, we devote the section \ref{sec:RF} to revisit the motion in the rotating frame in the circular case in order to describe the family $f$ as well as its reachable part in the averaged problem and characterize its neighbourhood.
Through this study, we show how the quasi-satellite domain reachable in the averaged problem shrinks by increasing $\eps$. 

At last, in the section \ref{sec:EC}, we come back to the averaged problem with the aim to extend in the eccentric case (i.e. planet on an eccentric orbit) a result on co-orbital frozen ellipses that has been highlighted in section \ref{sec:FOP}.


\section{The averaged problem}
\label{sec:AP}


In the framework of the planar restricted three-body problem, we consider a primary with a mass $1-\eps$ (the Sun or a star), a secondary (a planet) with a mass $\eps$ small with respect to $1$  and a massless third body (particle or asteroid).
We assume that the planet is in elliptic Keplerian motion whose eccentricity is denoted $e'$.
Without loss of generality, we set that its semi-major axis is equal to 1 and that the argument of the periaster is equal to zero.
Likewise, we fix its orbital period to $2\pi$ (and therefore its mean motion to $1$) which imposes the gravitational constant to be equal to $1$.

In an heliocentric frame, the Hamiltonian of the problem reads
\begin{align}
\cH(\br, \brd, t) = \cH_K(\br,\brd) + \cH_P(\br, t) \label{eq:Ham1}
\end{align}
with
$$\cH_K(\br,\brd) := \frac{1}{2}\norm{\brd}^2 - \frac{1}{\norm{\br}}$$
and
$$ \cH_P(\br,t) := \eps\bigg(-\frac{1}{\norm{\br - \br'(t)}}
					 +\frac{1}{\norm{\br}} 
                     +\frac{\br \cdot \br'(t)}{\norm{\br'(t)}^3}\bigg).$$       
In this expression, $\br$ is the heliocentric position of the particle, $\brd$ its conjugated variable and $\brp(t)$ is the position of the planet at the time $t$.

In order to work with an autonomous Hamiltonian, we extend the phase space by introducing $\Lam'$, the conjugated variable of $\lam':=t$ that corresponds to the mean longitude of the planet.
As a consequence the Hamiltonian becomes, on the extended phase space, equal to $\Lam' + \cH$.

In order to define a canonical coordinate system related to the elliptic elements
$(a, e, \lam, \omega)$ (respectively semi-major axis, eccentricity, mean longitude and  argument of the periaster) and adapted to the co-orbital resonance,  we introduce the canonical coordinates $(\theta, u, -i\xb, x, \lam', \tLam')$ where  
\be
\theta := \lam - \lam' \qtext{and}
u := \sqrt{a} - 1 
\ee
are the resonant variables,
\be
x := \sqrt{\Gam}\exp(i\omega) \qtext{with} \Gam := \sqrt{a}\big(1- \sqrt{1-e^2}\big)
\ee
is the Poincar\'e's variable associated with the eccentricity $e$, and $\tLam'$ that  is the conjugated variable of $\lam'$ such as
\be
\Lam' = \tLam' - u.
\ee
If we denote $\Phi$, the canonical transformation such that
\be
\Phi : \quad \Bigg\{
	\begin{array}{ccc}
\TT\times\RR\times\CC^2\times\TT\times\RR  &\longrightarrow & \RR^{4}\times\TT\times\RR \\
(\theta,u, -i\xb,x,\lam', \tLam' )      &\longmapsto     &(\br ,\brd ,\lam', \Lam')
	\end{array}
	\nnb,
\ee
the Hamiltonian of the problem reads $\tLam' + H$ with
\be
H:=\big(\Lam' + \cH\big)\circ\Phi -\tLam'= H_K - u + H_P
\ee
where
\be
H_K := -\frac{1}{2(1+u)^2} \qtext{and} H_P := \cH_P\circ\Phi \nnb.
\ee

In these variables, the Hamiltonian possesses 3 degrees of freedom, each one corresponding to a particular component of the dynamics inside the co-orbital resonance.
Indeed, the resonant angle $\theta$ varies slowly with respect to the fast angle $\lam'$.
Thus the degree of freedom $(\theta,u)$ is generally known as the ``semi-fast" component of the dynamics while the degree of freedom $(-i\xb, x)$ is associated with the ``secular" variations of the trajectory.
As a consequence, a natural way to reduce the dimension of the problem in order to study the ``semi-fast" and ``secular" dynamics of the co-orbital motion is to average the  Hamiltonian over $\lam'$. In the following, this averaged Hamiltonian will be denoted $\overline{H}$.


	\subsection{The averaged Hamiltonian}

According to the perturbation theory, there exists a canonical transformation 
\be
\cC : \quad \Bigg\{
\begin{array}{lcl}
\TT\times\RR\times\CC^2\times\TT\times\RR                &\longrightarrow & \TT\times\RR\times\CC^2\times\TT\times\RR\\
(\und{\theta}, \und{u},-i\overline{\und{x}},\und{x}, \und{\lam'},\und{\tLam'})  &\longmapsto     &(\theta,u,-i\xb,x, \lam', \tLam') ,
\end{array}\nnb
\ee
such as, in the averaged variables $(\und{\theta}, \und{u},-i\overline{\und{x}},\und{x}, \und{\lam'},\und{\tLam'})$, the Hamiltonian reads
\be
\und{\tLam'} + {\bf{H}} = \big(\tLam' + H\big)\circ\cC \qtext{with } {\bf{H}}:= \Hb + H_*
\label{eq:canon_C}
\ee
 where
\be
\Hb:= H_K -\und{u} + \Hb_P\nnb
\ee
with
\be
\Hb_P(\und{\theta}, \und{u}, -i\und{\overline{x}}, \und{x}) :=\frac{1}{2\pi}{\displaystyle{\int_0^{2\pi}}}H_P (\und{\theta}, \und{u}, -i\und{\overline{x}}, \und{x},\lam')d\lam'\nnb.
\label{eq:moy_pert}
\ee
$H_*$ is a remainder that is supposed to be small with respect to $\Hb_P$.
More precisely, the transformation $\cC$ is close to the identity and could be construct with the time-one map of the Hamiltonian flow generated by some auxiliary function $\chi$ \citep[for further details, see][]{RoNiPo2015}.
As a consequence, if $\{f,g\}$ represents the Poisson bracket of the two functions $f$ and $g$ and if $y$ stands for one of the variables $(\theta, u, -i\xb, x, \lam', \tLam')$, then the two coordinate systems are related by
\be
y = \und{y} + \{\chi, \und{y}\} + \ode\label{eq:Tr}
\ee
with 
\be
\chi(\theta, u, -i\xb, x, \lam') = \int_0^{\lam'}\Big[\Hb_P(\theta, u, -i\xb, x) - H_P(\theta, u, -i\xb, x, \tau)\Big] \,d\tau\nnb.
\ee


In this paper, we only consider the restriction  at first order in $\eps$ of the Hamiltonian in the equation \eqref{eq:canon_C}.
This approximation of the initial problem that is described by $\Hb$ is generally known as the ``averaged problem" (AP). 
Thus, the averaged problem possesses two degrees of freedom and two parameters, $\eps$ and $e'$, respectively the planetary mass ratio and eccentricity of the planet.. \\
For the sake of clarity, the ``underdot" used to denote the averaged coordinates will be omitted below.


	\subsection{Numerical averaging}
	
	
There exists at least two classical averaging techniques  adapted to the co-orbital resonance: an analytical one based on an expansion  of the Hamiltonian in power series of the  eccentricity \citep[e.g.][]{Mo2001, RoPo2013}, and  a numerical one consisting on a numerical evaluation of $\overline{H}$ and its derivatives \citep[e.g.][]{NeThFe2002, GiBeMi2010, BeRo2001, MiInWi2006, SiNeAr2014}.
Whereas for low eccentricities the analytical technique is very efficient, reaching  higher values of eccentricity requires high order expansions which generate very heavy expressions. 
Thus, in this case, the use of numerical methods may be more convenient.
Then in order to explore the phase space of the co-orbital resonance for all eccentricities lower than one, we use the numerical averaging method developed by \cite{NeThFe2002}.\\
%
%
%
%
This method consists on a numerical evaluation of the integral (\ref{eq:moy_pert}). More generally, let $F$ be a generic function  depending on $(\theta, u, -i\xb, x,E, E')$ where $E$ and $E'$ are the eccentric anomaly of the particle and the planet.  
As its average over the mean longitude $\lam'$ is computed for a given fixed value of $\theta$,  we have $d\lam' = d\lam = \big(1-e(x)\cos E\big) dE$. 
As 
\be
\theta = \lam - \lam' = E + \omega(x) - E' - e(x)\sin E + e'\sin E', 
\ee
the eccentric anomalies $E'$ can be expressed in terms of $(\theta,E,x,e')$.
Eventually, the integrals reads
\be
\overline{F}(\theta, u, -i\xb,x) = \frac{1}{2\pi}\int_0^{2\pi} F\big(\theta, u, -i\xb,x, E, E'(\theta,E,x, e')\big)\big(1 - e(x)\cos E\big)dE,
\ee
which can be computed by discretizing the variable $E$ as $E_k = \frac{k2\pi}{N}$ and $100\leq N\leq 300$ \citep[see][for more details]{NeThFe2002}.


\section{The co-orbital resonance in the circular case ($e'=0$)}
\label{sec:CC}

In the circular case -- that is the case  where the planet gravitates on a circle --, the averaged problem defined by $\Hb$ is invariant under the action of the symmetry group $SO(2)$ associated with the rotations around the vertical axis.
Thereby, in the vicinity of the quasi-circular orbits ($|x|\ll1$), the expansion of $\Hb$ in  power series of $x$ and $\xb$ reads
\be
\sum_{(p,\pb) \in\NN^2} \Psi_{p,\pb}(\theta,u) x^p \xb^\pb
\ee
where the integers occurring in these summations satisfy the relation
\be
p - \pb = 0
\ee
that results from the d'Alembert rule.
Hence, we have
\be
\dron{\Hb}{\omega}(\theta,u,-i\xb,x) = 0 = \dot{x}{\xb} + x\dot{\xb} =\dGam,
\ee
which imposes $\Gam$ to be a first integral.
As a consequence, in the averaged problem, the two degrees of freedom of the problem are separable and a reduction is possible.\\
By fixing the value of the parameter $\Gam=|x|^2$ and eliminating the cyclic variable $\omega=\arg(x)$, we remove one degree of freedom.
We call this new problem the ``reduced averaged problem" (RAP).
However, instead of using $\Gam$ as a parameter, we introduce $e_0$ such as
\be
\Gam = (1 + u)\big(1-\sqrt{1-e^2}\big) = 1-\sqrt{1-e_0^2}  .
\ee
Then, if $u\ll1$, the parameter $e_0$ that is equal to $e + \cO(u)$ provides an approximation of the eccentricity value $e$ of the trajectory.


	\subsection{The reduced Hamiltonian}


For a given value $e_0=a$ such that $0\leq a<1$, let us define $\MmoyI\subset\TT\times\RR\times\CC^2$ the intersection of  the phase space of the averaged problem (denoted $\Mmoy\subset\TT\times\RR\times\CC^2$)  with the hyperplane $\{e_0 = a\}$, and  $\MmoyIQ$, the quotient space of this  section by the symmetry group $SO(2)$. 
Under the action of the application 
\begin{align}
\psi_{e_0}: \quad \Bigg\{
\begin{array}{ccc}
\MmoyI & \longrightarrow & \MmoyIQ\\
 (\theta, u, -i\xb, x) &\longmapsto &(\theta, u)
\end{array},\label{eq:psie0}
\end{align}
the problem is reduced to one degree of freedom and is associated with the reduced Hamiltonian 
\be\Hb_{e_0}:= \Hb\big(\,\cdot, \, \cdot, -i\xb(e_0),x(e_0)\big).\ee
%
%
Thus, for a fixed $e_0$, a trajectory in the RAP is generally a periodic orbit, but can also be a fixed point.
As a consequence,  the description of the RAP's phase portrait obtained for various values of $e_0$ allows to understand the global dynamics of the co-orbital resonance in the circular case.

The AP being more usual to illustrate the semi-fast and secular variations of the orbital elements and the rotating frame (RF) more classic to understand the dynamics of the restricted three-body problem, we will see in the next section how a given orbit is represented in these three different points of view.


	\subsection{Correspondence between the RAP, the AP and the RF}\label{sec:Intp}


For a given value of $e_0$, let us consider a periodic trajectory of frequency $\nu$ in the RAP.
The correspondence between the RAP and the AP consists in the pullback  of a trajectory belonging to $\MmoyIQ$ by the application $\psi_{e_0}^{-1}$.
However, $\omega=\arg(x)$ being ignorable in the RAP,  $\psi_{e_0}^{-1}$ is not an injection, which implies that a set of orbits in the AP parametrized by $\omega_0:= \omega(t=0)\in\TT$ is mapped by $\psi_{e_0}$ to the initial trajectory.
Furthermore, as
\be
\dot{\omega}(t) = -\dron{}{\Gam}\Hb\big(\theta(t), u(t)\big),\nnb
\ee
then $\dot{\omega}(t)$ is $2\pi/\nu$-periodic and could be decomposed such as
\be
\dot{\omega}(t) = g -\Big[\dron{}{\Gam}\Hb\big(\theta(t), u(t)\big) + g\Big]
\ee 
where
\be 
g :=\frac{\nu}{2\pi} \int_0^{2\pi/\nu} -\dron{}{\Gamma}\Hb_{e_0}\big(\theta(t),u(t)\big) dt\nnb
\ee is the secular precession frequency of $\omega$.
Thus, for each orbits of the family, the temporal evolution of the argument of its periaster is given by
\be
\omega(t) = \omega_0 + gt - \int_0^t \left[\dron{}{\Gamma}\Hb_{e_0} \big((\theta(\tau),u(\tau)\big)+ g\right] d\tau .
\ee
As a consequence, a given periodic trajectory in the RAP generally corresponds, in the AP, to a set of quasi-periodic orbits of frequencies $\nu$ and $g$.
Nevertheless, $\omega$ being ignorable when the osculating ellipses are circles (i.e. $e_0=0$), the trajectories are fixed points or periodic orbits of frequency $\nu$ in both approaches.
When $e_0>0$ and $g=0$, a periodic trajectory of the RAP provides a set of periodic orbits of frequency $\nu$ in the AP.
Likewise a fixed point corresponds to a set of degenerated fixed points. 
These fixed points being distributed along a circle in the phase space represented by the variables $(x, -i\xb)$. Their set will be describe as a ``circle of fixed points" in what follows.

Next, to connect the AP with the RF, we firstly have to apply $\cC$ to the trajectory which adds the fast frequency in the variations of the orbital elements, i.e. the planet mean motion. 
In the circular case, the d'Alembert rule  implies that $\tLam'+ H$ only depends on the angles $\lam'-\omega$ and $\theta$.
Consequently, by defining the canonical transformation 
\be
\widehat{\psi}\ : \quad \Bigg\{
	\begin{array}{ccc}
\sM      &\longrightarrow & \widehat{\psi}(\sM) \\
(\theta,u,-i\xb,x,\lam', \tLam') &\longmapsto     &(\theta,u,-i\overline{\xi},\xi,\lam', \tLam'-\Gam)
	\end{array}\nnb
\ee
with $\sM$ that corresponds to the non-averaged phase space\footnote{As we have to take into account the degree of freedom $(\lam', \tLam')$, we have $\sM\subset\TT\times\RR\times\CC^2\times\TT\times\RR$.},  $\xi=\sqrt{\Gam}\exp(i\varphi)$ and $\varphi= \lam'-\omega$,  the Hamiltonian $(\tLam' + H)\circ\widehat{\psi}^{-1}$ becomes autonomous with two degrees of freedom associated with the frequencies $\nu$ and $1-g$.
Moreover, this Hamiltonian is related to those in the RF by the pullback by $\Phi^{-1}$, that is the canonical transformation in Cartesian coordinates.
Thus, a trajectory in the RF is generally quasi-periodic with two frequencies. 
As a consequence, a given trajectory of the RAP generally corresponds to a set of orbits in the RF parametrized by $\varphi_0:=\varphi(t=0)\in\TT$ with one more frequency.

For the sake of clarity, we summarize the status of the remarkable orbits in the three different approaches in the table \ref{tab:Orb}.

	\begin{table}
\begin{center}
\small
\begin{tabular}{|C{1.6cm}||C{1.5cm}|C{1.5cm}|C{1.5cm}|C{1.5cm}|C{1.5cm}|C{1.5cm}|}
\hline \multirow{2}*{\textbf{Approach}}  &    \multicolumn{2}{c|}{$e_0=0$} & \multicolumn{4}{c|}{$e_0>0$}  \\
\cline{4-7}               &  \multicolumn{2}{c|}{}             & \multicolumn{2}{c|}{$g\neq 0$} & \multicolumn{2}{c|}{$g= 0$} \\  
\hline  \multirow{2}*{\textbf{RAP}}     & FP        & PO     & FP       &  PO         &  FP     & PO     \\
\multirow{2}*{$\downarrow  $ }                               &           & $(\nu)$&          & $(\nu)$     &         & $(\nu)$\\
\cline{2-7}  \multirow{2}*{\textbf{AP}}      & FP        & PO     & $S_{\omega_0}$PO       &  $S_{\omega_0}$QPO        &  $S_{\omega_0}$FP     & $S_{\omega_0}$PO     \\
\multirow{2}*{$\downarrow  $ }                           &           & $(\nu)$&$(g)$     & $(\nu,g)$   &         & $(\nu)$\\
\cline{2-7}  \multirow{2}*{\textbf{RF}}      & FP        & PO     & $S_{\varphi_0}$PO       &  $S_{\varphi_0}$QPO        &  $S_{\varphi_0}$PO     & $S_{\varphi_0}$QPO     \\
                               &           & $(\nu)$&  $(1-g)$        & $(\nu,1-g)$& $(1)$ & $(\nu, 1)$\\ \hline
\end{tabular}
\smallskip
\caption{\small Correspondence between the three approaches for a given trajectory in the RAP. $S_{\omega_0}$,$S_{\varphi_0}$: set of solutions parametrized by  $\omega_0$ and $\varphi_0\in\TT$. \textbf{FP}: Fixed point. \textbf{PO}: Periodic orbit. \textbf{QPO}: Quasi-periodic orbit. Parenthesis: associated frequencies. }\label{tab:Orb}
\end{center}
\end{table}


	\subsection{Phase portraits of the RAP}\label{sec:Phase}
	
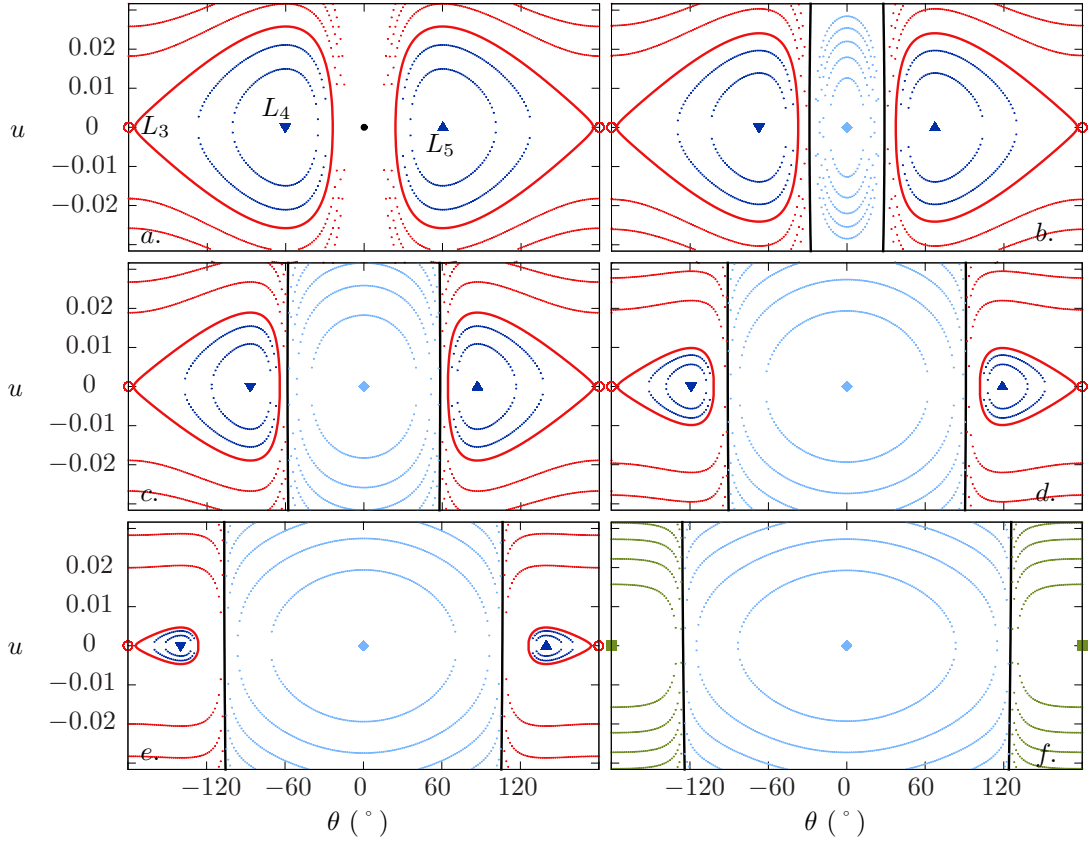
\begin{figure}
\begin{center}
\small
\def\svgwidth{1.\textwidth}
\begingroup%
  \makeatletter%
  \providecommand\color[2][]{%
    \errmessage{(Inkscape) Color is used for the text in Inkscape, but the package 'color.sty' is not loaded}%
    \renewcommand\color[2][]{}%
  }%
  \providecommand\transparent[1]{%
    \errmessage{(Inkscape) Transparency is used (non-zero) for the text in Inkscape, but the package 'transparent.sty' is not loaded}%
    \renewcommand\transparent[1]{}%
  }%
  \providecommand\rotatebox[2]{#2}%
  \ifx\svgwidth\undefined%
    \setlength{\unitlength}{752bp}%
    \ifx\svgscale\undefined%
      \relax%
    \else%
      \setlength{\unitlength}{\unitlength * \real{\svgscale}}%
    \fi%
  \else%
    \setlength{\unitlength}{\svgwidth}%
  \fi%
  \global\let\svgwidth\undefined%
  \global\let\svgscale\undefined%
  \makeatother%
  \begin{picture}(1,0.78723404)%
    \put(0,0){\includegraphics[width=\unitlength]{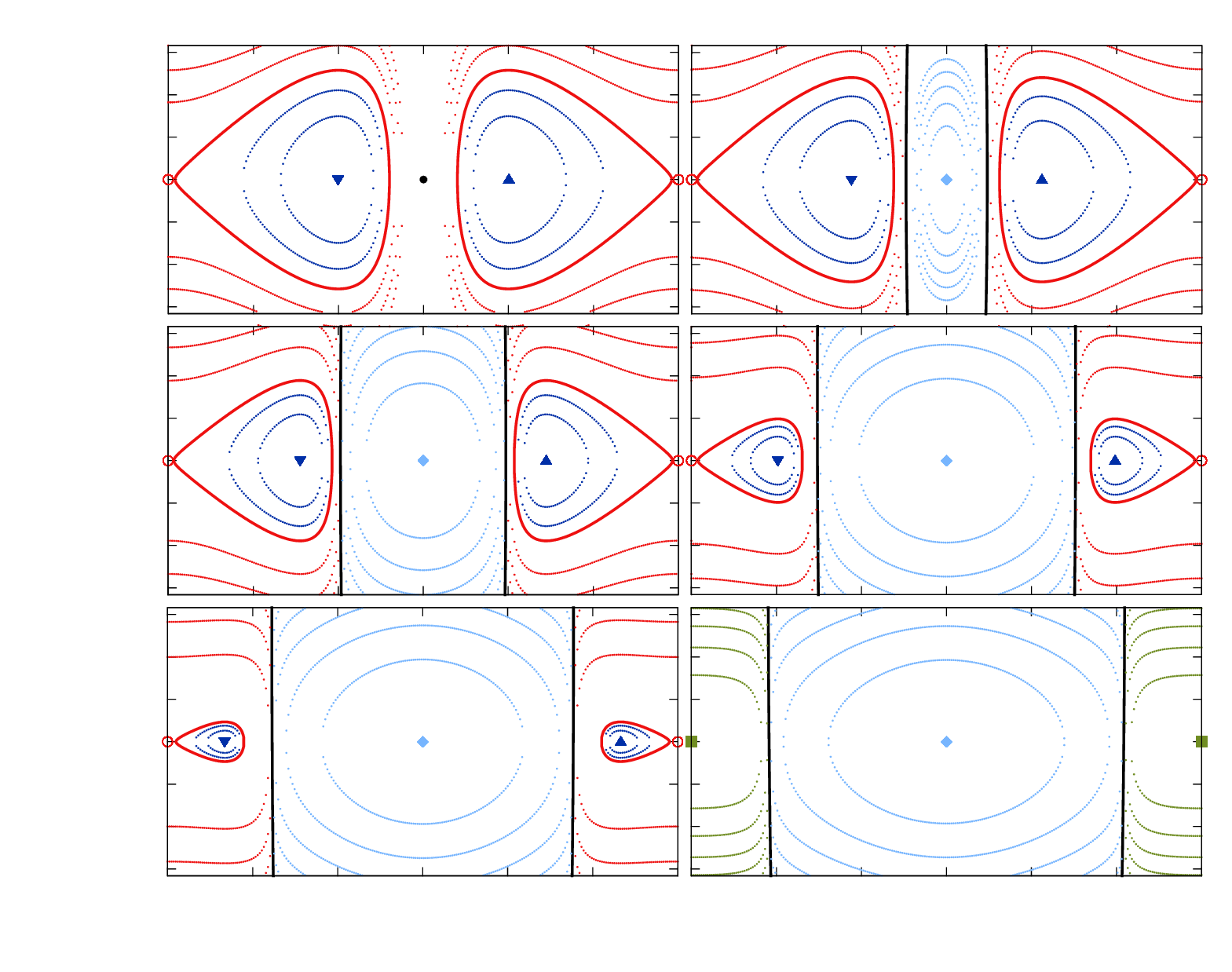}}%
    \put(0.06783473,0.56884895){\makebox(0,0)[lb]{\smash{$-0.02$}}}%
    \put(0.06783473,0.60328837){\makebox(0,0)[lb]{\smash{$-0.01$}}}%
    \put(0.08374001,0.67223991){\makebox(0,0)[lb]{\smash{$0.01$}}}%
    \put(0.08374001,0.70667932){\makebox(0,0)[lb]{\smash{$0.02$}}}%
    \put(0.25355303,0.65395404){\color[rgb]{0,0,0}\makebox(0,0)[lb]{\smash{$L_4$}}}%
    \put(0.39784131,0.62203915){\color[rgb]{0,0,0}\makebox(0,0)[lb]{\smash{$L_5$}}}%
    \put(0.18235921,0.05609273){\makebox(0,0)[lb]{\smash{$-120$}}}%
    \put(0.25609487,0.05609273){\makebox(0,0)[lb]{\smash{$-60$}}}%
    \put(0.33478066,0.05609273){\makebox(0,0)[lb]{\smash{$0$}}}%
    \put(0.39682894,0.05609273){\makebox(0,0)[lb]{\smash{$60$}}}%
    \put(0.46139115,0.05609273){\makebox(0,0)[lb]{\smash{$120$}}}%
    \put(0.60776163,0.05612821){\makebox(0,0)[lb]{\smash{$-120$}}}%
    \put(0.68149729,0.05612821){\makebox(0,0)[lb]{\smash{$-60$}}}%
    \put(0.76018307,0.05612821){\makebox(0,0)[lb]{\smash{$0$}}}%
    \put(0.82223135,0.05612821){\makebox(0,0)[lb]{\smash{$60$}}}%
    \put(0.88679356,0.05612821){\makebox(0,0)[lb]{\smash{$120$}}}%
    \put(0.14717004,0.5419235){\color[rgb]{0,0,0}\makebox(0,0)[lb]{\smash{$a.$}}}%
    \put(0.93448523,0.54352821){\color[rgb]{0,0,0}\makebox(0,0)[lb]{\smash{$b.$}}}%
    \put(0.14717004,0.31426351){\color[rgb]{0,0,0}\makebox(0,0)[lb]{\smash{$c.$}}}%
    \put(0.93440439,0.31480386){\color[rgb]{0,0,0}\makebox(0,0)[lb]{\smash{$d.$}}}%
    \put(0.14717004,0.08660403){\color[rgb]{0,0,0}\makebox(0,0)[lb]{\smash{$e.$}}}%
    \put(0.93440439,0.08501672){\color[rgb]{0,0,0}\makebox(0,0)[lb]{\smash{$f.$}}}%
    \put(0.09849547,0.63779531){\makebox(0,0)[lb]{\smash{$0$}}}%
    \put(0.03248916,0.63489088){\color[rgb]{0,0,0}\makebox(0,0)[lb]{\smash{$u$}}}%
    \put(0.06570707,0.34118936){\makebox(0,0)[lb]{\smash{$-0.02$}}}%
    \put(0.06570707,0.37562878){\makebox(0,0)[lb]{\smash{$-0.01$}}}%
    \put(0.08161235,0.44458034){\makebox(0,0)[lb]{\smash{$0.01$}}}%
    \put(0.08161235,0.47901975){\makebox(0,0)[lb]{\smash{$0.02$}}}%
    \put(0.09636779,0.41013572){\makebox(0,0)[lb]{\smash{$0$}}}%
    \put(0.0303615,0.40723127){\color[rgb]{0,0,0}\makebox(0,0)[lb]{\smash{$u$}}}%
    \put(0.31106141,0.02448512){\color[rgb]{0,0,0}\makebox(0,0)[lb]{\smash{$\theta$ $(\degre)$}}}%
    \put(0.73110061,0.02474781){\color[rgb]{0,0,0}\makebox(0,0)[lb]{\smash{$\theta$ $(\degre)$}}}%
    \put(0.14717005,0.63799659){\color[rgb]{0,0,0}\makebox(0,0)[lb]{\smash{$L_3$}}}%
    \put(0.06570707,0.11352912){\makebox(0,0)[lb]{\smash{$-0.02$}}}%
    \put(0.06570707,0.14796868){\makebox(0,0)[lb]{\smash{$-0.01$}}}%
    \put(0.08161235,0.21692055){\makebox(0,0)[lb]{\smash{$0.01$}}}%
    \put(0.08161235,0.25136008){\makebox(0,0)[lb]{\smash{$0.02$}}}%
    \put(0.09636779,0.18247576){\makebox(0,0)[lb]{\smash{$0$}}}%
    \put(0.0303615,0.17957131){\color[rgb]{0,0,0}\makebox(0,0)[lb]{\smash{$u$}}}%
  \end{picture}%
\endgroup%
\caption{\small Phase portraits of a Sun-Jupiter like system  in the circular case. For a, b, c, d, e and f, $e_0$ is equal to $0$, $0.25$, $0.5$, $0.75$, $0.85$ and $0.95$ . 
The black dot (a.) and curves represent the collision with the planet. 
The blue, sky blue and red dots are level curves of TP, QS and HS orbits.
For $e_0=0$, the blue triangles and red circles represents $L_4$, $L_5$ and $L_3$, while for $e_0>0$ they form the families $\GLQ$, $\GLC$ and $\GLT$.
From $L_3$ and the unstable part of $\GLT$ originates a separatrix  that is represented by a red curve.
The sky blue diamonds form the family $\GQS$.
Eventually the green squares represents the stable part of $\GLT$ around which trajectories represented by green dots librate.}\label{fig:phase}
\end{center}
\end{figure}	
	
	
The figure \ref{fig:phase} displays the phase portraits of the RAP associated with  six different values of the parameter $e_0$ for a Sun-Jupiter like system ($\eps =0.001$). 

%
%
%
%
In Fig.\ref{fig:phase}a, $e_0$ is equal to zero: the osculating ellipses of all the orbits are circles.  
The singular point located at $\theta = u = 0$ corresponds to the collision between the asteroid and the planet, where $\Hb$ is not defined (the integral  (\ref{eq:moy_pert}) is divergent).  
The two elliptic fixed points, in $(\theta, u) =  (\pm 60 \degre,0)$,  correspond to the Lagrangian equilateral configurations $L_4$ and $L_5$ whereas the hyperbolic fixed point, close to $(\theta, u) =  (180 \degre,0)$, is associated with the Eulerian aligned configuration $L_3$. \\
On the phase portraits described by \cite{NeThFe2002} two additional  equilibria appears located at $\theta=0\degre$: the Eulerian aligned configurations $L_1$ and $L_2$. 
But as it has been shown in \cite{RoPo2013}, there exists a neighbourhood of the collision singularity inside which the averaged Hamiltonian does not reflect properly the dynamics of the ``initial" problem.
Indeed,  a remainder which depends on the fast variable and that is supposed to be small with respect to $\Hb_P$  is generated by the averaging process; 
we denoted it $H_*$ in the expression (\ref{eq:canon_C}).
Although $H_*$ is equal to $\ode$ in the major part of the phase space, when the distance to the collision is of order  $\eps^{1/3}$ and less, $H_*$ is at least of the same order than the perturbation  $\Hb_P$ \citep{RoNiPo2015}.
Thus, this define an ``exclusion zone" inside which the trajectories, and especially the equilibria $L_1$ and $L_2$, fall outside the scope of the averaged Hamiltonian.

%
The orbits that librate around $L_4$ or $L_5$ lying inside the separatrix originated from $L_3$ correspond to the tadpoles (TP) orbits.
For $e_0=0$, these two domains form two families of $2\pi/\nu$-periodic orbits originating in $L_4$ and $L_5$ and that are parametrized by $u\geq 0$. 
We denote them $\NLQ$ and $\NLC$. 
More precisely, they are the Lyapounov families of the Lagrangian equilateral configurations associated with the libration and generally known as the long period families $\LQl$ and $\LCl$  in the RF \citep[see][]{MeHa1992}.
Eventually, outside the separatrix lies the horseshoe (HS) domain: the orbits that encompass the three equilibria $L_3$, $L_4$ and $L_5$.

%
%
%
\begin{figure}
\begin{center}
\small
\def\svgwidth{0.85\textwidth}
\begingroup%
  \makeatletter%
  \providecommand\color[2][]{%
    \errmessage{(Inkscape) Color is used for the text in Inkscape, but the package 'color.sty' is not loaded}%
    \renewcommand\color[2][]{}%
  }%
  \providecommand\transparent[1]{%
    \errmessage{(Inkscape) Transparency is used (non-zero) for the text in Inkscape, but the package 'transparent.sty' is not loaded}%
    \renewcommand\transparent[1]{}%
  }%
  \providecommand\rotatebox[2]{#2}%
  \ifx\svgwidth\undefined%
    \setlength{\unitlength}{785.43127415bp}%
    \ifx\svgscale\undefined%
      \relax%
    \else%
      \setlength{\unitlength}{\unitlength * \real{\svgscale}}%
    \fi%
  \else%
    \setlength{\unitlength}{\svgwidth}%
  \fi%
  \global\let\svgwidth\undefined%
  \global\let\svgscale\undefined%
  \makeatother%
  \begin{picture}(1,0.52477539)%
    \put(0,0){\includegraphics[width=\unitlength]{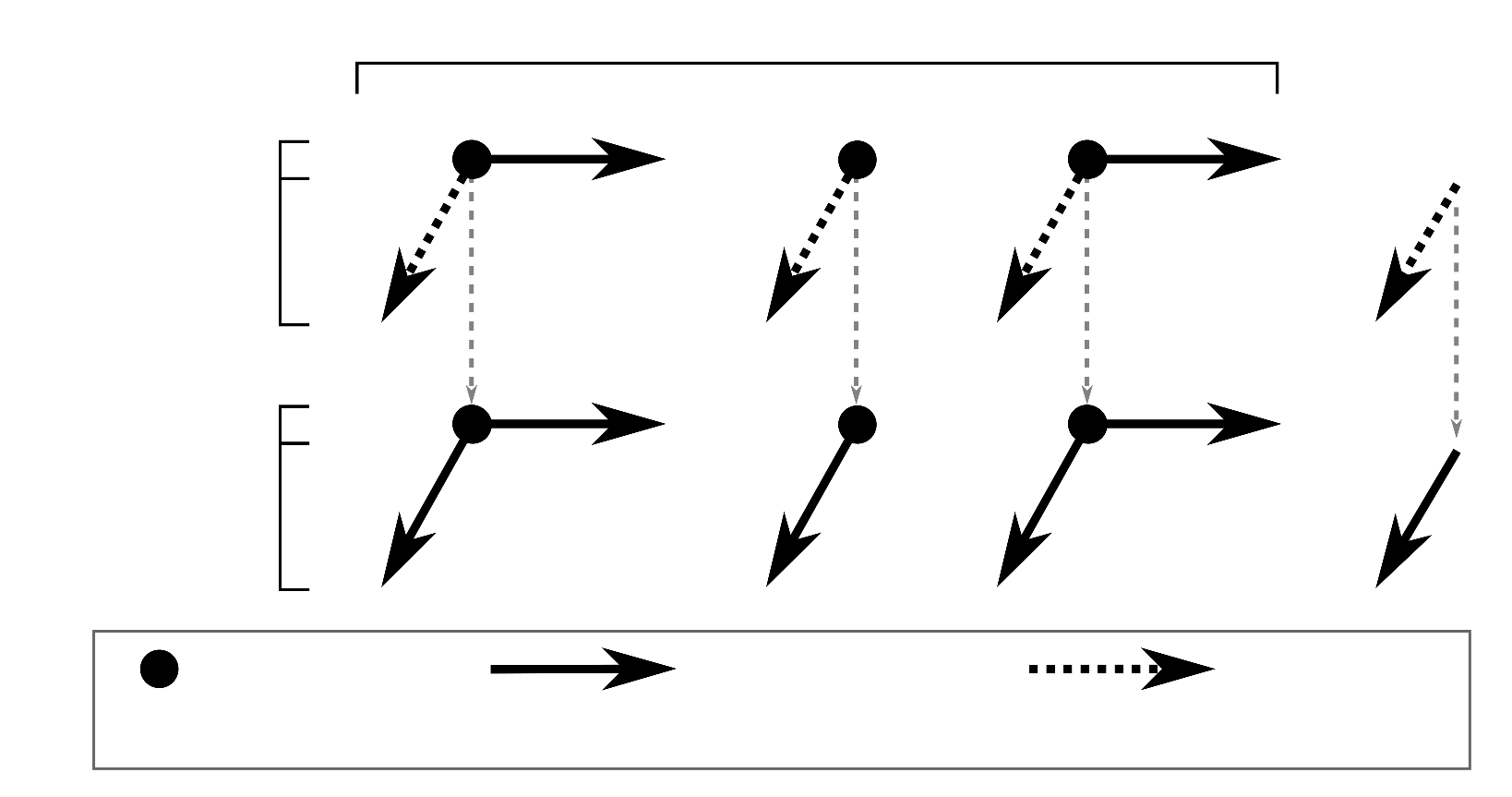}}%
    \put(0.0908876,0.03439255){\color[rgb]{0,0,0}\makebox(0,0)[lb]{\smash{Fixed point}}}%
    \put(0.32006105,0.03834142){\color[rgb]{0,0,0}\makebox(0,0)[lb]{\smash{Periodic orbit family}}}%
    \put(0.6765531,0.03834142){\color[rgb]{0,0,0}\makebox(0,0)[lb]{\smash{Fixed point family}}}%
    \put(0.29321474,0.44718322){\color[rgb]{0,0,0}\makebox(0,0)[lb]{\smash{$L_4$}}}%
    \put(0.54785191,0.44718322){\color[rgb]{0,0,0}\makebox(0,0)[lb]{\smash{$L_3$}}}%
    \put(0.70063421,0.44718322){\color[rgb]{0,0,0}\makebox(0,0)[lb]{\smash{$L_5$}}}%
    \put(0.74212705,0.38839565){\color[rgb]{0,0,0}\makebox(0,0)[lb]{\smash{$\NLC$}}}%
    \put(0.20732249,0.19690869){\color[rgb]{0,0,0}\makebox(0,0)[lb]{\smash{$(\LQs)$}}}%
    \put(0.32922253,0.2111683){\color[rgb]{0,0,0}\makebox(0,0)[lb]{\smash{$(\LQl)$}}}%
    \put(0.46198858,0.19690869){\color[rgb]{0,0,0}\makebox(0,0)[lb]{\smash{$(\LT)$}}}%
    \put(0.61478821,0.19690869){\color[rgb]{0,0,0}\makebox(0,0)[lb]{\smash{$(\LCs)$}}}%
    \put(0.73805312,0.20913119){\color[rgb]{0,0,0}\makebox(0,0)[lb]{\smash{$(\LCl)$}}}%
    \put(0.88982548,0.19509823){\color[rgb]{0,0,0}\makebox(0,0)[lb]{\smash{$(f)$}}}%
    \put(0,0.18803035){\color[rgb]{0,0,0}\makebox(0,0)[lb]{\smash{\textbf{AP \& RF}
}}}%
    \put(0.33466177,0.38839565){\color[rgb]{0,0,0}\makebox(0,0)[lb]{\smash{$\NLQ$}}}%
    \put(0.2113968,0.37617334){\color[rgb]{0,0,0}\makebox(0,0)[lb]{\smash{$\GLQ$}}}%
    \put(0.46606279,0.3761734){\color[rgb]{0,0,0}\makebox(0,0)[lb]{\smash{$\GLT$}}}%
    \put(0.6188624,0.37617334){\color[rgb]{0,0,0}\makebox(0,0)[lb]{\smash{$\GLC$}}}%
    \put(0.87352857,0.37617334){\color[rgb]{0,0,0}\makebox(0,0)[lb]{\smash{$\GQS$}}}%
    \put(0.03844179,0.36525775){\color[rgb]{0,0,0}\makebox(0,0)[lb]{\smash{\textbf{RAP}
}}}%
    \put(0.0861032,0.41284109){\color[rgb]{0,0,0}\makebox(0,0)[lb]{\smash{$e_0=0$}}}%
    \put(0.0861032,0.32015316){\color[rgb]{0,0,0}\makebox(0,0)[lb]{\smash{$e_0>0$}}}%
    \put(0.44737964,0.49273941){\color[rgb]{0,0,0}\makebox(0,0)[lb]{\smash{Lyapounov families}}}%
    \put(0.0861032,0.23765071){\color[rgb]{0,0,0}\makebox(0,0)[lb]{\smash{$e_0=0$}}}%
    \put(0.0861032,0.14088858){\color[rgb]{0,0,0}\makebox(0,0)[lb]{\smash{$e_0>0$}}}%
  \end{picture}%
\endgroup%
\caption{\small Representation of the co-orbital families of periodic orbits from the three different points of view. From each Lagrangian triangular equilibrium originates two Lyapounov families that correspond to a periodic orbit family and a fixed point family in the RAP. These families are associated with the long and short period families in the AP and the RF. $L_3$ being a saddle center type in the RAP, only one Lyapounov family emanates from this equilibrium that is a fixed point family in the RAP and a periodic orbit family in the AP and RF. Eventually, for $e_0>0$, there exists a family of fixed points in the RAP that is not a Lyapounov family: $\GQS$. This family is associated with a periodic orbit family in the RF: the family $f$.}\label{fig:POF}
\end{center}
\end{figure}
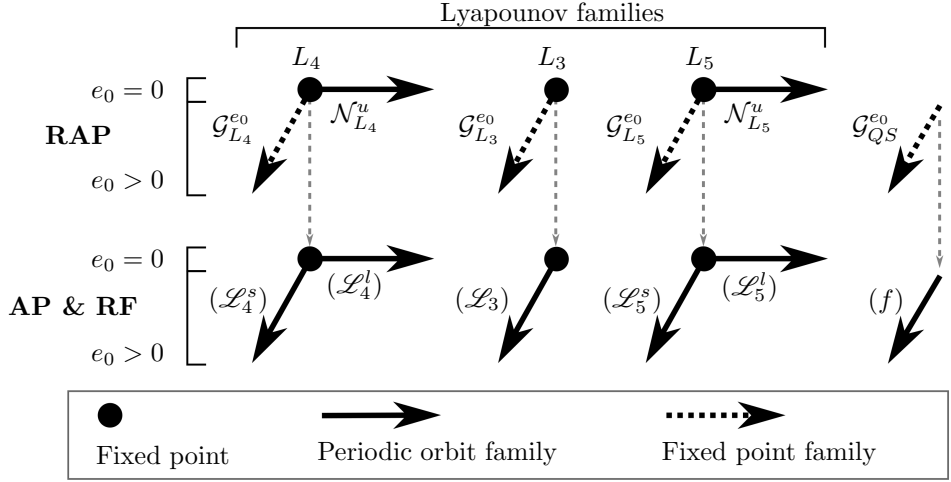
%
%
If, when $e_0=0$, the domain of definition of $\Hb_{e_0}$ excludes the origin $\theta=u = 0$,  the location of its singularities (associated with the collision with the planet) evolves with the parameter $e_0$. 
Indeed, as soon as $e_0>0$, the origin becomes a regular point while the set of singular points describes a curve that surrounds the origin.  
The phase space is now divided in two different domains.\\
For small $e_0$ (for example $e_0=0.25$ represented in Fig.\ref{fig:phase}b), the domain outside the collision curve has the same topology as for $e_0=0$: two stable equilibria close to the $L_4$ and $L_5$'s location and a separatrix emerging from an hyperbolic fixed point close to $L_3$ that bounds the TP and the HS domains.
However, contrarily to $e_0=0$, the fixed points do not correspond to equilibria in the AP and the RF but to periodic orbits of frequency respectively $g$ and $1-g$.
Consequently, orbits in their vicinity correspond to quasi-periodic orbits.
Thus, by varying $e_0$, these fixed points form three one-parameter families that we denote $\GLT$, $\GLQ$ and $\GLC$.
In the RF, these ones are known as the short period families $\LQs$, $\LCs$ and $\LT$, the Lyapounov families associated with the precession, that emanate from $L_4$, $L_5$ and $L_3$ \citep[see][]{MeHa1992}.\\
Inside the collision curve appears a new domain containing orbits that librate around a fixed point of coordinates close to the origin: the QS domain.
By varying $e_0$, the fixed points form a one-parameter family characterized by $\theta=0\degre$ and that originates from the singular point for $e_0=0$;
we denote it $\GQS$.
In the RF, these fixed points correspond\footnote{See the section \ref{sec:Intp}.} to periodic retrograde satellite orbits of frequency $1-g$ .
As a consequence, the family $\GQS$ is related to the family $f$ that is\footnote{See the section \ref{sec:RF} for further details on the family $f$.} the one-parameter family of simple-periodic symmetrical retrograde satellite orbits.\\
%
%
%
Thus, for small eccentricities, TP, HS and QS domains are structured around two periodic orbit families ($\NLQ$ and $\NLC$) and  four fixed point families ($\GLT$, $\GLQ$, $\GLC$ and $\GQS$) that we outline in Fig.\ref{fig:POF} to clarify their representations in the different approaches.
%
%
%
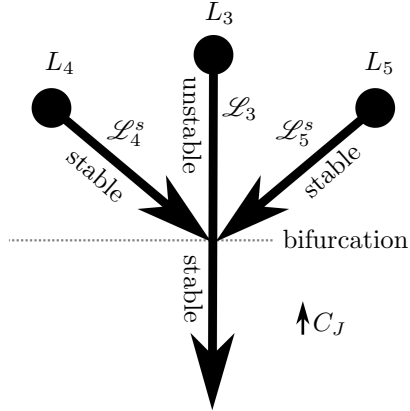
\begin{figure}
\begin{center}
\small
\def\svgwidth{0.5\textwidth}
\begingroup%
  \makeatletter%
  \providecommand\color[2][]{%
    \errmessage{(Inkscape) Color is used for the text in Inkscape, but the package 'color.sty' is not loaded}%
    \renewcommand\color[2][]{}%
  }%
  \providecommand\transparent[1]{%
    \errmessage{(Inkscape) Transparency is used (non-zero) for the text in Inkscape, but the package 'transparent.sty' is not loaded}%
    \renewcommand\transparent[1]{}%
  }%
  \providecommand\rotatebox[2]{#2}%
  \ifx\svgwidth\undefined%
    \setlength{\unitlength}{512bp}%
    \ifx\svgscale\undefined%
      \relax%
    \else%
      \setlength{\unitlength}{\unitlength * \real{\svgscale}}%
    \fi%
  \else%
    \setlength{\unitlength}{\svgwidth}%
  \fi%
  \global\let\svgwidth\undefined%
  \global\let\svgscale\undefined%
  \makeatother%
  \begin{picture}(1,0.78125)%
    \put(0,0){\includegraphics[width=\unitlength]{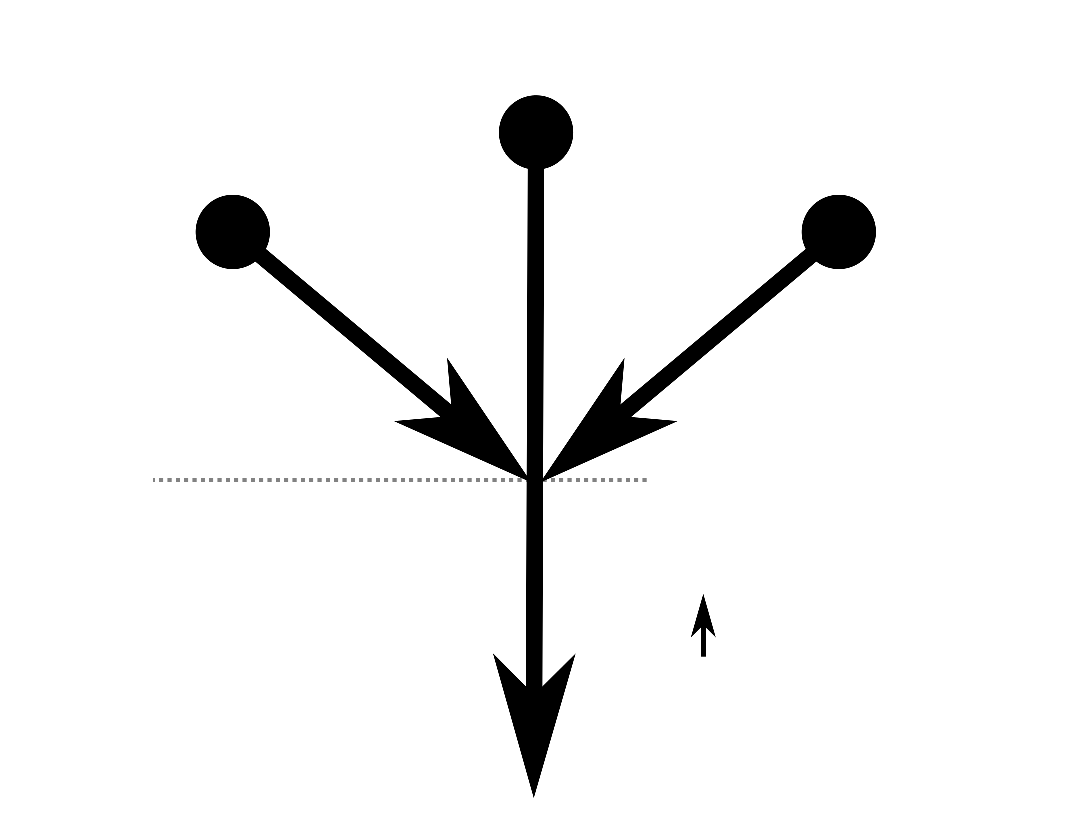}}%
    \put(0.315625,0.509375){\color[rgb]{0,0,0}\makebox(0,0)[lb]{\smash{$\LQs$}}}%
    \put(0.51875,0.55){\color[rgb]{0,0,0}\makebox(0,0)[lb]{\smash{$\LT$}}}%
    \put(0.4875,0.71875){\color[rgb]{0,0,0}\makebox(0,0)[lb]{\smash{$L_3$}}}%
    \put(0.20625,0.63125){\color[rgb]{0,0,0}\makebox(0,0)[lb]{\smash{$L_4$}}}%
    \put(0.76875,0.63125){\color[rgb]{0,0,0}\makebox(0,0)[lb]{\smash{$L_5$}}}%
    \put(0.6125,0.509375){\color[rgb]{0,0,0}\makebox(0,0)[lb]{\smash{$\LCs$}}}%
    \put(0.24203809,0.47512563){\color[rgb]{0,0,0}\rotatebox{-41.14297817}{\makebox(0,0)[lb]{\smash{stable}}}}%
    \put(0.67528726,0.40428322){\color[rgb]{0,0,0}\rotatebox{41.43048196}{\makebox(0,0)[lb]{\smash{stable}}}}%
    \put(0.45138968,0.30618701){\color[rgb]{0,0,0}\rotatebox{-89.60275888}{\makebox(0,0)[lb]{\smash{stable}}}}%
    \put(0.45056681,0.62040441){\color[rgb]{0,0,0}\rotatebox{-90.33965996}{\makebox(0,0)[lb]{\smash{unstable}}}}%
    \put(0.678125,0.171875){\color[rgb]{0,0,0}\makebox(0,0)[lb]{\smash{$C_{J}$}}}%
    \put(0.625,0.31875){\color[rgb]{0,0,0}\makebox(0,0)[lb]{\smash{bifurcation}}}%
  \end{picture}%
\endgroup%
\caption{\small Representation of the result of \cite{DeJaPa1967} in the RF: the merge of the short period families $\LQs$ and $\LCs$ with $\LT$ and bifurcation of the latter that becomes stable.}\label{fig:bif}
\end{center}
\end{figure}
	
For higher values of $e_0$ (see Fig.\ref{fig:phase}c, d, e and f), the topology of the phase portraits does not change inside the collision curve: the QS domain is always present, but its size increases until it dominates the phase portrait for high eccentricity values.
Outside the collision curve, the situation is different.
As $e_0$ increases, the two stable equilibria get closer  to the hyperbolic fixed point, which implies that the TP domains shrink and vanish when the three merge. 
This bifurcation generates a new domain new domain inside of which the orbits librate around the fixed point close to $(\theta,u)=(180\degre,0)$ (see Fig.\ref{fig:phase}f).
A similar result was found by \cite{DeJaPa1967} for an Earth-Moon like system in the circular case ($\eps = 1/81$).
In the RF, the authors showed that the short period families $\LQs$ and $\LCs$ terminate on a periodic orbit of $\LT$ (see the outline in Fig.\ref{fig:bif}).


Now, let us focus on the QS domain.
As mentioned above, there exists an exclusion zone in the vicinity of the collision curve such that the QS orbits does not represent ``real" trajectories of the initial problem. 
For high eccentricities, the QS dominates the phase portraits; the size of the intersection between the QS domain and the exclusion zone is small relatively to the whole domain.  
However by decreasing $e_0$, the QS domain shrinks with  the collision curve.
As a consequence, the relative size of the intersection increases until a critical value of $e_0$ where the exclusion zone contains all the QS orbits. 
In this case, the AP and a fortiori the RAP are not relevant to study the QS motion.\\
%
%
%
A simple way to estimate  a validity limit of theses two approaches is to consider that the whole QS domain is excluded if and only if $\GQS$ is inside  the exclusion zone.
Thus the study of the fixed points family $\GQS$ allows to determinate the eccentricity value under which the averaging method  cannot be applied to QS motion.


	\subsection{Fixed point families of the RAP}
	\label{sec:FOP}

For a given value of $e_0$, let us consider a fixed point of the RAP, denoted $(\theta_0,u_0)$, such as
\be \dron{}{\theta}\Hb_{e_0}(\theta_0,u_0) = 0 \qtext{and} \dron{}{u}\Hb_{e_0}(\theta_0,u_0) = 0.\ee
The linear stability of this fixed point, is deduced from the eigenvalues of the matrix\footnote{In practice, the matrix $\cM(\theta_0,u_0)$ is provided by a numerical differentiation of the equations of motion at the fixed point $(\theta_0,u_0)$.} 
\be
\cM :=\begin{pmatrix}
\dronss{\Hb_{e_0}}{\theta}{u} & \drons{\Hb_{e_0}}{u}\\
-\drons{\Hb_{e_0}}{\theta} & -\dronss{\Hb_{e_0}}{\theta}{u}
\end{pmatrix}\nnb
\ee
that comes from the variational equations
\be
\begin{pmatrix}
\dot{\theta}\\\dot{u}
\end{pmatrix} = \cM(\theta_0,u_0)\begin{pmatrix}
\theta\\u
\end{pmatrix}\nnb
\ee
associated with the linearization of the equations of motion in the vicinity of $(\theta_0,u_0)$.
When this fixed point is elliptic, its eigenvalues are equal to $\pm i\nu$, where the real number $\nu$ is the rotation frequency around the equilibrium.
Moreover, the secular precession frequency of its corresponding orbits in the AP is equal to 
\be g=-\dron{}{\Gamma}\Hb_{e_0}(\theta_0,u_0) .\ee

%
%
%
%
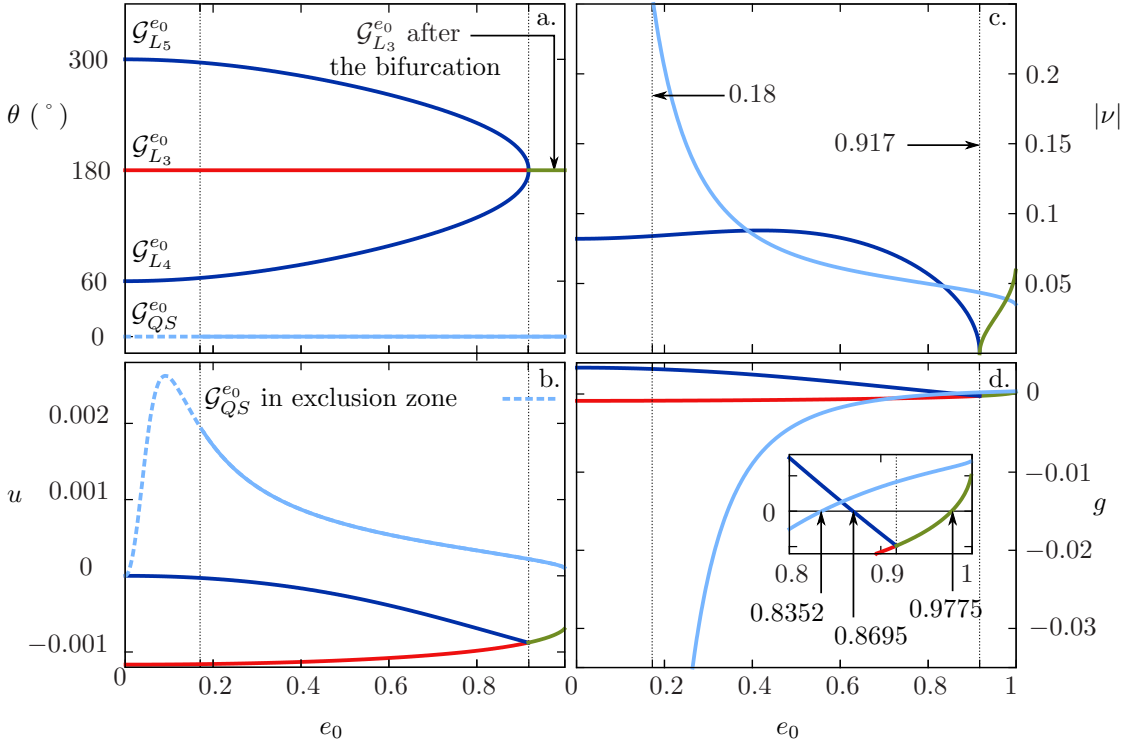
\begin{figure}
\begin{center}
\small
\def\svgwidth{1.05\textwidth}
\begingroup%
  \makeatletter%
  \providecommand\color[2][]{%
    \errmessage{(Inkscape) Color is used for the text in Inkscape, but the package 'color.sty' is not loaded}%
    \renewcommand\color[2][]{}%
  }%
  \providecommand\transparent[1]{%
    \errmessage{(Inkscape) Transparency is used (non-zero) for the text in Inkscape, but the package 'transparent.sty' is not loaded}%
    \renewcommand\transparent[1]{}%
  }%
  \providecommand\rotatebox[2]{#2}%
  \ifx\svgwidth\undefined%
    \setlength{\unitlength}{827.96875bp}%
    \ifx\svgscale\undefined%
      \relax%
    \else%
      \setlength{\unitlength}{\unitlength * \real{\svgscale}}%
    \fi%
  \else%
    \setlength{\unitlength}{\svgwidth}%
  \fi%
  \global\let\svgwidth\undefined%
  \global\let\svgscale\undefined%
  \makeatother%
  \begin{picture}(1,0.63038309)%
    \put(0,0){\includegraphics[width=\unitlength]{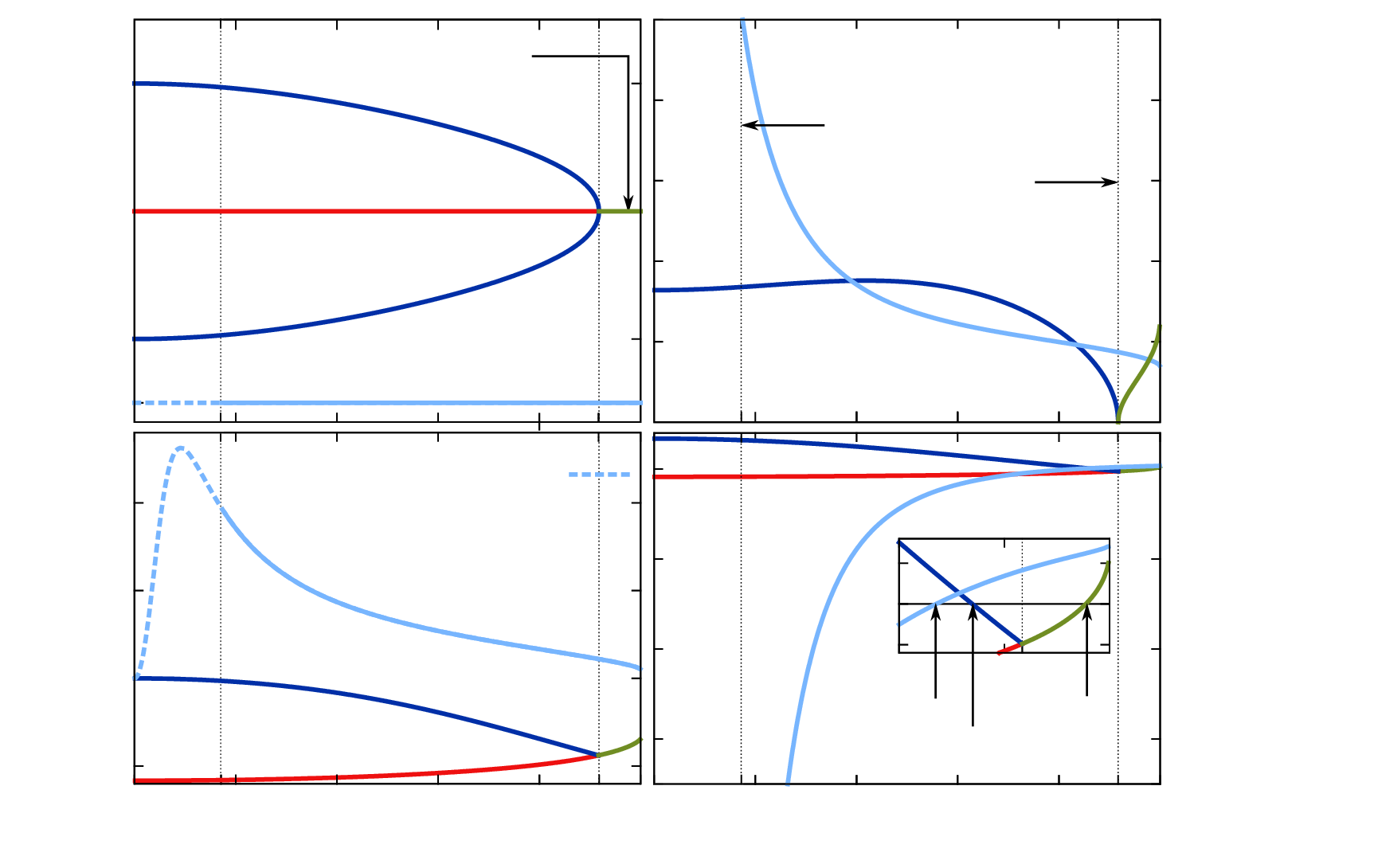}}%
    \put(0.85153734,0.37580836){\makebox(0,0)[lb]{\smash{$0.05$}}}%
    \put(0.85226201,0.43438545){\makebox(0,0)[lb]{\smash{$0.1$}}}%
    \put(0.85153734,0.49302387){\makebox(0,0)[lb]{\smash{$0.15$}}}%
    \put(0.85226201,0.55160096){\makebox(0,0)[lb]{\smash{$0.2$}}}%
    \put(0.85280641,0.28534685){\makebox(0,0)[lb]{\smash{$0$}}}%
    \put(0.85096445,0.217096){\makebox(0,0)[lb]{\smash{$-0.01$}}}%
    \put(0.85153734,0.15321248){\makebox(0,0)[lb]{\smash{$-0.02$}}}%
    \put(0.85491029,0.08858373){\makebox(0,0)[lb]{\smash{$-0.03$}}}%
    \put(0.46618748,0.03895989){\makebox(0,0)[lb]{\smash{$0$}}}%
    \put(0.53739056,0.03895989){\makebox(0,0)[lb]{\smash{$0.2$}}}%
    \put(0.61106482,0.03895989){\makebox(0,0)[lb]{\smash{$0.4$}}}%
    \put(0.68467768,0.03895989){\makebox(0,0)[lb]{\smash{$0.6$}}}%
    \put(0.75835196,0.03895989){\makebox(0,0)[lb]{\smash{$0.8$}}}%
    \put(0.83449742,0.03895989){\makebox(0,0)[lb]{\smash{$1$}}}%
    \put(0.63168904,0.18410828){\makebox(0,0)[lb]{\smash{$0$}}}%
    \put(0.64207571,0.13326206){\makebox(0,0)[lb]{\smash{$0.8$}}}%
    \put(0.72251633,0.13326234){\makebox(0,0)[lb]{\smash{$0.9$}}}%
    \put(0.79763708,0.13326234){\makebox(0,0)[lb]{\smash{$1$}}}%
    \put(0.62170003,0.10038044){\color[rgb]{0,0,0}\makebox(0,0)[lb]{\smash{$0.8352$}}}%
    \put(0.69175264,0.08066474){\color[rgb]{0,0,0}\makebox(0,0)[lb]{\smash{$0.8695$}}}%
    \put(0.75460781,0.10383025){\color[rgb]{0,0,0}\makebox(0,0)[lb]{\smash{$0.9775$}}}%
    \put(1.16859355,0.7773549){\color[rgb]{0,0,0}\makebox(0,0)[lt]{\begin{minipage}{0.69983355\unitlength}\raggedright \end{minipage}}}%
    \put(0.06091892,0.37777717){\makebox(0,0)[lb]{\smash{$60$}}}%
    \put(0.05246922,0.47077585){\makebox(0,0)[lb]{\smash{$180$}}}%
    \put(0.05053678,0.56377453){\makebox(0,0)[lb]{\smash{$300$}}}%
    \put(0.06936371,0.33127783){\makebox(0,0)[lb]{\smash{$0$}}}%
    \put(0.69229601,0.49387145){\makebox(0,0)[lb]{\smash{$0.917$}}}%
    \put(0.01567453,0.07456431){\makebox(0,0)[lb]{\smash{$-0.001$}}}%
    \put(0.05970151,0.13845372){\makebox(0,0)[lb]{\smash{$0$}}}%
    \put(0.03398533,0.20234784){\makebox(0,0)[lb]{\smash{$0.001$}}}%
    \put(0.03398533,0.26623725){\makebox(0,0)[lb]{\smash{$0.002$}}}%
    \put(0.08801911,0.04678548){\makebox(0,0)[lb]{\smash{$0$}}}%
    \put(0.15922216,0.03905572){\makebox(0,0)[lb]{\smash{$0.2$}}}%
    \put(0.23289643,0.03905572){\makebox(0,0)[lb]{\smash{$0.4$}}}%
    \put(0.30650935,0.03905572){\makebox(0,0)[lb]{\smash{$0.6$}}}%
    \put(0.38018365,0.03905572){\makebox(0,0)[lb]{\smash{$0.8$}}}%
    \put(0.29150392,0.58661087){\makebox(0,0)[lb]{\smash{$\GLT$ after}}}%
    \put(0.10383759,0.49162283){\color[rgb]{0,0,0}\makebox(0,0)[lb]{\smash{$\GLT$}}}%
    \put(0.10383759,0.58727863){\color[rgb]{0,0,0}\makebox(0,0)[lb]{\smash{$\GLC$}}}%
    \put(0.10383759,0.40369682){\color[rgb]{0,0,0}\makebox(0,0)[lb]{\smash{$\GLQ$}}}%
    \put(0.10318539,0.35152092){\color[rgb]{0,0,0}\makebox(0,0)[lb]{\smash{$\GQS$}}}%
    \put(0.16433689,0.28056263){\color[rgb]{0,0,0}\makebox(0,0)[lb]{\smash{$\GQS$ in exclusion zone}}}%
    \put(0.26057214,0.00517645){\color[rgb]{0,0,0}\makebox(0,0)[lb]{\smash{$e_0$}}}%
    \put(0.64319528,0.00517645){\color[rgb]{0,0,0}\makebox(0,0)[lb]{\smash{$e_0$}}}%
    \put(-0.00116638,0.19960186){\color[rgb]{0,0,0}\makebox(0,0)[lb]{\smash{$u$}}}%
    \put(-0.00116638,0.5177016){\color[rgb]{0,0,0}\makebox(0,0)[lb]{\smash{$\theta$ $(\degre)$}}}%
    \put(0.90901335,0.19305153){\color[rgb]{0,0,0}\makebox(0,0)[lb]{\smash{$g$}}}%
    \put(0.90901335,0.51823966){\color[rgb]{0,0,0}\makebox(0,0)[lb]{\smash{$|\nu|$}}}%
    \put(0.26909721,0.55827924){\color[rgb]{0,0,0}\makebox(0,0)[lb]{\smash{the bifurcation}}}%
    \put(0.4445321,0.2964186){\color[rgb]{0,0,0}\makebox(0,0)[lb]{\smash{b.}}}%
    \put(0.60365867,0.53543796){\makebox(0,0)[lb]{\smash{$0.18$}}}%
    \put(0.44259965,0.59981283){\color[rgb]{0,0,0}\makebox(0,0)[lb]{\smash{a.}}}%
    \put(0.82135574,0.59981283){\color[rgb]{0,0,0}\makebox(0,0)[lb]{\smash{c.}}}%
    \put(0.82135574,0.2964186){\color[rgb]{0,0,0}\makebox(0,0)[lb]{\smash{d.}}}%
  \end{picture}%
\endgroup%
\caption{\small Location in $\theta$ (a.) and $u$ (b.) and frequencies $|\nu|$ (c.) and $g$ (d.) of the fixed points families for a Sun-Jupiter like system ($\eps = 10^{-3}$). $\cF_{L_4}$ and $\cF_{L_5}$ (blue curves) merge with $\GLT$ (red curve) which gives rise to a stable family of fixed points (green curve). The AP is relevant for QS motion when $\GQS$ (sky blue curve) is a continuous curve. There are particular orbits without precession on each family which correspond to degenerated fixed points of the AP.}\label{fig:CIF}
\end{center}
\end{figure}

The evolution of the location and of the frequencies of the orbits associated with the families $\GQS$, $\GLT$, $\GLQ$ and $\GLC$ versus $e_0$ are described in  Fig.\ref{fig:CIF} for a mass ratio equal to $\eps = 10^{-3}$ (a Sun-Jupiter like system).

The red curve close to $(\theta,u) = (180\degre,0)$ represents the family $\GLT$ while the two blue curves that start in $L_4$ and $L_5$ correspond to $\GLQ$ and $\GLC$.
As described in section \ref{sec:Phase}, by increasing $e_0$ these two last families merge with $\GLT$ for $e_0 \simeq 0.917$ (vertical dashed line). Above this critical value, the last family becomes stable (green curves in Fig.\ref{fig:CIF}).\\
%
%
%
The sky blue curve located nearby $(\theta,u) = (0\degre,0)$ represents the family $\GQS$.
Along this family, for  $0.4\leq e_0<1$, the frequencies $|\nu|$ and $|g|$ are of the same order as those of the TP equilibria, but the sign of $g$ is different.
Then, by decreasing $e_0$, the moduli of the frequencies increase and tend to infinity.
When the frequencies reach values of the same order or higher than the fast frequency, $\GQS$ enters the exclusion zone and the averaged problem does not describe accurately the quasi-satellite's motion.\\
In order to estimate an eccentricity range where the averaged problem is adapted to QS motion, we consider that $\GQS$ is outside the exclusion zone when $\vert g\vert$ and $\vert\nu\vert$ are lower  than $1/4$.
Fig.\ref{fig:CIF} shows that this quantity is given by $e_0=0.18$ (vertical dashed line).
Therefore, the AP and RAP are relevant to study $\GQS$ and thus the QS motion for $e_0\geq 0.18$ in the Sun-Jupiter system.


Now, we focus on the variations of $g$ along each families of fixed points.
For each of them, the frequency  is monotonous and crosses zero for a critical value of eccentricity: $e_0 \simeq 0.8352$ for $\GQS$, $e_0 \simeq 0.8695$ for $\GLQ$ and $\GLC$, and    $e_0 \simeq 0.9775$ for $\GLT$.
According to the section \ref{sec:Intp}, these particular trajectories in the RAP correspond to circles of fixed points in the AP, and $2\pi$-periodic orbits in the RF, i.e. frozen ellipses in the heliocentric frame.
We denote them $\GdQS$, $\GdLQ$, $\GdLC$ and $\GdLT$.


\begin{figure}
\begin{center}
\small 
\def\svgwidth{0.7\textwidth}
\begingroup%
  \makeatletter%
  \providecommand\color[2][]{%
    \errmessage{(Inkscape) Color is used for the text in Inkscape, but the package 'color.sty' is not loaded}%
    \renewcommand\color[2][]{}%
  }%
  \providecommand\transparent[1]{%
    \errmessage{(Inkscape) Transparency is used (non-zero) for the text in Inkscape, but the package 'transparent.sty' is not loaded}%
    \renewcommand\transparent[1]{}%
  }%
  \providecommand\rotatebox[2]{#2}%
  \ifx\svgwidth\undefined%
    \setlength{\unitlength}{639.09375bp}%
    \ifx\svgscale\undefined%
      \relax%
    \else%
      \setlength{\unitlength}{\unitlength * \real{\svgscale}}%
    \fi%
  \else%
    \setlength{\unitlength}{\svgwidth}%
  \fi%
  \global\let\svgwidth\undefined%
  \global\let\svgscale\undefined%
  \makeatother%
  \begin{picture}(1,1.02122146)%
    \put(0,0){\includegraphics[width=\unitlength]{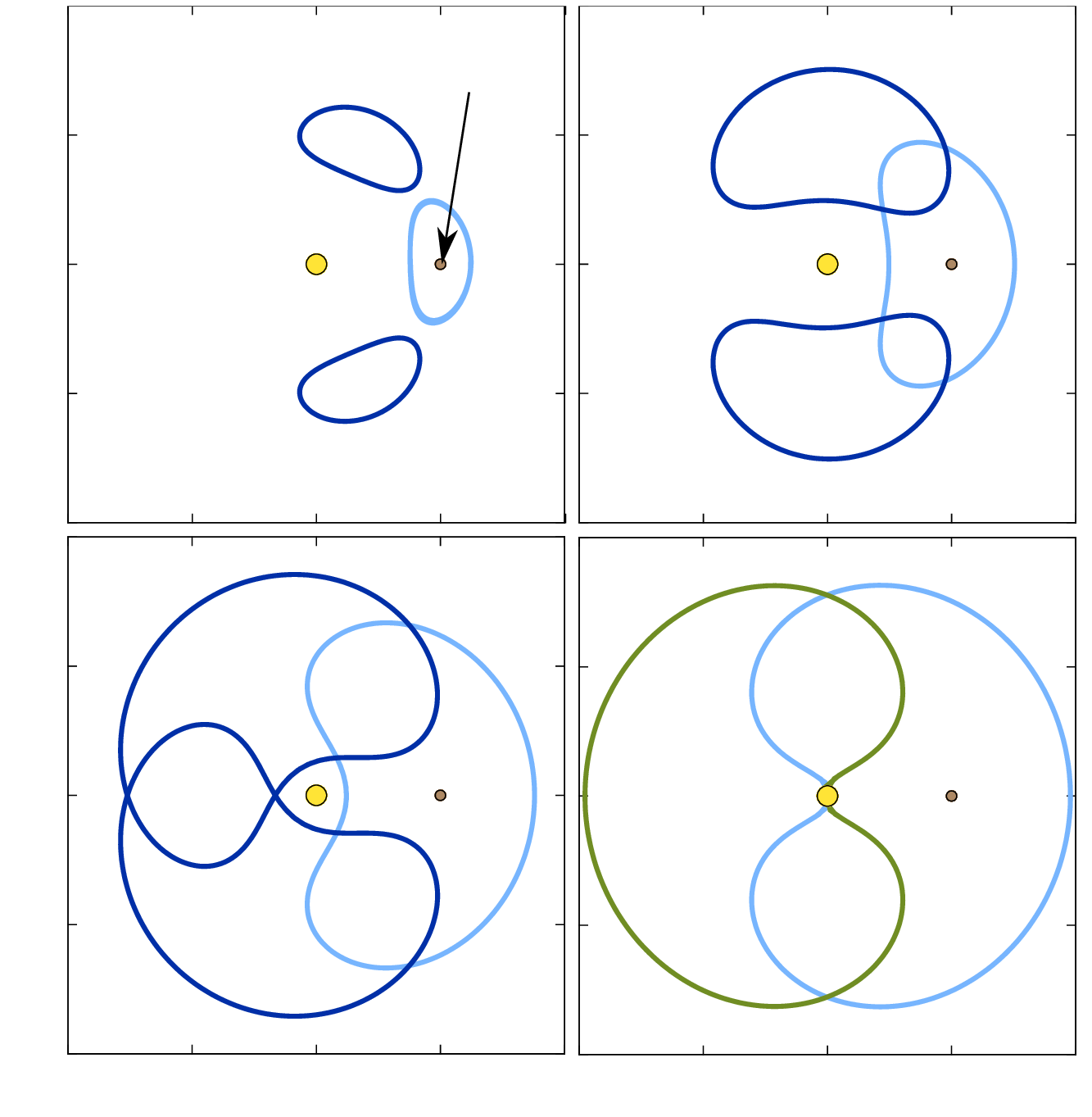}}%
    \put(0.27860358,0.00990187){\color[rgb]{0,0,0}\makebox(0,0)[lb]{\smash{$0$}}}%
    \put(0.39126302,0.00990187){\color[rgb]{0,0,0}\makebox(0,0)[lb]{\smash{$1$}}}%
    \put(0.14341169,0.00990187){\color[rgb]{0,0,0}\makebox(0,0)[lb]{\smash{$-1$}}}%
    \put(0.74676661,0.00990187){\color[rgb]{0,0,0}\makebox(0,0)[lb]{\smash{$0$}}}%
    \put(0.85942656,0.00990187){\color[rgb]{0,0,0}\makebox(0,0)[lb]{\smash{$1$}}}%
    \put(0.60907133,0.00990187){\color[rgb]{0,0,0}\makebox(0,0)[lb]{\smash{$-1$}}}%
    \put(0.03090878,0.28386084){\color[rgb]{0,0,0}\makebox(0,0)[lb]{\smash{$0$}}}%
    \put(0.00210408,0.40276581){\color[rgb]{0,0,0}\makebox(0,0)[lb]{\smash{$-1$}}}%
    \put(0.02949482,0.16728391){\color[rgb]{0,0,0}\makebox(0,0)[lb]{\smash{$1$}}}%
    \put(0.03090878,0.76955043){\color[rgb]{0,0,0}\makebox(0,0)[lb]{\smash{$0$}}}%
    \put(-0.00039947,0.88845411){\color[rgb]{0,0,0}\makebox(0,0)[lb]{\smash{$-1$}}}%
    \put(0.02949482,0.65297329){\color[rgb]{0,0,0}\makebox(0,0)[lb]{\smash{$1$}}}%
    \put(0.19677151,0.77257128){\color[rgb]{0,0,0}\makebox(0,0)[lb]{\smash{Sun}}}%
    \put(0.3797436,0.94306139){\color[rgb]{0,0,0}\makebox(0,0)[lb]{\smash{Jupiter}}}%
    \put(0.43846932,0.77155349){\color[rgb]{0,0,0}\makebox(0,0)[lb]{\smash{ $f$}}}%
    \put(0.20691014,0.90160073){\color[rgb]{0,0,0}\makebox(0,0)[lb]{\smash{$\LQs$}}}%
    \put(0.0895764,0.97357608){\color[rgb]{0,0,0}\makebox(0,0)[lb]{\smash{a.}}}%
    \put(0.0895764,0.08732071){\color[rgb]{0,0,0}\makebox(0,0)[lb]{\smash{c.}}}%
    \put(0.93840074,0.08732071){\color[rgb]{0,0,0}\makebox(0,0)[lb]{\smash{d.}}}%
    \put(0.93840074,0.97357608){\color[rgb]{0,0,0}\makebox(0,0)[lb]{\smash{b.}}}%
    \put(0.19924422,0.64542226){\color[rgb]{0,0,0}\makebox(0,0)[lb]{\smash{$\LCs$}}}%
    \put(0.55047455,0.29488564){\color[rgb]{0,0,0}\makebox(0,0)[lb]{\smash{$\LT$}}}%
  \end{picture}%
\endgroup%
\caption{\small Periodic orbits in the rotating frame associated with stable orbits of $\GLQ$, $\GLC$, $\GLT$ and $\GQS$ for $e_0= 0.25$ (a.), $0.5$ (b.), $0.75$ (c.) and $0.95$ (d.) (see Fig.\ref{fig:phase} b, c, d and f). The blue curves are associated with $\LQs$; the sky blue curve with the family $f$ and the green curve corresponds to $\LT$ after the bifurcation.}\label{fig:RF}
\end{center}
\end{figure}  

To conclude this section, we connect the fixed points families in the RAP to the corresponding trajectory in the RF.
Outside the exclusion zone, the transformation of these ones by $\Phi\circ\widehat{\psi}\circ\cC\circ\psi_{e_0}^{-1}$ provides us a first order approximation of their initial conditions in the RF.
Therefore, by improving them with an iterative algorithm that removes the frequency $\nu$ \citep{CoLaCo2010}, we integrated the corresponding trajectories in the RF.
An example of stable trajectories is represented on the figure \ref{fig:RF} for several values of $e_0$. \\
For a Sun-Jupiter like system, the families $\GLQ$, $\GLC$ provide the entire short period families, from their respective equilibrium to their merge with $\GLT$ and its collision orbit with the Sun. 
On the contrary, $\GQS$ provide only a part of the family $f$, from the collision with the Sun to the orbit with an eccentricity $e\simeq 0.18$.
The figure \ref{fig:RF} shows that by increasing $e_0$, the size of the periodic trajectories in the RF increases.
As expected, the libration center of the family $f$ is located close to the planet, while those of $\LQs$ and $\LCs$ shift from $L_4$ and $L_5$ towards $L_3$ where they merge with those of $\LT$.
After the bifurcation, only trajectories of the $f$ and $\LT$  families remain. 


\section{Quasi-satellite's domains in the rotating frame with the planet}
\label{sec:RF}


\subsection{The family $f$ in the RF}

The RAP seems to be the most adapted approach to understand the co-orbital motion in the circular case.
However, the averaged approaches have the drawback to be poorly significant in the exclusion zone  that surround the singularity associated with the collision with the planet.
For the QS motion, we showed in the section \ref{sec:CC} that the whole domain could not be reachable by low eccentric orbits, that is when the trajectories get closer to the planet.
As a consequence, to understand the QS nearby the planet and connect our results in the averaged approaches, we chose to revisit the classical works in the RF  \citep{HeGu1970, Be1975} on the simple-periodic symmetrical family of retrograde satellite orbits, generally known as the family $f$.


\begin{figure}
\begin{center}
\small
\def\svgwidth{0.52\textwidth}
\begingroup%
  \makeatletter%
  \providecommand\color[2][]{%
    \errmessage{(Inkscape) Color is used for the text in Inkscape, but the package 'color.sty' is not loaded}%
    \renewcommand\color[2][]{}%
  }%
  \providecommand\transparent[1]{%
    \errmessage{(Inkscape) Transparency is used (non-zero) for the text in Inkscape, but the package 'transparent.sty' is not loaded}%
    \renewcommand\transparent[1]{}%
  }%
  \providecommand\rotatebox[2]{#2}%
  \ifx\svgwidth\undefined%
    \setlength{\unitlength}{147.65891278bp}%
    \ifx\svgscale\undefined%
      \relax%
    \else%
      \setlength{\unitlength}{\unitlength * \real{\svgscale}}%
    \fi%
  \else%
    \setlength{\unitlength}{\svgwidth}%
  \fi%
  \global\let\svgwidth\undefined%
  \global\let\svgscale\undefined%
  \makeatother%
  \begin{picture}(1,0.67030245)%
    \put(0,0){\includegraphics[width=\unitlength]{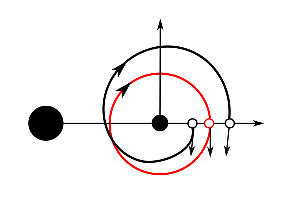}}%
    \put(0.5558573,0.61464801){\color[rgb]{0,0,0}\makebox(0,0)[lb]{\smash{$Y$}}}%
    \put(0.85536281,0.31652393){\color[rgb]{0,0,0}\makebox(0,0)[lb]{\smash{$X$}}}%
    \put(0.08430112,0.18092022){\color[rgb]{0,0,0}\makebox(0,0)[lb]{\smash{Sun}}}%
    \put(0.45457689,0.21804835){\color[rgb]{0,0,0}\makebox(0,0)[lb]{\smash{Planet}}}%
  \end{picture}%
\endgroup%

\caption{\small Representation of retrograde satellite trajectories in the RF. Red curve: simple-periodic symmetrical retrograte satellite trajectory that belongs to the family $f$. Black curve: trajectory in the neighbourhood of the previous one and that intersects the Poincar\'e section (black circles).}  \label{fig:Sec}
\end{center}
\end{figure}
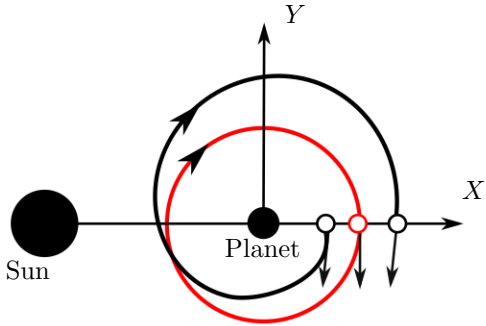

In the RF with the planet on a circular orbit, the problem has two degrees of freedom that we represent by the position $\br = (X,Y)$ and the velocity $\brd=(\dot{X}, \dot{Y})$ in the frame whose origin is the planet position, the horizontal axis is the Sun-planet alignment and the vertical axis, its perpendicular (see Fig.\ref{fig:Sec}).
This problem is autonomous and possesses a first integral $C_{J}$ generally known as the Jacobi constant. \\
%
%
For a given value of $C_{J}$, a simple-periodic symmetrical retrograde satellite orbit crosses the axes $\{Y = 0\}$ with $\dot{Y}<0$ and $\dot{X} = 0$ when $X>0$.
By defining the Poincar\'e map $\Pi_T$ associated with the section $\{Y=0; \dot{Y}<0\}$ where $T$ is the time between two consecutive crossings, the problem could be reduced to one degree of freedom represented by $(X,\dot{X})$ and $\dot{Y}=\dot{Y}(X,Y,\dot{X},C_{J})$.  
As a consequence, an orbit of the family $f$ corresponds to a fixed point in this Poincar\'e section whose coordinates  in the RF are $(X,0,0,\dot{Y})$ with \be T=2\pi/(1-g)\label{eq:T} \ee
where $g$ is the precession frequency of the periaster argument $\omega$.\\
Moreover the stability of the fixed point is deduced from the trace of the monodromy matrix $d\Pi_T(X,0)$ evaluated at the fixed point.
When the fixed point is stable, the frequency $\nu$ that characterized the oscillation of the resonant angle $\theta$ is obtained\footnote{Floquet theory; for further details, see \cite{MeHa1992}.}  from its two conjugated eigenvalues $(\kappa, \overline{\kappa})$ such as \be \kappa = \exp (i\nu T).\label{eq:nu}\ee


\subsection{Application to a Sun-Jupiter like system}


 \begin{figure}
 \begin{center}
 \small
 \centering
\def\svgwidth{0.62\textwidth}
\begingroup%
  \makeatletter%
  \providecommand\color[2][]{%
    \errmessage{(Inkscape) Color is used for the text in Inkscape, but the package 'color.sty' is not loaded}%
    \renewcommand\color[2][]{}%
  }%
  \providecommand\transparent[1]{%
    \errmessage{(Inkscape) Transparency is used (non-zero) for the text in Inkscape, but the package 'transparent.sty' is not loaded}%
    \renewcommand\transparent[1]{}%
  }%
  \providecommand\rotatebox[2]{#2}%
  \ifx\svgwidth\undefined%
    \setlength{\unitlength}{379.19092064bp}%
    \ifx\svgscale\undefined%
      \relax%
    \else%
      \setlength{\unitlength}{\unitlength * \real{\svgscale}}%
    \fi%
  \else%
    \setlength{\unitlength}{\svgwidth}%
  \fi%
  \global\let\svgwidth\undefined%
  \global\let\svgscale\undefined%
  \makeatother%
  \begin{picture}(1,0.61737122)%
    \put(0,0){\includegraphics[width=\unitlength]{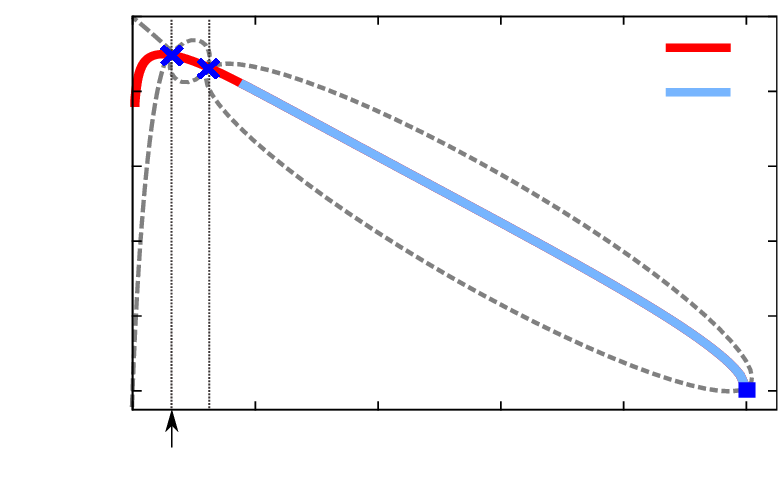}}%
    \put(0.53399487,0.49931262){\makebox(0,0)[lb]{\smash{family $f$ via $\GQS$}}}%
    \put(0.65889174,0.55184543){\color[rgb]{0,0,0}\makebox(0,0)[lb]{\smash{family $f$}}}%
    \put(0.15984561,0.06979731){\color[rgb]{0,0,0}\makebox(0,0)[lb]{\smash{$0$}}}%
    \put(0.30346915,0.06864153){\color[rgb]{0,0,0}\makebox(0,0)[lb]{\smash{$0.2$}}}%
    \put(0.45985668,0.06710268){\color[rgb]{0,0,0}\makebox(0,0)[lb]{\smash{$0.4$}}}%
    \put(0.61485288,0.06787538){\color[rgb]{0,0,0}\makebox(0,0)[lb]{\smash{$0.6$}}}%
    \put(0.77465865,0.068265){\color[rgb]{0,0,0}\makebox(0,0)[lb]{\smash{$0.8$}}}%
    \put(0.93755909,0.06710922){\color[rgb]{0,0,0}\makebox(0,0)[lb]{\smash{$1$}}}%
    \put(0.12205965,0.58645557){\color[rgb]{0,0,0}\makebox(0,0)[lb]{\smash{$0$}}}%
    \put(0.07848528,0.49630646){\color[rgb]{0,0,0}\makebox(0,0)[lb]{\smash{$-0.4$}}}%
    \put(0.08576092,0.11753509){\color[rgb]{0,0,0}\makebox(0,0)[lb]{\smash{$-2$}}}%
    \put(0.07892432,0.21006102){\color[rgb]{0,0,0}\makebox(0,0)[lb]{\smash{$-1.6$}}}%
    \put(0.07848528,0.40069589){\color[rgb]{0,0,0}\makebox(0,0)[lb]{\smash{$-0.8$}}}%
    \put(0.07848528,0.30816962){\color[rgb]{0,0,0}\makebox(0,0)[lb]{\smash{$-1.2$}}}%
    \put(0.5398519,0.02438055){\color[rgb]{0,0,0}\makebox(0,0)[lb]{\smash{$X$}}}%
    \put(0.01326252,0.35223936){\color[rgb]{0,0,0}\makebox(0,0)[lb]{\smash{$\dot{Y}$}}}%
    \put(0.1883645,0.02567297){\color[rgb]{0,0,0}\makebox(0,0)[lb]{\smash{$X_{L_2}$}}}%
  \end{picture}%
\endgroup%
\caption{\small The family $f$ in the $(X,\dot{Y})$ plane  (red curve) and its reachable part in the AP via $\GQS$ (sky blue curve).
The two blue crosses indicate the particular orbits (whose fundamental frequencies are in $1:3$ resonance) that split the neighbourhood of the family $f$. The blue square indicates the collision orbit with the Sun while $\{X = 0\}$ corresponds to the collision with the planet.
The grey outline schematizes the three connected domains of the family $f$ neighbourhood.}
\label{fig:f1}

\small 
 \centering
  \def\svgwidth{0.62\textwidth}

\begingroup%
  \makeatletter%
  \providecommand\color[2][]{%
    \errmessage{(Inkscape) Color is used for the text in Inkscape, but the package 'color.sty' is not loaded}%
    \renewcommand\color[2][]{}%
  }%
  \providecommand\transparent[1]{%
    \errmessage{(Inkscape) Transparency is used (non-zero) for the text in Inkscape, but the package 'transparent.sty' is not loaded}%
    \renewcommand\transparent[1]{}%
  }%
  \providecommand\rotatebox[2]{#2}%
  \ifx\svgwidth\undefined%
    \setlength{\unitlength}{328.68303223bp}%
    \ifx\svgscale\undefined%
      \relax%
    \else%
      \setlength{\unitlength}{\unitlength * \real{\svgscale}}%
    \fi%
  \else%
    \setlength{\unitlength}{\svgwidth}%
  \fi%
  \global\let\svgwidth\undefined%
  \global\let\svgscale\undefined%
  \makeatother%
  \begin{picture}(1,0.85389052)%
    \put(0,0){\includegraphics[width=\unitlength]{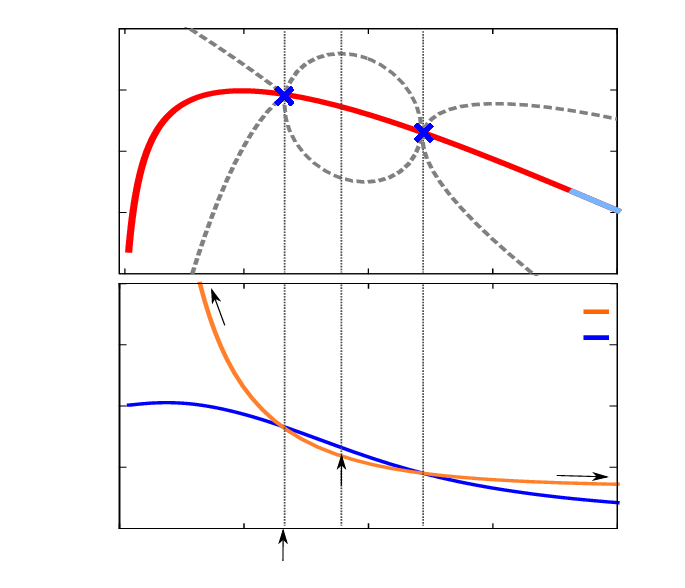}}%
    \put(0.24075658,0.74868859){\color[rgb]{0,0,0}\makebox(0,0)[lb]{\smash{sRS}}}%
    \put(0.78814849,0.62575826){\color[rgb]{0,0,0}\makebox(0,0)[lb]{\smash{\QSh}}}%
    \put(0.5064953,0.70581997){\color[rgb]{0,0,0}\makebox(0,0)[lb]{\smash{\QSb}}}%
    \put(0.32972998,0.35896169){\color[rgb]{0,0,0}\makebox(0,0)[lb]{\smash{$1\ll |g|$}}}%
    \put(0.72869788,0.17692188){\color[rgb]{0,0,0}\makebox(0,0)[lb]{\smash{$1\gg |g|$}}}%
    \put(0.43776211,0.1256974){\color[rgb]{0,0,0}\makebox(0,0)[lb]{\smash{$g=-1$}}}%
    \put(0.37460493,0.01335134){\color[rgb]{0,0,0}\makebox(0,0)[lb]{\smash{$X_{L_2}$}}}%
    \put(0.14573275,0.04708358){\color[rgb]{0,0,0}\makebox(0,0)[lb]{\smash{$0$}}}%
    \put(0.32719817,0.04708358){\color[rgb]{0,0,0}\makebox(0,0)[lb]{\smash{$0.05$}}}%
    \put(0.50813542,0.04708358){\color[rgb]{0,0,0}\makebox(0,0)[lb]{\smash{$0.1$}}}%
    \put(0.69675529,0.04708358){\color[rgb]{0,0,0}\makebox(0,0)[lb]{\smash{$0.15$}}}%
    \put(0.86832303,0.04673653){\color[rgb]{0,0,0}\makebox(0,0)[lb]{\smash{$0.2$}}}%
    \put(0.10412061,0.16374701){\color[rgb]{0,0,0}\makebox(0,0)[lb]{\smash{$0.5$}}}%
    \put(0.12467937,0.25491536){\color[rgb]{0,0,0}\makebox(0,0)[lb]{\smash{$1$}}}%
    \put(0.10716015,0.34302926){\color[rgb]{0,0,0}\makebox(0,0)[lb]{\smash{$1.5$}}}%
    \put(0.7994271,0.35823074){\color[rgb]{0,0,0}\makebox(0,0)[lb]{\smash{$\nu$}}}%
    \put(0.67677142,0.39372494){\color[rgb]{0,0,0}\makebox(0,0)[lb]{\smash{$(1-g)/3$}}}%
    \put(0.5059089,0.00938983){\color[rgb]{0,0,0}\makebox(0,0)[lb]{\smash{$X$}}}%
    \put(0.08354252,0.80522298){\color[rgb]{0,0,0}\makebox(0,0)[lb]{\smash{$-0.1$}}}%
    \put(0.08354252,0.71648718){\color[rgb]{0,0,0}\makebox(0,0)[lb]{\smash{$-0.2$}}}%
    \put(0.08354252,0.62775138){\color[rgb]{0,0,0}\makebox(0,0)[lb]{\smash{$-0.3$}}}%
    \put(0.08354252,0.53679751){\color[rgb]{0,0,0}\makebox(0,0)[lb]{\smash{$-0.4$}}}%
    \put(0.08354252,0.45249883){\color[rgb]{0,0,0}\makebox(0,0)[lb]{\smash{$-0.5$}}}%
    \put(0.01438875,0.66271825){\color[rgb]{0,0,0}\makebox(0,0)[lb]{\smash{$\dot{Y}$}}}%
    \put(0.19259551,0.46330583){\color[rgb]{0,0,0}\makebox(0,0)[lb]{\smash{a.}}}%
    \put(0.1925955,0.09599568){\color[rgb]{0,0,0}\makebox(0,0)[lb]{\smash{b.}}}%
  \end{picture}%
\endgroup%
\caption{\small (a.) Zoom in of Fig.\ref{fig:f1} on the two particular orbits whose fundamental frequencies are in $1:3$ resonance. (b.) Variation of the frequencies of the system along the family $f$. Comparing to Fig.\ref{fig:CIF}b, $\nu$ does not tend to infinity when the periodic orbits get closer to the planet ($\{X=0\}$), but increases and tends to $1$. The $1:3$ resonance splits the neighbourhood in three domains neatly defined in terms of frequencies: ``satellized" retrograde satellite (sRS), binary quasi-satellite (\QSb ) and heliocentric quasi-satellite (\QSh )}\label{fig:f2}
\end{center}
\end{figure}	

The figure \ref{fig:f1} and \ref{fig:f2}a represent the family $f$ in the $(X,\dot{Y})$ plane (red curve) and its reachable part in the averaged approaches (sky blue curve).

Fig.\ref{fig:f1} shows that the family $f$ extends from the orbits in an infinitesimal neighbourhood of the planet ($X\simeq 0$) to the collision orbit with the Sun.
Although, the whole family is linearly stable, we cannot predict the size of the stable region surrounding it.
Indeed, this domain depends strongly on the position of the resonances between the fundamental frequencies $1-g$ and $\nu$, which are themselves conditioned by the value of $X$.
This is what occurs in  particular orbits of the family $f$ (blue crosses and dashed lines) where the stability domain's diameter tends to zero.
Consequently, these two orbits  divide the neighbourhood of the family $f$ in three connected domains that we outlined in grey in Fig.\ref{fig:f1} and Fig.\ref{fig:f2}a.

The figure \ref{fig:f2}b exhibits the variations of the frequencies\footnote{In practice, the numerical algorithm of the Poincar\'e map provides $g$ as in the equation \eqref{eq:T} while the frequency $\nu$ is obtained via the monodromy matrix $d\Pi_T$ (see equation \eqref{eq:nu}) that is calculated with a numerical differentiation algorithm on the Poincar\'e map. } $\nu$ and $1-g$.
Comparing to Fig.\ref{fig:CIF}b, we remark that $\nu$ does not tend to infinity when the periodic orbits get closer to the planet, but increases and tends to $1$.
Likewise, Fig.\ref{fig:f2}b highlights that the resonance between the frequencies of the system is $\nu/(1-g) = 1/3$ and that the three domains are neatly defined in terms of frequencies as follows:
\begin{align}
\mbox{sRS}:\left\{\begin{array}{l}
3\nu < 1-g\\
|g|>1
\end{array}\right., \quad
&\mbox{\QSb}:\quad
3\nu > 1-g
\qtext{and}
&\mbox{\QSh}:\left\{\begin{array}{l}
3\nu < 1-g\\
|g|<1
\end{array}\right.\mbox{.}\nnb
\end{align}
The closest domain to the planet corresponds to the ``satellized" retrograde satellite orbits (sRS).
Indeed, as the upper bound of this domain matches with the $L_2$ position, we recovered the notion of Hill's sphere in the context of the retrograde satellite trajectories. 
Hence, this domain  consists of trajectories dominated by the gravitational influence of the planet whereas the star acts as a perturbator.
Therefore the planetocentric osculating ellipses are the most relevant variables to represent the motion and perturbative treatments are possible.\\
The domain outside the Hill's sphere corresponds to the QS that is divided in two others domains. \\
The domain of \QSh orbits, that is the heliocentric QS, corresponds to the farthest domain to the planet,  which implies that this body acts as a perturbator whereas the influence of the star dominates the dynamics.
Therefore, the heliocentric orbital elements are well suited to the problem, and the perturbative treatment as well as the averaging over the fast angle are natural.
As a consequence, it is the \QSh trajectories that are reachable in the averaged problem.
As the orbits of the family $f$ included in the \QSh domain cross the  Poincar\'e section at their aphelion, the $X$ coordinates is related to $e_0$ by the expression  \be X= e = e_0 + \oue .\label{eq:Xe0}\ee
The third domain, that we called the binary QS domain (\QSb),
is intermediate between the sRS and \QSh ones.
In this region, none of the two massive bodies has a dominant influence on the massless one.
As a consequence, the frequencies $g$ could be of the same order or even equal to $1$, making inappropriate any method of averaging.

Remark that in the planetary problem, \cite{HaVo2011} highlight a family of periodic orbits that corresponds to the family $f$.
Indeed, along this family that ranges from orbits for which the two planets collide with the star to the orbits where the two planets are mutually satellized, all trajectories are stable and satisfy $\theta = 0\degre$. These authors also decomposed the family in three domains, denoted $A$, $B$ and planetary, which seem to correspond to our sRS, \QSb and \QSh domains.

\subsection{Extension  to arbitrary mass ratio}
\begin{figure}
\begin{center}
\small 
\def\svgwidth{0.8\textwidth}
\begingroup%
  \makeatletter%
  \providecommand\color[2][]{%
    \errmessage{(Inkscape) Color is used for the text in Inkscape, but the package 'color.sty' is not loaded}%
    \renewcommand\color[2][]{}%
  }%
  \providecommand\transparent[1]{%
    \errmessage{(Inkscape) Transparency is used (non-zero) for the text in Inkscape, but the package 'transparent.sty' is not loaded}%
    \renewcommand\transparent[1]{}%
  }%
  \providecommand\rotatebox[2]{#2}%
  \ifx\svgwidth\undefined%
    \setlength{\unitlength}{624.34375bp}%
    \ifx\svgscale\undefined%
      \relax%
    \else%
      \setlength{\unitlength}{\unitlength * \real{\svgscale}}%
    \fi%
  \else%
    \setlength{\unitlength}{\svgwidth}%
  \fi%
  \global\let\svgwidth\undefined%
  \global\let\svgscale\undefined%
  \makeatother%
  \begin{picture}(1,1.03308474)%
    \put(0,0){\includegraphics[width=\unitlength]{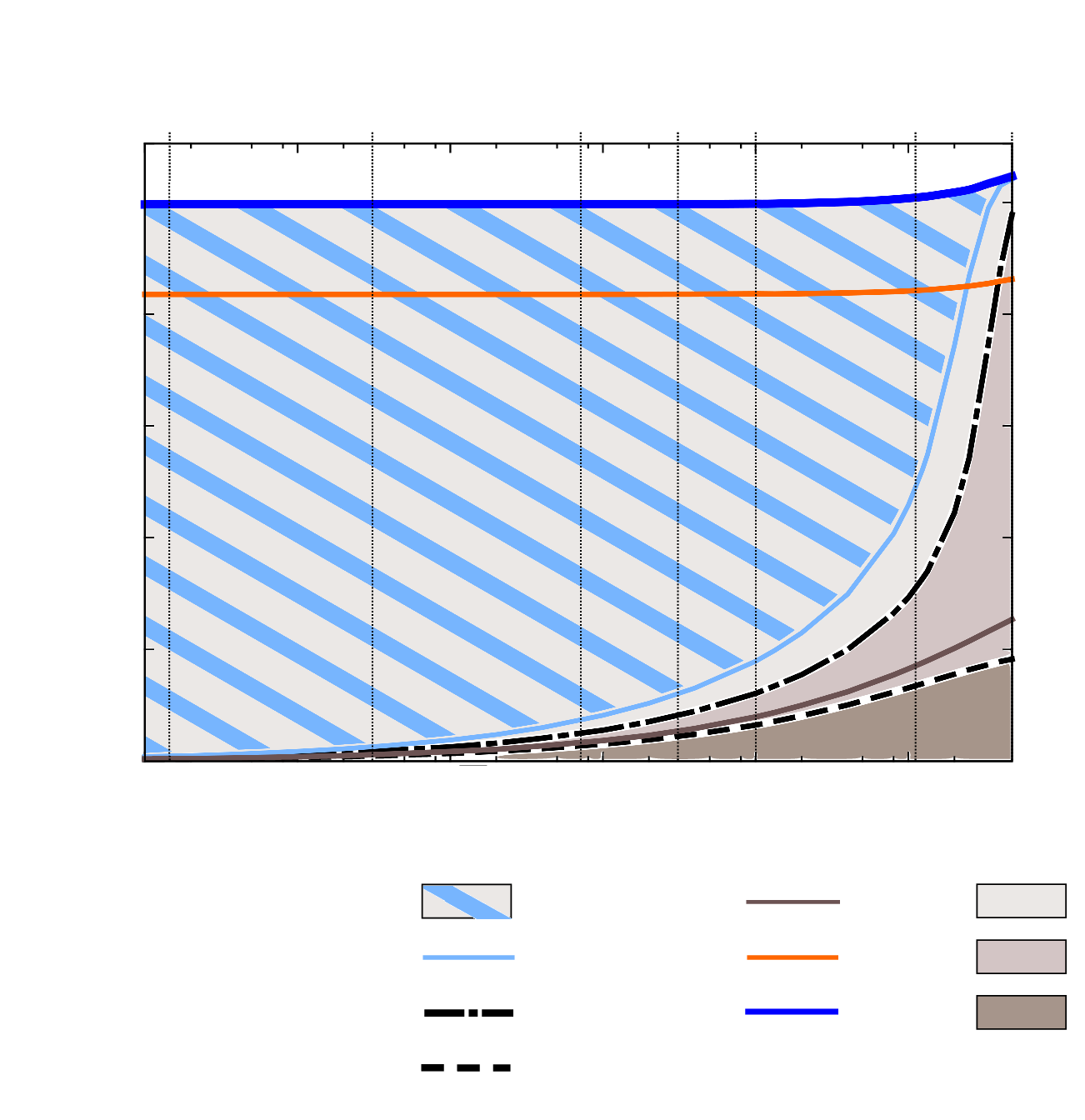}}%
    \put(0.09327766,0.3255121){\makebox(0,0)[lb]{\smash{0}}}%
    \put(0.07646626,0.42857656){\makebox(0,0)[lb]{\smash{0.2}}}%
    \put(0.07646626,0.53164729){\makebox(0,0)[lb]{\smash{0.4}}}%
    \put(0.07646626,0.63479309){\makebox(0,0)[lb]{\smash{0.6}}}%
    \put(0.07646626,0.73786382){\makebox(0,0)[lb]{\smash{0.8}}}%
    \put(0.09327766,0.84092828){\makebox(0,0)[lb]{\smash{1}}}%
    \put(0.11614793,0.29795942){\makebox(0,0)[lb]{\smash{1e-07}}}%
    \put(0.25701458,0.29795942){\makebox(0,0)[lb]{\smash{1e-06}}}%
    \put(0.39788751,0.29795942){\makebox(0,0)[lb]{\smash{1e-05}}}%
    \put(0.53334224,0.29795942){\makebox(0,0)[lb]{\smash{0.0001}}}%
    \put(0.67981478,0.29795942){\makebox(0,0)[lb]{\smash{0.001}}}%
    \put(0.82628733,0.29795942){\makebox(0,0)[lb]{\smash{0.01}}}%
    \put(0.03375426,0.14450183){\makebox(0,0)[lb]{\smash{AP validity limit}}}%
    \put(0.51029479,0.14497735){\makebox(0,0)[lb]{\smash{$g=0$ ($\GdQS$)}}}%
    \put(0.51065039,0.19575565){\makebox(0,0)[lb]{\smash{$g=-1$}}}%
    \put(0.80670202,0.19440419){\makebox(0,0)[lb]{\smash{\QSh}}}%
    \put(0.81078115,0.14315038){\makebox(0,0)[lb]{\smash{\QSb}}}%
    \put(0.80973489,0.0899361){\makebox(0,0)[lb]{\smash{sRS}}}%
    \put(0.03476742,0.09324802){\makebox(0,0)[lb]{\smash{\QSb-\QSh  boundary}}}%
    \put(0.03989281,0.04199421){\makebox(0,0)[lb]{\smash{sRS-QS  boundary}}}%
    \put(0.03492593,0.19440419){\makebox(0,0)[lb]{\smash{\QSh reachable in the AP}}}%
    \put(0.49063695,0.25595875){\color[rgb]{0,0,0}\makebox(0,0)[lb]{\smash{$\eps$}}}%
    \put(0.0192851,0.58589849){\color[rgb]{0,0,0}\makebox(0,0)[lb]{\smash{$X$}}}%
    \put(0.50825114,0.0899361){\makebox(0,0)[lb]{\smash{Sun collision}}}%
    \put(0.08886184,1.00712277){\color[rgb]{0,0,0}\rotatebox{-57.27471102}{\makebox(0,0)[lb]{\smash{Sun-Mer}}}}%
    \put(0.27850202,1.00712183){\color[rgb]{0,0,0}\rotatebox{-57.27471102}{\makebox(0,0)[lb]{\smash{Sun-Ear}}}}%
    \put(0.47070456,1.00712141){\color[rgb]{0,0,0}\rotatebox{-57.27471102}{\makebox(0,0)[lb]{\smash{Sun-Ura}}}}%
    \put(0.56039878,1.00712168){\color[rgb]{0,0,0}\rotatebox{-57.27471102}{\makebox(0,0)[lb]{\smash{Sun-Sat}}}}%
    \put(0.63215339,1.00712138){\color[rgb]{0,0,0}\rotatebox{-57.27471102}{\makebox(0,0)[lb]{\smash{Sun-Jup}}}}%
    \put(0.77822451,1.01224573){\color[rgb]{0,0,0}\rotatebox{-57.27471102}{\makebox(0,0)[lb]{\smash{Ear-Moo}}}}%
    \put(0.8704801,0.99430573){\color[rgb]{0,0,0}\rotatebox{-57.27471102}{\makebox(0,0)[lb]{\smash{0.0477}}}}%
  \end{picture}%
\endgroup%

\caption{\small Evolution of the sRS, \QSb and \QSh boundaries along the family $f$ by varying the mass ratio $\eps$. For small $\eps$, the \QSh domain dominates the family $f$ implying that the AP and the RAP are fully adapted to study the QS motion. By increasing $\eps$, the size of the part associated with the sRS and the \QSb trajectories increases making not reachable the orbits with small eccentricities in the averaged problem. Eventually, for $\eps>0.01$, the sRS and the \QSb domains become dominant while the \QSh one  is reduced so much that the averaged problem becomes useless for all values of $e_0$.}\label{fig:map}
\end{center}
\end{figure}
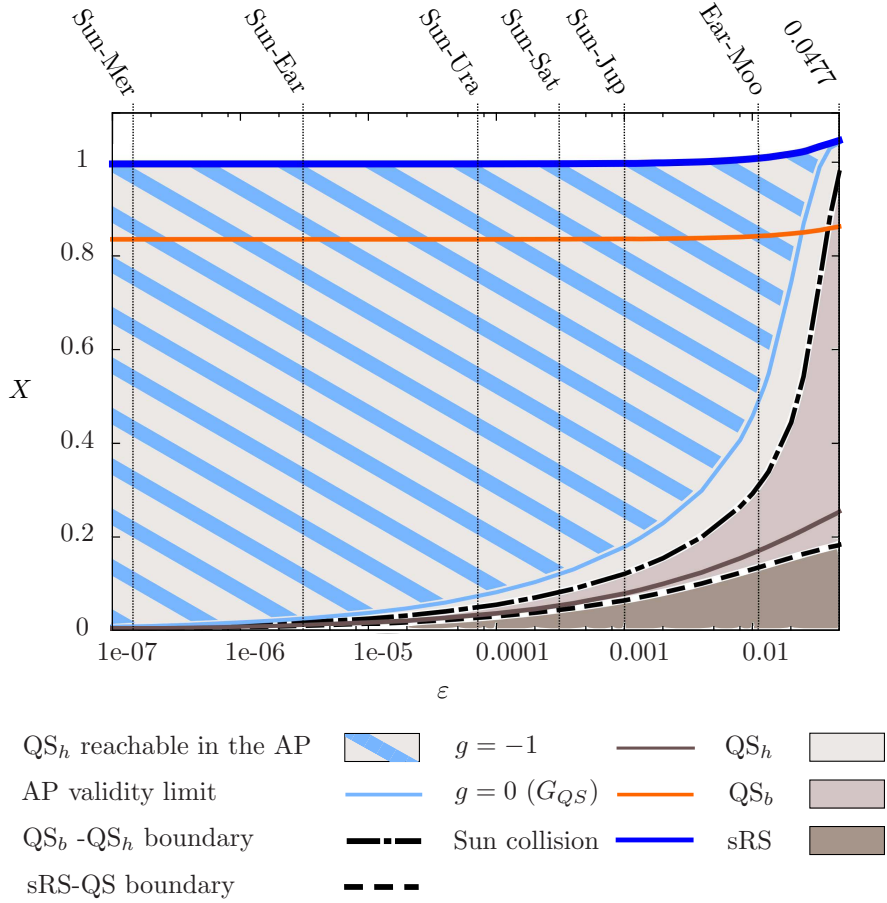

By varying the mass ratio $\eps$, we follow the evolution of the boundaries of the three domains along the family $f$ as well as the validity limit of the averaged problem.
In Fig.\ref{fig:map}, the parameter $\eps$ ranges from $10^{-7}$ to    $0.0477$ which is the critical mass ratio where a part of the family $f$ becomes unstable \citep[see][]{HeGu1970}.
For Sun-terrestrial planet systems, the size of the \QSb and sRS domains is negligible with respect to the \QSh one.
As a consequence, for these systems, the AP and RAP are fully adapted to describe the main part of the family $f$ and its neighbourhood (except for very small eccentricities).
For Sun-giant planet systems as well as the Earth-Moon system, the gravitational influence of the planet being stronger, the size of the \QSb domain $f$ increases until to be of the same order than those of the sRS one while the size of the \QSh decreases.
By the equation \eqref{eq:Xe0}, we established that for the Sun-Uranus, Sun-Saturn, Sun-Jupiter and Earth-Moon systems, the \QSh orbits are reachable in the averaged problem by $e_0$ greater than $0.08$, $0.13$, $0.18$ and $0.5$.
Then, by increasing $\eps$, the \QSb domain becomes dominant while the \QSh one is reduced so much that the averaged problem becomes useless for all values of $e_0$ ($\eps\simeq 0.04$).
Consequently, for the Pluto-Charon system ($\eps\simeq 1/10$) none \QSh trajectory could be described in the averaged approaches.
Moreover, according to the stability map of the family $f$ in \cite{Be1975}, this system could not harbour a \QSh companion: only \QSb and sRS trajectories exist for this value of mass ratio.

	\section{On the frozen ellipses: an extension to the eccentric case ($e'\geq 0$)}
\label{sec:EC}
An important result of our study in the circular case has been to highlight the particular orbits $\GdQS$, $\GdLT$, $\GdLQ$ and $\GdLC$, that correspond\footnote{See the figure \ref{fig:CIF}.} to circles of fixed points in the averaged problem and therefore frozen ellipses in the heliocentric frame.

A natural question is to know if these structures are preserved when a small eccentricity is given to the planetary orbit.
This question can be addressed in a perturbative way.
Indeed, for sufficiently low values of planet's eccentricity, the Hamiltonian of the problem reads $\Hb|_{e'=0} + e'R$, i.e. the perturbation of the Hamiltonian in the circular case by the first order term in planetary eccentricity.
However, as $\omega=\arg(x)$ is no longer an ignorable variable in this Hamiltonian, the dimension of the phase space could not been reduced as in the section \ref{sec:CC} and  the persistence of the set of degenerated fixed points is not necessary guaranteed.

In the present paper, we limit our approach to numerical explorations of the phase space associated with $\Hb|_{e'\geq 0}$.
For a very low value of $e'$ in a Sun-Jupiter like system, the (numerical) solving of the equations of motion in the averaged problem,
\be\left\{
\begin{split}
\dron{}{\theta}\Hb(\theta,u,-i\xb, x) = 0\\
\dron{}{u}\Hb(\theta,u,-i\xb, x) = 0\\
\dron{}{x}\Hb(\theta,u,-i\xb, x) = 0
\end{split}\right. ,
\ee
 shows that although each circle of fixed points is destroyed, two fixed points survived to the perturbation.  
One is stable and the other unstable.
We denote these fixed points  $G_{X,1}^{e'}$ and $G_{X,2}^{e'}$ with $X$ corresponding to QS, $L_4$, $L_5$ and $L_3$.
By varying $e'$, we followed them and show families of fixed points of the averaged problem that originate from $\GdLT$, $\GdLQ$, $\GdLC$ and $\GdQS$.

\begin{figure}
\begin{center}
\small 
\def\svgwidth{0.85\textwidth}
\begingroup%
  \makeatletter%
  \providecommand\color[2][]{%
    \errmessage{(Inkscape) Color is used for the text in Inkscape, but the package 'color.sty' is not loaded}%
    \renewcommand\color[2][]{}%
  }%
  \providecommand\transparent[1]{%
    \errmessage{(Inkscape) Transparency is used (non-zero) for the text in Inkscape, but the package 'transparent.sty' is not loaded}%
    \renewcommand\transparent[1]{}%
  }%
  \providecommand\rotatebox[2]{#2}%
  \ifx\svgwidth\undefined%
    \setlength{\unitlength}{699.99995117bp}%
    \ifx\svgscale\undefined%
      \relax%
    \else%
      \setlength{\unitlength}{\unitlength * \real{\svgscale}}%
    \fi%
  \else%
    \setlength{\unitlength}{\svgwidth}%
  \fi%
  \global\let\svgwidth\undefined%
  \global\let\svgscale\undefined%
  \makeatother%
  \begin{picture}(1,0.73330313)%
    \put(0,0){\includegraphics[width=\unitlength]{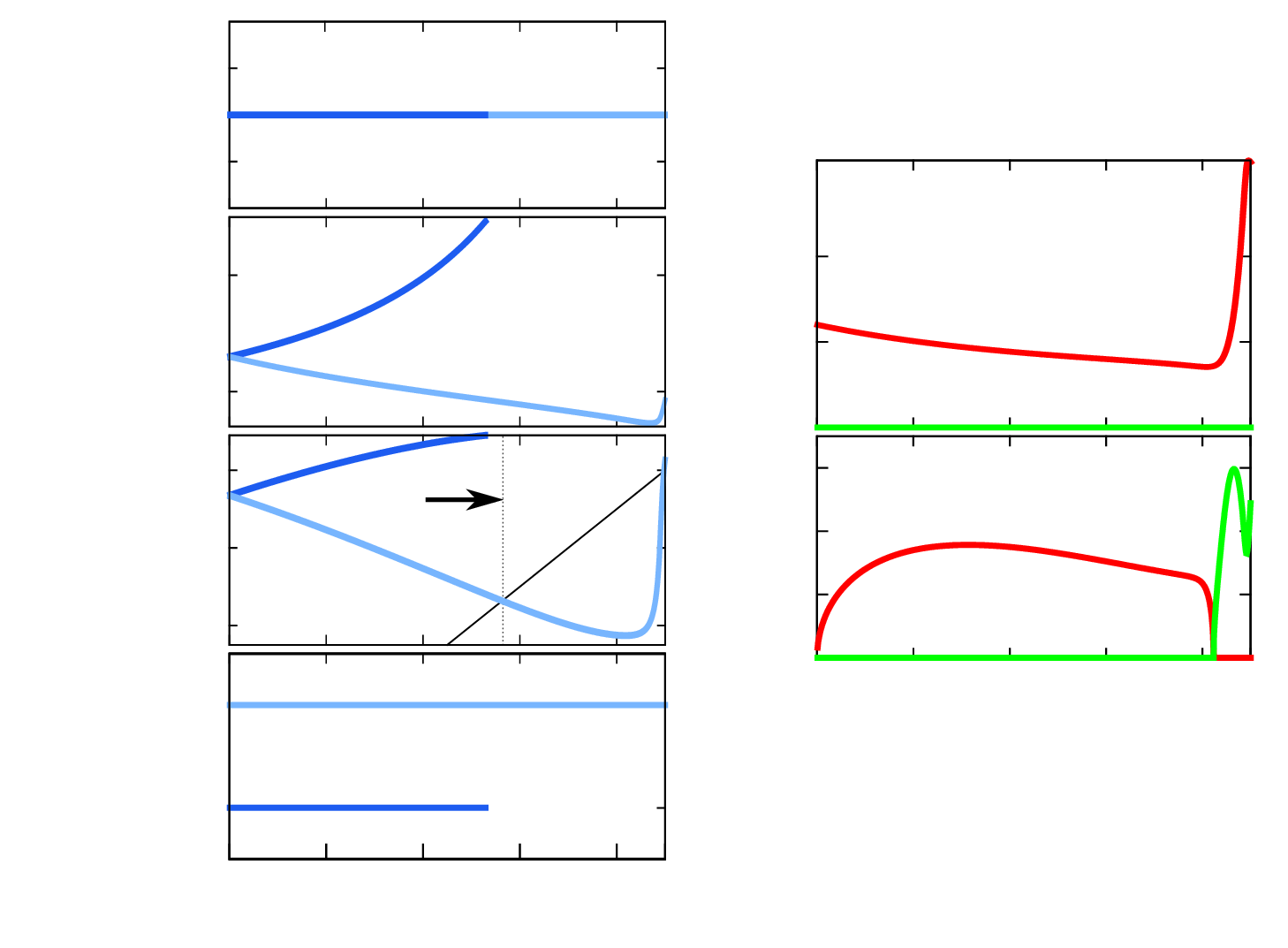}}%
    \put(0.13324807,0.64010945){\makebox(0,0)[lb]{\smash{0}}}%
    \put(0.09334182,0.67610834){\color[rgb]{0,0,0}\makebox(0,0)[lb]{\smash{1e-10}}}%
    \put(0.13324805,0.42639401){\makebox(0,0)[lb]{\smash{0}}}%
    \put(0.09825925,0.51625451){\makebox(0,0)[lb]{\smash{0.001}}}%
    \put(0.12511078,0.2435401){\makebox(0,0)[lb]{\smash{0.5}}}%
    \put(0.12511078,0.30346763){\makebox(0,0)[lb]{\smash{0.7}}}%
    \put(0.1273965,0.36339509){\makebox(0,0)[lb]{\smash{0.9}}}%
    \put(0.25396054,0.33970541){\makebox(0,0)[lb]{\smash{$0.565$}}}%
    \put(0.13553377,0.10567965){\makebox(0,0)[lb]{\smash{0}}}%
    \put(0.1178251,0.18304123){\makebox(0,0)[lb]{\smash{180}}}%
    \put(0.08191325,0.60753664){\makebox(0,0)[lb]{\smash{-1e-10}}}%
    \put(1.03833511,0.68419386){\makebox(0,0)[lb]{\smash{}}}%
    \put(0.15761856,0.03923992){\makebox(0,0)[lb]{\smash{0}}}%
    \put(0.22598325,0.03923991){\makebox(0,0)[lb]{\smash{0.2}}}%
    \put(0.30291938,0.03923991){\makebox(0,0)[lb]{\smash{0.4}}}%
    \put(0.37995209,0.03923991){\makebox(0,0)[lb]{\smash{0.6}}}%
    \put(0.4550706,0.03923991){\makebox(0,0)[lb]{\smash{0.8}}}%
    \put(1.36154127,0.27933396){\color[rgb]{0,0,0}\makebox(0,0)[lt]{\begin{minipage}{0.28218421\unitlength}\raggedright \end{minipage}}}%
    \put(0.29665837,0.00267315){\makebox(0,0)[lb]{\smash{$e'$}}}%
    \put(0.56439382,0.46216946){\makebox(0,0)[lb]{\smash{0.04}}}%
    \put(0.56439382,0.52859813){\makebox(0,0)[lb]{\smash{0.08}}}%
    \put(0.62043416,0.19458238){\makebox(0,0)[lb]{\smash{0}}}%
    \put(0.67960337,0.19458238){\makebox(0,0)[lb]{\smash{0.2}}}%
    \put(0.75286659,0.19458238){\makebox(0,0)[lb]{\smash{0.4}}}%
    \put(0.82605195,0.19458238){\makebox(0,0)[lb]{\smash{0.6}}}%
    \put(0.89932153,0.19458238){\makebox(0,0)[lb]{\smash{0.8}}}%
    \put(0.64563656,-0.25541861){\makebox(0,0)[lb]{\smash{}}}%
    \put(0.68277109,0.32705256){\color[rgb]{0,0,0}\makebox(0,0)[lb]{\smash{$|g|$}}}%
    \put(0.68518721,0.23982497){\color[rgb]{0,0,0}\makebox(0,0)[lb]{\smash{$|f|$}}}%
    \put(0.55354233,0.26602654){\makebox(0,0)[lb]{\smash{2e-02}}}%
    \put(0.55354233,0.31516717){\makebox(0,0)[lb]{\smash{4e-04}}}%
    \put(0.55354233,0.36431338){\makebox(0,0)[lb]{\smash{6e-04}}}%
    \put(0.59896248,0.21948864){\makebox(0,0)[lb]{\smash{0}}}%
    \put(0.60102058,0.40048382){\makebox(0,0)[lb]{\smash{0}}}%
    \put(0.6788059,0.48332936){\color[rgb]{0,0,0}\makebox(0,0)[lb]{\smash{$|\nu|$}}}%
    \put(0.67907777,0.41867705){\color[rgb]{0,0,0}\makebox(0,0)[lb]{\smash{$|\mu|$}}}%
    \put(0.752021,0.16263602){\color[rgb]{0,0,0}\makebox(0,0)[lb]{\smash{$e'$}}}%
    \put(0.75957746,0.62963153){\color[rgb]{0,0,0}\makebox(0,0)[lb]{\smash{$\GdQSs$}}}%
    \put(0.03023607,-0.26468991){\color[rgb]{0,0,0}\makebox(0,0)[lb]{\smash{}}}%
    \put(1.42151651,0.61728121){\color[rgb]{0,0,0}\makebox(0,0)[lb]{\smash{}}}%
    \put(0.01227473,0.60813832){\color[rgb]{0,0,0}\makebox(0,0)[lb]{\smash{}}}%
    \put(0.0099887,0.44871993){\color[rgb]{0,0,0}\makebox(0,0)[lb]{\smash{$u$}}}%
    \put(0.0099887,0.30126125){\color[rgb]{0,0,0}\makebox(0,0)[lb]{\smash{$e$}}}%
    \put(0.18980308,0.50788522){\makebox(0,0)[lb]{\smash{$\GdQSu$}}}%
    \put(0.3231338,0.4452602){\makebox(0,0)[lb]{\smash{$\GdQSs$}}}%
    \put(0.41771786,0.29841087){\rotatebox{38.15209713}{\makebox(0,0)[lb]{\smash{$e=e'$}}}}%
    \put(0.38123178,0.69485446){\color[rgb]{0,0,0}\makebox(0,0)[lb]{\smash{}}}%
    \put(0.0099887,0.64013844){\color[rgb]{0,0,0}\makebox(0,0)[lb]{\smash{$\theta~(\degre)$}}}%
    \put(0.01227473,0.13042267){\color[rgb]{0,0,0}\makebox(0,0)[lb]{\smash{}}}%
    \put(0.0099887,0.14592195){\color[rgb]{0,0,0}\makebox(0,0)[lb]{\smash{$\omega~(\degre)$}}}%
    \put(0.1853162,0.20650571){\color[rgb]{0,0,0}\makebox(0,0)[lb]{\smash{d.}}}%
    \put(0.64602004,0.58366627){\color[rgb]{0,0,0}\makebox(0,0)[lb]{\smash{e.}}}%
    \put(0.1853162,0.37107765){\color[rgb]{0,0,0}\makebox(0,0)[lb]{\smash{c.}}}%
    \put(0.1853162,0.54250686){\color[rgb]{0,0,0}\makebox(0,0)[lb]{\smash{b.}}}%
    \put(0.1853162,0.69107887){\color[rgb]{0,0,0}\makebox(0,0)[lb]{\smash{a.}}}%
    \put(0.64602004,0.37109402){\color[rgb]{0,0,0}\makebox(0,0)[lb]{\smash{f.}}}%
  \end{picture}%
\endgroup%

\caption{\small a, b, c and d: orbital elements of the families of fixed points $\GdQSu$ and $\GdQSs$ versus $e'$. e and f: variations of the moduli of the real and imaginary part of the eigenvalues of the Hessian matrix along $\GdQSs$. $\GdQSs$ describe a configuration of two ellipses with opposite periaster and  $\theta=0\degre$. The fixed points are stable until $e'<0.8$. 
Moreover, this family possesses a particular orbit where $e=e'\simeq 0.565$.
On the contrary, the family $\GdQSu$ is unstable and describe a configuration of two ellipses with aligned periaster and  $\theta=0\degre$.}\label{fig:FPQS}
\end{center}
\end{figure}

\begin{figure}
\begin{center}
\small
\def\svgwidth{0.85\textwidth}
\begingroup%
  \makeatletter%
  \providecommand\color[2][]{%
    \errmessage{(Inkscape) Color is used for the text in Inkscape, but the package 'color.sty' is not loaded}%
    \renewcommand\color[2][]{}%
  }%
  \providecommand\transparent[1]{%
    \errmessage{(Inkscape) Transparency is used (non-zero) for the text in Inkscape, but the package 'transparent.sty' is not loaded}%
    \renewcommand\transparent[1]{}%
  }%
  \providecommand\rotatebox[2]{#2}%
  \ifx\svgwidth\undefined%
    \setlength{\unitlength}{699.99995117bp}%
    \ifx\svgscale\undefined%
      \relax%
    \else%
      \setlength{\unitlength}{\unitlength * \real{\svgscale}}%
    \fi%
  \else%
    \setlength{\unitlength}{\svgwidth}%
  \fi%
  \global\let\svgwidth\undefined%
  \global\let\svgscale\undefined%
  \makeatother%
  \begin{picture}(1,0.72907706)%
    \put(0,0){\includegraphics[width=\unitlength]{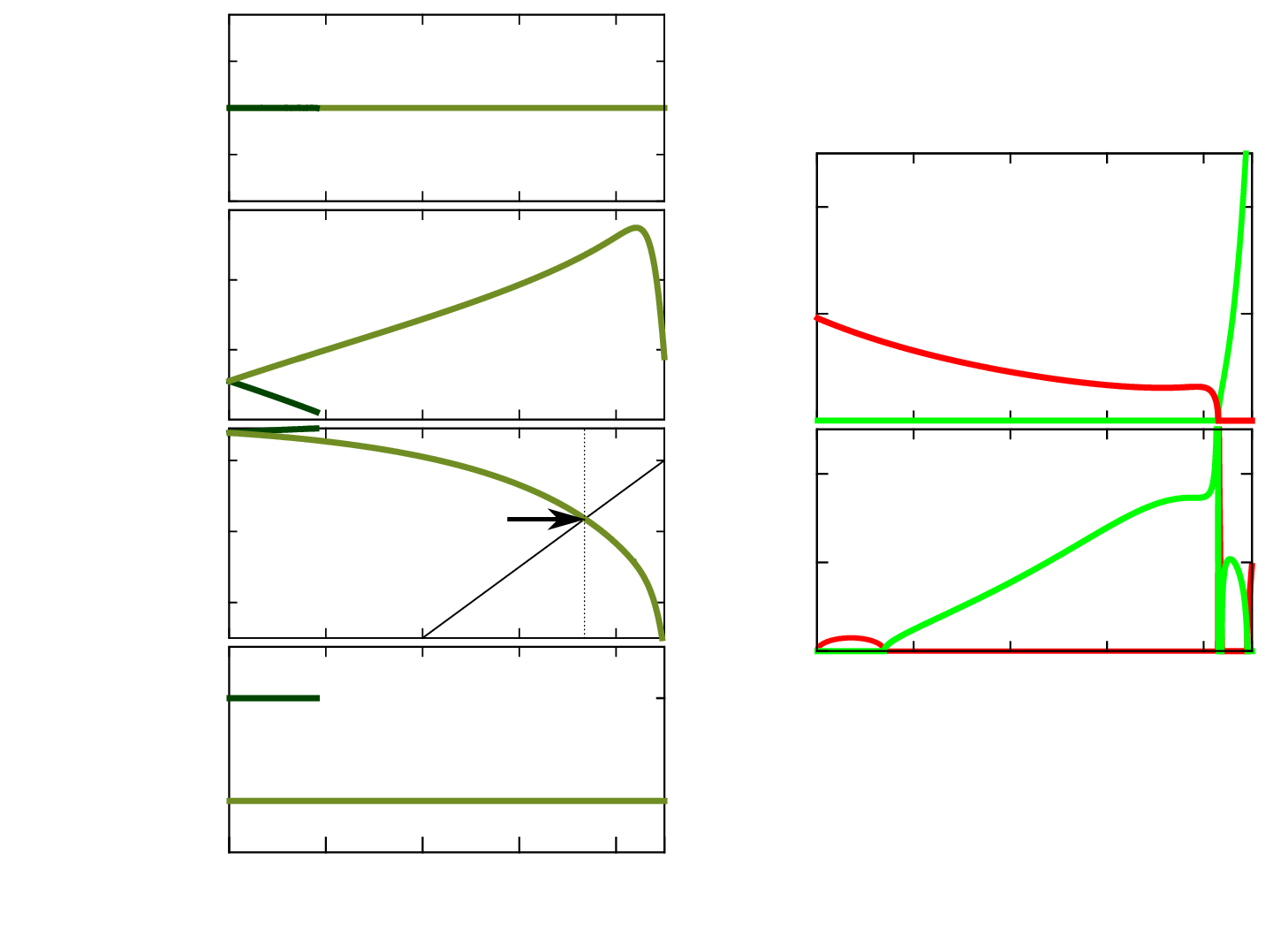}}%
    \put(0.14691418,0.64305703){\makebox(0,0)[lb]{\smash{0}}}%
    \put(0.09329375,0.67905594){\makebox(0,0)[lb]{\smash{1e-10}}}%
    \put(1.03833512,1.40648129){\makebox(0,0)[lb]{\smash{}}}%
    \put(0.06455831,0.44948784){\makebox(0,0)[lb]{\smash{-7.5e-04}}}%
    \put(0.06540983,0.50334387){\makebox(0,0)[lb]{\smash{-7.0e-04}}}%
    \put(0.12734831,0.25662959){\makebox(0,0)[lb]{\smash{0.5}}}%
    \put(0.12734831,0.31148442){\makebox(0,0)[lb]{\smash{0.7}}}%
    \put(0.12506264,0.36627241){\makebox(0,0)[lb]{\smash{0.9}}}%
    \put(0.13548561,0.10405586){\makebox(0,0)[lb]{\smash{0}}}%
    \put(0.11777696,0.18370317){\makebox(0,0)[lb]{\smash{180}}}%
    \put(0.07957941,0.60134139){\makebox(0,0)[lb]{\smash{-1e-10}}}%
    \put(0.16353257,0.04063414){\makebox(0,0)[lb]{\smash{0}}}%
    \put(0.22863353,0.04062984){\makebox(0,0)[lb]{\smash{0.2}}}%
    \put(0.30548212,0.04062984){\makebox(0,0)[lb]{\smash{0.4}}}%
    \put(0.38242746,0.04062984){\makebox(0,0)[lb]{\smash{0.6}}}%
    \put(0.45745988,0.04062984){\makebox(0,0)[lb]{\smash{0.8}}}%
    \put(1.36154128,1.00162139){\color[rgb]{0,0,0}\makebox(0,0)[lt]{\begin{minipage}{0.28218421\unitlength}\raggedright \end{minipage}}}%
    \put(0.30548212,0.00406237){\makebox(0,0)[lb]{\smash{$e'$}}}%
    \put(0.19101435,0.65688437){\color[rgb]{0,0,0}\rotatebox{0.37200928}{\makebox(0,0)[lb]{\smash{+180}}}}%
    \put(0.64563658,0.46686882){\makebox(0,0)[lb]{\smash{}}}%
    \put(0.56591071,0.47966255){\makebox(0,0)[lb]{\smash{0.04}}}%
    \put(0.56591071,0.56258666){\makebox(0,0)[lb]{\smash{0.08}}}%
    \put(0.55505918,0.2889471){\makebox(0,0)[lb]{\smash{4e-04}}}%
    \put(0.55505918,0.35544486){\makebox(0,0)[lb]{\smash{8e-04}}}%
    \put(0.59848777,0.22030316){\makebox(0,0)[lb]{\smash{0}}}%
    \put(0.59807669,0.4009665){\makebox(0,0)[lb]{\smash{0}}}%
    \put(0.62199655,0.1957577){\makebox(0,0)[lb]{\smash{0}}}%
    \put(0.68090768,0.19574187){\makebox(0,0)[lb]{\smash{0.2}}}%
    \put(0.75649367,0.19574187){\makebox(0,0)[lb]{\smash{0.4}}}%
    \put(0.83199931,0.19574187){\makebox(0,0)[lb]{\smash{0.6}}}%
    \put(0.90759184,0.19574187){\makebox(0,0)[lb]{\smash{0.8}}}%
    \put(0.75562145,0.16379677){\color[rgb]{0,0,0}\makebox(0,0)[lb]{\smash{$e'$}}}%
    \put(0.76027911,0.63081726){\color[rgb]{0,0,0}\makebox(0,0)[lb]{\smash{$\GdLTs$}}}%
    \put(0.03023608,0.45759751){\color[rgb]{0,0,0}\makebox(0,0)[lb]{\smash{}}}%
    \put(1.42151652,1.33956864){\color[rgb]{0,0,0}\makebox(0,0)[lb]{\smash{}}}%
    \put(0.33116892,0.32282687){\makebox(0,0)[lb]{\smash{$0.73$}}}%
    \put(0.24734016,0.42166593){\makebox(0,0)[lb]{\smash{$\GdLTu$}}}%
    \put(0.20369423,0.48779382){\makebox(0,0)[lb]{\smash{$\GdLTs$}}}%
    \put(0.33412174,0.248689){\rotatebox{38.18524256}{\makebox(0,0)[lb]{\smash{$e=e'$}}}}%
    \put(0.38123179,1.41714188){\color[rgb]{0,0,0}\makebox(0,0)[lb]{\smash{}}}%
    \put(0.77716022,0.24448478){\color[rgb]{0,0,0}\makebox(0,0)[lb]{\smash{$|g|$}}}%
    \put(0.74288075,0.30267063){\color[rgb]{0,0,0}\makebox(0,0)[lb]{\smash{$|f|$}}}%
    \put(0.65594886,0.49475936){\color[rgb]{0,0,0}\makebox(0,0)[lb]{\smash{$|\nu|$}}}%
    \put(0.65622073,0.41639238){\color[rgb]{0,0,0}\makebox(0,0)[lb]{\smash{$|\mu|$}}}%
    \put(0.01227476,0.60722011){\color[rgb]{0,0,0}\makebox(0,0)[lb]{\smash{}}}%
    \put(0.01456018,0.44323029){\color[rgb]{0,0,0}\makebox(0,0)[lb]{\smash{$u$}}}%
    \put(0.01456018,0.30034336){\color[rgb]{0,0,0}\makebox(0,0)[lb]{\smash{$e$}}}%
    \put(0.01456018,0.63464883){\color[rgb]{0,0,0}\makebox(0,0)[lb]{\smash{$\theta~(\degre)$}}}%
    \put(0.01227476,0.12950508){\color[rgb]{0,0,0}\makebox(0,0)[lb]{\smash{}}}%
    \put(0.01456018,0.14500426){\color[rgb]{0,0,0}\makebox(0,0)[lb]{\smash{$\omega~(\degre)$}}}%
    \put(0.10742923,0.66057307){\color[rgb]{0,0,0}\makebox(0,0)[lb]{\smash{}}}%
    \put(0.18927895,0.69495931){\color[rgb]{0,0,0}\makebox(0,0)[lb]{\smash{a.}}}%
    \put(0.58521013,0.1794866){\color[rgb]{0,0,0}\makebox(0,0)[lb]{\smash{b.}}}%
    \put(0.18927895,0.54181629){\color[rgb]{0,0,0}\makebox(0,0)[lb]{\smash{b.}}}%
    \put(0.19156467,0.37038705){\color[rgb]{0,0,0}\makebox(0,0)[lb]{\smash{c.}}}%
    \put(0.18927895,0.20124354){\color[rgb]{0,0,0}\makebox(0,0)[lb]{\smash{d.}}}%
    \put(0.6445721,0.58057303){\color[rgb]{0,0,0}\makebox(0,0)[lb]{\smash{e.}}}%
    \put(0.6445721,0.37028652){\color[rgb]{0,0,0}\makebox(0,0)[lb]{\smash{f.}}}%
  \end{picture}%
\endgroup%

\caption{\small a, b, c and d: orbital elements of the families of fixed points $\GdLTu$ and $\GdLTs$ versus $e'$. e and f: variations of the moduli of the real and imaginary part of the eigenvalues of the Hessian matrix along $\GdLTs$.
$\GdLTs$ describe a configuration of two ellipses with aligned periaster and $\theta=180\degre$. The fixed points are stable for $ 0 \leq e'\leq 0.15$. Moreover, this family possesses a particular orbit with $e'=e\simeq 0.73$ where the planet and the particle share the same ellipses.
On the contrary, $\GdLTu$ is unstable and describe a configuration of two ellipses with opposite periaster and $\theta=180\degre$.  }\label{fig:FPL3}
\end{center}
\end{figure}

\begin{figure}
\begin{center}
\small
\def\svgwidth{0.85\textwidth}
\begingroup%
  \makeatletter%
  \providecommand\color[2][]{%
    \errmessage{(Inkscape) Color is used for the text in Inkscape, but the package 'color.sty' is not loaded}%
    \renewcommand\color[2][]{}%
  }%
  \providecommand\transparent[1]{%
    \errmessage{(Inkscape) Transparency is used (non-zero) for the text in Inkscape, but the package 'transparent.sty' is not loaded}%
    \renewcommand\transparent[1]{}%
  }%
  \providecommand\rotatebox[2]{#2}%
  \ifx\svgwidth\undefined%
    \setlength{\unitlength}{700bp}%
    \ifx\svgscale\undefined%
      \relax%
    \else%
      \setlength{\unitlength}{\unitlength * \real{\svgscale}}%
    \fi%
  \else%
    \setlength{\unitlength}{\svgwidth}%
  \fi%
  \global\let\svgwidth\undefined%
  \global\let\svgscale\undefined%
  \makeatother%
  \begin{picture}(1,0.74118039)%
    \put(0,0){\includegraphics[width=\unitlength]{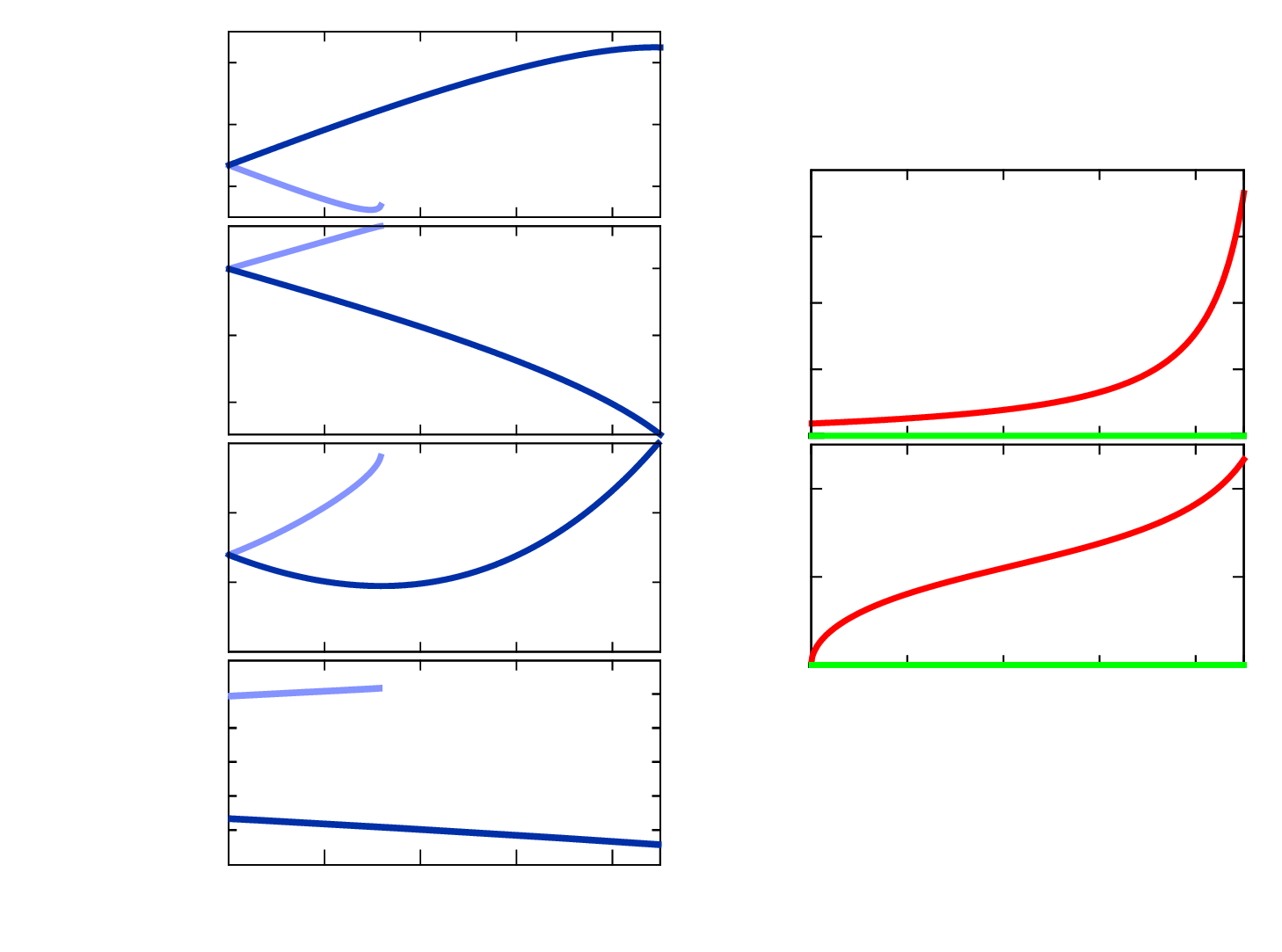}}%
    \put(1.03833504,2.16076764){\makebox(0,0)[lb]{\smash{}}}%
    \put(0.11847354,0.59085965){\makebox(0,0)[lb]{\smash{140}}}%
    \put(0.11847354,0.63657063){\makebox(0,0)[lb]{\smash{160}}}%
    \put(0.11847354,0.68228709){\makebox(0,0)[lb]{\smash{180}}}%
    \put(0.06570621,0.42100265){\makebox(0,0)[lb]{\smash{-1.0e-03}}}%
    \put(0.06496385,0.47521355){\makebox(0,0)[lb]{\smash{-0.9e-03}}}%
    \put(0.06496385,0.52485868){\makebox(0,0)[lb]{\smash{-0.8e-03}}}%
    \put(0.11347354,0.28121584){\makebox(0,0)[lb]{\smash{0.85}}}%
    \put(0.12347354,0.33743017){\makebox(0,0)[lb]{\smash{0.9}}}%
    \put(0.1616451,0.03822494){\makebox(0,0)[lb]{\smash{0}}}%
    \put(0.22583068,0.03822061){\makebox(0,0)[lb]{\smash{0.2}}}%
    \put(0.30244348,0.03822061){\makebox(0,0)[lb]{\smash{0.4}}}%
    \put(0.3791528,0.03822061){\makebox(0,0)[lb]{\smash{0.6}}}%
    \put(0.45395504,0.03822061){\makebox(0,0)[lb]{\smash{0.8}}}%
    \put(1.36154118,1.75590777){\color[rgb]{0,0,0}\makebox(0,0)[lt]{\begin{minipage}{0.28218419\unitlength}\raggedright \end{minipage}}}%
    \put(0.30244348,0.00165331){\makebox(0,0)[lb]{\smash{$e'$}}}%
    \put(0.64563652,1.22115524){\makebox(0,0)[lb]{\smash{}}}%
    \put(0.57930072,0.44954103){\makebox(0,0)[lb]{\smash{0.2}}}%
    \put(0.57930072,0.50139371){\makebox(0,0)[lb]{\smash{0.4}}}%
    \put(0.57930072,0.55089375){\makebox(0,0)[lb]{\smash{0.6}}}%
    \put(0.59690954,0.2201336){\makebox(0,0)[lb]{\smash{0}}}%
    \put(0.54890933,0.2864918){\makebox(0,0)[lb]{\smash{4e-04}}}%
    \put(0.54890933,0.35298952){\makebox(0,0)[lb]{\smash{8e-04}}}%
    \put(0.59670115,0.39918912){\makebox(0,0)[lb]{\smash{0}}}%
    \put(0.62270361,0.19330237){\makebox(0,0)[lb]{\smash{0}}}%
    \put(0.68204409,0.19329774){\makebox(0,0)[lb]{\smash{0.2}}}%
    \put(0.7549259,0.19329774){\makebox(0,0)[lb]{\smash{0.4}}}%
    \put(0.83046789,0.19329774){\makebox(0,0)[lb]{\smash{0.6}}}%
    \put(0.90335609,0.19329774){\makebox(0,0)[lb]{\smash{0.8}}}%
    \put(0.75953462,0.16134253){\color[rgb]{0,0,0}\makebox(0,0)[lb]{\smash{$e'$}}}%
    \put(0.74643399,0.62618224){\color[rgb]{0,0,0}\makebox(0,0)[lb]{\smash{$\GdLQs$}}}%
    \put(0.03023606,1.21188393){\color[rgb]{0,0,0}\makebox(0,0)[lb]{\smash{}}}%
    \put(1.42151641,2.093855){\color[rgb]{0,0,0}\makebox(0,0)[lb]{\smash{}}}%
    \put(0.38123176,2.17142823){\color[rgb]{0,0,0}\makebox(0,0)[lb]{\smash{}}}%
    \put(0.28461408,0.52518707){\makebox(0,0)[lb]{\smash{$\GdLQu$}}}%
    \put(0.20179709,0.47133693){\makebox(0,0)[lb]{\smash{$\GdLQs$}}}%
    \put(0.67296213,0.29365017){\color[rgb]{0,0,0}\makebox(0,0)[lb]{\smash{$|g|$}}}%
    \put(0.67238539,0.23542523){\color[rgb]{0,0,0}\makebox(0,0)[lb]{\smash{$|f|$}}}%
    \put(0.64359641,0.4318199){\color[rgb]{0,0,0}\makebox(0,0)[lb]{\smash{$|\nu|$}}}%
    \put(0.86421986,0.41128898){\color[rgb]{0,0,0}\makebox(0,0)[lb]{\smash{$|\mu|$}}}%
    \put(0.01227444,0.44525767){\color[rgb]{0,0,0}\makebox(0,0)[lb]{\smash{$u$}}}%
    \put(0.01227444,0.29779944){\color[rgb]{0,0,0}\makebox(0,0)[lb]{\smash{$e$}}}%
    \put(0.01227444,0.63667604){\color[rgb]{0,0,0}\makebox(0,0)[lb]{\smash{$\theta~(\degre)$}}}%
    \put(0.01227444,0.14246052){\color[rgb]{0,0,0}\makebox(0,0)[lb]{\smash{$\omega~(\degre)$}}}%
    \put(0.18468928,0.20378515){\color[rgb]{0,0,0}\makebox(0,0)[lb]{\smash{d.}}}%
    \put(0.64328176,0.37033057){\color[rgb]{0,0,0}\makebox(0,0)[lb]{\smash{f.}}}%
    \put(0.10599225,0.08584916){\color[rgb]{0,0,0}\makebox(0,0)[lb]{\smash{-150}}}%
    \put(0.10599225,0.11240159){\color[rgb]{0,0,0}\makebox(0,0)[lb]{\smash{-100}}}%
    \put(0.11388084,0.13895402){\color[rgb]{0,0,0}\makebox(0,0)[lb]{\smash{-50}}}%
    \put(0.12623994,0.16539959){\color[rgb]{0,0,0}\makebox(0,0)[lb]{\smash{0}}}%
    \put(0.11834694,0.19195215){\color[rgb]{0,0,0}\makebox(0,0)[lb]{\smash{50}}}%
    \put(0.64328176,0.58289875){\color[rgb]{0,0,0}\makebox(0,0)[lb]{\smash{e.}}}%
    \put(0.18468928,0.37749679){\color[rgb]{0,0,0}\makebox(0,0)[lb]{\smash{c.}}}%
    \put(0.18468928,0.54435139){\color[rgb]{0,0,0}\makebox(0,0)[lb]{\smash{b.}}}%
    \put(0.18468928,0.69749191){\color[rgb]{0,0,0}\makebox(0,0)[lb]{\smash{a.}}}%
    \put(-1.35581003,0.78547104){\color[rgb]{0,0,0}\makebox(0,0)[lb]{\smash{140}}}%
    \put(-1.35581003,0.86352017){\color[rgb]{0,0,0}\makebox(0,0)[lb]{\smash{160}}}%
    \put(-1.35581003,0.94157836){\color[rgb]{0,0,0}\makebox(0,0)[lb]{\smash{180}}}%
    \put(-1.37438498,0.835242){\color[rgb]{0,0,0}\rotatebox{90}{\makebox(0,0)[lb]{\smash{$\theta$}}}}%
    \put(-1.37438498,0.85895644){\color[rgb]{0,0,0}\rotatebox{90}{\makebox(0,0)[lb]{\smash{($\degre$) }}}}%
    \put(-1.09138078,0.80062533){\color[rgb]{0,0,0}\makebox(0,0)[lb]{\smash{F}}}%
    \put(-1.0842448,0.79505){\color[rgb]{0,0,0}\makebox(0,0)[lb]{\smash{L4}}}%
    \put(-1.07213158,0.80062533){\color[rgb]{0,0,0}\makebox(0,0)[lb]{\smash{FP}}}%
    \put(-1.05751432,0.79505){\color[rgb]{0,0,0}\makebox(0,0)[lb]{\smash{+}}}%
    \put(-1.0851028,0.7718231){\color[rgb]{0,0,0}\makebox(0,0)[lb]{\smash{F}}}%
    \put(-1.07796681,0.76624778){\color[rgb]{0,0,0}\makebox(0,0)[lb]{\smash{L}}}%
    \put(-1.07231166,0.76178758){\color[rgb]{0,0,0}\makebox(0,0)[lb]{\smash{3}}}%
    \put(-1.06728871,0.7718231){\color[rgb]{0,0,0}\makebox(0,0)[lb]{\smash{FP}}}%
    \put(-1.05267146,0.76624778){\color[rgb]{0,0,0}\makebox(0,0)[lb]{\smash{-}}}%
    \put(-1.37211499,0.52412762){\color[rgb]{0,0,0}\makebox(0,0)[lb]{\smash{-0.001}}}%
    \put(-1.38000984,0.60856515){\color[rgb]{0,0,0}\makebox(0,0)[lb]{\smash{-0.0009}}}%
    \put(-1.38000984,0.69301173){\color[rgb]{0 ,0,0}\makebox(0,0)[lb]{\smash{-0.0008}}}%
    \put(-1.39832505,0.61081333){\color[rgb]{0,0,0}\rotatebox{90}{\makebox(0,0)[lb]{\smash{u }}}}%
    \put(-1.35975745,0.30796577){\color[rgb]{0,0,0}\makebox(0,0)[lb]{\smash{0.85}}}%
    \put(-1.3518626,0.395661){\color[rgb]{0,0,0}\makebox(0,0)[lb]{\smash{0.9}}}%
    \put(-1.38253535,0.34796358){\color[rgb]{0,0,0}\rotatebox{90}{\makebox(0,0)[lb]{\smash{e }}}}%
    \put(-1.36027712,0.05800169){\color[rgb]{0,0,0}\makebox(0,0)[lb]{\smash{-200}}}%
    \put(-1.36027712,0.0850707){\color[rgb]{0,0,0}\makebox(0,0)[lb]{\smash{-150}}}%
    \put(-1.36027712,0.11213063){\color[rgb]{0,0,0}\makebox(0,0)[lb]{\smash{-100}}}%
    \put(-1.35238667,0.13919056){\color[rgb]{0,0,0}\makebox(0,0)[lb]{\smash{-50}}}%
    \put(-1.34002473,0.1661416){\color[rgb]{0,0,0}\makebox(0,0)[lb]{\smash{0}}}%
    \put(-1.34791958,0.19320153){\color[rgb]{0,0,0}\makebox(0,0)[lb]{\smash{50}}}%
    \put(-1.35581003,0.22027053){\color[rgb]{0,0,0}\makebox(0,0)[lb]{\smash{100}}}%
    \put(-1.32621315,0.02919947){\color[rgb]{0,0,0}\makebox(0,0)[lb]{\smash{0}}}%
    \put(-1.29480116,0.02919947){\color[rgb]{0,0,0}\makebox(0,0)[lb]{\smash{0.1}}}%
    \put(-1.25746803,0.02919947){\color[rgb]{0,0,0}\makebox(0,0)[lb]{\smash{0.2}}}%
    \put(-1.22013489,0.02919947){\color[rgb]{0,0,0}\makebox(0,0)[lb]{\smash{0.3}}}%
    \put(-1.18280618,0.02919947){\color[rgb]{0,0,0}\makebox(0,0)[lb]{\smash{0.4}}}%
    \put(-1.14541577,0.02919947){\color[rgb]{0,0,0}\makebox(0,0)[lb]{\smash{0.5}}}%
    \put(-1.10808705,0.02919947){\color[rgb]{0,0,0}\makebox(0,0)[lb]{\smash{0.6}}}%
    \put(-1.07075392,0.02919947){\color[rgb]{0,0,0}\makebox(0,0)[lb]{\smash{0.7}}}%
    \put(-1.0334252,0.02919947){\color[rgb]{0,0,0}\makebox(0,0)[lb]{\smash{0.8}}}%
    \put(-0.99609207,0.02919947){\color[rgb]{0,0,0}\makebox(0,0)[lb]{\smash{0.9}}}%
    \put(-1.37438498,0.11603944){\color[rgb]{0,0,0}\rotatebox{90}{\makebox(0,0)[lb]{\smash{$\omega$}}}}%
    \put(-1.37438498,0.13742456){\color[rgb]{0,0,0}\rotatebox{90}{\makebox(0,0)[lb]{\smash{($\degre$) }}}}%
    \put(-1.17609623,-0.01453018){\color[rgb]{0,0,0}\makebox(0,0)[lb]{\smash{e}}}%
    \put(-1.16846352,-0.02010551){\color[rgb]{0,0,0}\makebox(0,0)[lb]{\smash{planet}}}%
  \end{picture}%
\endgroup%

\caption{\small a, b, c and d: orbital elements of the families of fixed points $\GdLQu$ and $\GdLQs$ versus $e'$. e and f: variations of the moduli of the real and imaginary part of the eigenvalues of the Hessian matrix along $\GdLQs$. The whole family $\GdLQs$ is stable whereas $\GdLQu$ is unstable.}\label{fig:FPL4}
\end{center}
\end{figure}

For a given $e'$ in the AP, the linear dynamics in the vicinity of a fixed point is given by two couples of eigenvalues: $\pm\mu$ or $\pm i\nu$ and $\pm f$ or $\pm ig $ where $\mu$, $f$, $\nu$ and $g$ are real.
If these eigenvalues are all imaginary then they characterized an elliptic fixed point with libration and secular precession frequencies $\nu$ and $g$.
Otherwise, the fixed point is unstable.
Thus, we also characterized the stability variations of these families of fixed points by varying $e'$.\\
Their initial conditions and the moduli of the real and imaginary part of the eigenvalues versus $e'$ are plotted on Fig.\ref{fig:FPQS}, \ref{fig:FPL3} and \ref{fig:FPL4}.

According to the figures \ref{fig:FPQS}, \ref{fig:FPL3} and \ref{fig:FPL4}, we find eight families of fixed points in the averaged problem that correspond to frozen ellipses in the heliocentric frame.
For $e'=0$, these equilibria of the averaged problem belong to the set of degenerated fixed points or ``circles of fixed points" that exist for $\omega\in\TT$ and that we denoted $\GdLT$, $\GdLQ$, $\GdLC$ and $\GdQS$.

Among these eight families of fixed points, two are more relevant: $\GdQSs$ and $\GdLTs$.
The fixed points of $\GdQSs$ originates from $\GdQS$ and  are stable until $e'\simeq 0.8$. 
It corresponds to a configuration of two ellipses with two opposite periaster ($\omega = 180\degre$), $\theta=0\degre$ and a very high eccentricity that decreases when $e'$ increases (the slope being close to $\der{e}{e'}= -1/2$). 
On the contrary, the fixed points of $\GdLTs$ originates from $\GLT$ and is only stable for  $0 \leq e'\leq 0.15$.
It describes a configuration of two ellipses with aligned periaster ($\omega = 0\degre$), $\theta=180\degre$ and a very high eccentricity that decreases when $e'$ increases.\\
Along these two families, there exists a critical value of $e'$ where the planet and the asteroid ellipses have the same eccentricities.
The dashed lines of the figures \ref{fig:FPQS} and \ref{fig:FPL3} show that these particular orbits exist for $e'=e\simeq 0.565$ along $\GdQSs$ and $e'=e\simeq 0.73$  along $\GdLTs$.

Let us notice that these two families of configurations have been highlighted in the planetary problem.
Indeed, these two families have certainly to do with the stable and unstable families of periodic orbits described in \cite{HaPsVo2009} and \cite{HaVo2011}.
As regard $\GdQSs$, it could also be associated with the QS fixed point family in \cite{GiBeMi2010}.
In \cite{GiBeMi2010} as well as in \cite{HaVo2011}, these authors remarked that the configuration described by $\GdQSs$ with two equal eccentricities exists with an eccentricity value close to $0.565$ for several planetary mass ratio.
In our study, we establish that this particular orbit also exists in the restricted three-body problem for $e=e'\simeq 0.565$ (see Fig.\ref{fig:FPQS}). 
Likewise, according to \cite{HaPsVo2009}, the configuration described by $\GdLTs$ with two equal eccentricities seems to exist in the planetary problem for an eccentricity value close to $0.73$. 
Consequently, this suggests that these two particular configurations are weakly dependent on the ratio of the planetary masses.

Eventually, we remark that the existence of some of these eight configurations has already been showed. 
Indeed, in the range $0.01 \leq e'\leq 0.5$, \cite{NeThFe2002} exhibit QS stable and unstable fixed points. 
In addition, these authors also shown very high eccentric fixed points that correspond to the configuration of $\GdLQs$ and $\GdLQu$.\\
Likewise, \cite{Bi1978} and \cite{Ed1985} highlighted some frozen ellipses in co-orbital motion in the Sun-Jupiter system with $e'=e'_{Jupiter}\simeq 0.048$.
The first author found six very high eccentric fixed points denoted $P_1$, $Q_1$, $P_2$, $Q_2$, $P_3$, $Q_3$ that correspond to $\GdLQs$, $\GdLQu$, $\GdLCs$, $\GdLCu$, $\GdQSs$ and $\GdQSu$.
The other found a frozen ellipse in co-orbital resonance with $e = 0.975$ and $\theta=180\degre$, that is an orbit of $\GdLTs$.

\section{Conclusions}
In this paper, we clarify the definition of quasi-satellite motion and estimate a validity limit of the averaged approach by revisiting the planar and circular restricted three-body problem . 

First of all, we focussed on the co-orbital resonance via the averaged problem and showed that the studies of the phase portraits of the reduced averaged problem parametrized by $e_0$ allow to understand its global dynamics.
Indeed, they reveal that tadpole, horseshoe and quasi-satellite domains are structured around four families of fixed points originating from $L_4$, $L_5$ ($\GLQ$ and $\GLC$), $L_3$ ($\GLT$) and the singularity point for $e_0=0$ ($\GQS$).
By increasing $e_0$, the quasi-satellite orbits appear inside the domain opened by the collision curve for $e_0>0$ and becomes dominant for high eccentricities.
On the contrary, tadpole and horseshoe domains shrink and vanish when $\GLQ$ and $\GLC$ get closer and merge $\GLT$.
Moreover, we showed that this remaining family bifurcates and generates a new domain of high eccentric orbits librating around $(\theta,u)=(180\degre,0)$. \\
%
However, the averaged approaches having the drawback to be poorly significant in the exclusion zone, we highlighted that for sufficiently small eccentricities, the whole quasi-satellite domain is contained inside it, which makes this type of motion unreachable by averaging process.  
The study of the evolution of the libration and secular precession frequencies along $\GQS$ allowed us to show that the family $f$ and a fortiori the quasi-satellite domain are not reachable by $0\leq e_0<0.18$ in a Sun-Jupiter like system.

In order to clarify the terminology to use between ``retrograde satellite" or ``quasi-satellite" when these orbits are close encountering trajectories with the planet, we revisited the works in the rotating frame on  the family of simple-periodic symmetrical retrograde satellite orbits, or family $f$.\\
We highlighted that the family $f$ possesses two particular orbits that divide its neighbourhood in three connected areas: ``satellized" retrograde satellite, binary quasi-satellite and heliocentric quasi-satellite domains.
We established that the last one is the only one reachable in the averaged approaches.

The study of the frequencies of the fixed point families of the reduced averaged problem has also shown some frozen ellipses in the heliocentric frame which are equivalent to sets of degenerated fixed points (also denoted ``circles of fixed points") in the averaged problem.
In order to exhibit fixed points when the planet's orbit is eccentric, we highlighted numerically that from each circles of fixed points originates at least two families of fixed points parametrized by the planet eccentricity.
Among them, $\GdQSs$ is the most interesting as it is in quasi-satellite motion with a configuration of two ellipses with opposite perihelia and connected to the stable family described in \cite{HaPsVo2009} in the planetary problem.
Moreover,  $\GdQSs$ as well as the family in the planetary problem possess a configuration with equal eccentricities for any mass ratio with an eccentricity value close to $0.565$. 
As a consequence, this suggests that this remarkable configuration is weakly dependent to the mass ratio.
Likewise, let us mentioned that this configuration is similar to those of the family ``A.1/1" described in \cite{Br1975} in the general three-body problem with three equal masses which suggests a connection between them.  

When $e_0>0.4$, we denoted that the moduli of the libration frequency $\nu$ and of the secular precession frequency $g$ along the family $f$ are of the same order than those of the two tadpole periodic orbit families.
Thus, in the framework of long-term dynamics of the Jovian quasi-satellite asteroids in the solar system, we can assume that a study of the global dynamics by means of the frequency map analysis will reveal resonant structures close to those of the trojans identified in \cite{RoGa2006}.
However, by remarking that the direction of the perihelion precession being the opposite of those of the planets in the solar system, resonances with these secular frequencies should be of higher order in comparison to the tadpoles orbits.
On the contrary, resonances  with their node precession should be of lower order.
These questions will be addressed in a forthcoming work.

\section*{Acknowledgement}
This work has been developed during the Ph.D thesis of Alexandre Pousse at the ``Astronomie et Syst\`emes Dynamiques", IMCCE, Observatoire de Paris.

\bibliographystyle{apalike}

\begin{thebibliography}{}

\bibitem[{Beaug{\'e}} and {Roig}, 2001]{BeRo2001}
{Beaug{\'e}}, C., {Roig}, F.:
\newblock {A Semianalytical Model for the Motion of the Trojan Asteroids: Proper Elements and Families}.
\newblock {Icarus} \textbf{153}, 391--415 (2001)

\bibitem[{Benest}, 1974]{Be1974}
{Benest}, D.:
\newblock {Effects of the Mass Ratio on the Existence of Retrograde Satellites  in the Circular Plane Restricted Problem}.
\newblock {Astron. Astrophys.} \textbf{32}, 39--46 (1974)

\bibitem[{Benest}, 1975]{Be1975}
{Benest}, D.:
\newblock {Effects of the mass ratio on the existence of retrograde satellites in the circular plane restricted problem. II}.
\newblock {Astron. Astrophys.} \textbf{45}, 353--363 (1975).

\bibitem[{Benest}, 1976]{Be1976}
{Benest}, D.:
\newblock {Effects of the mass ratio on the existence of retrograde satellites in the circular plane restricted problem. III}.
\newblock {Astron. Astrophys.} \textbf{53}, 231--236 (1976)

\bibitem[{Bien}, 1978]{Bi1978}
{Bien}, R.:
\newblock {Long-period effects in the motion of Trojan asteroids and of fictitious objects at the 1/1 resonance}.
\newblock {Astron. Astrophys.} \textbf{68}, 295--301  (1978)

\bibitem[Brasser et~al., 2004]{BrInCo2004}
Brasser, R., Innanen, K., Connors, M., Veillet, C., Wiegert, P.A., Mikkola, S., Chodas, P.:
\newblock {Transient co-orbital asteroids}.
\newblock {Icarus} \textbf{171}, 102--109 (2004)

\bibitem[{Broucke}, 1968]{Br1968}
{Broucke}, R.A.:
\newblock {Periodic orbits in the restricted three-body problem with  earth-moon masses}.
\newblock{JPL Technical report}, 32--1168 (1968)

\bibitem[{Broucke}, 1975]{Br1975}
{Broucke}, R.A.:
\newblock {On relative periodic solutions of the planar general three-body problem}.
\newblock {Celest. Mech.} \textbf{12}, 439--462  (1975)


\bibitem[Connors et~al., 2002]{CoChMi2002}
Connors, M., Chodas, P., Mikkola, S., Wiegert, P.A., Veillet, C., Innanen, K.:
\newblock {Discovery of an asteroid and quasi-satellite in an Earth-like horseshoe orbit}.
\newblock {Meteorit. Planet. Sci.} \textbf{37},1435--1441 (2002).

\bibitem[Connors et~al., 2004]{CoVeBr2004}
Connors, M., Veillet, C., Brasser, R., Wiegert, P.A., Chodas, P., Mikkola, S.,  and Innanen, K.:
\newblock {Discovery of Earth's quasi-satellite}.
\newblock {Meteorit. Planet. Sci.} \textbf{39}:1251--1255 (2004).

\bibitem[{Couetdic} et~al., 2010]{CoLaCo2010}
{Couetdic}, J., {Laskar}, J., {Correia}, A.C.~M., {Mayor}, M., {Udry}, S.:
\newblock {Dynamical stability analysis of the HD 202206 system and constraints to the planetary orbits}.
\newblock {Astron. Astrophys.}, \textbf{519}, A10 (2010).

\bibitem[{Danielsson} and {Ip}, 1972]{DaIp1972}
{Danielsson}, L., {Ip}, W.-H.:
\newblock {Capture Resonance of the Asteroid 1685 Toro by the Earth}.
\newblock {Science} \textbf{176}, 906--907 (1972)

\bibitem[{de la Fuente Marcos} and {de la Fuente Marcos}, 2012]{DeDe2012}
{de la Fuente Marcos}, C., {de la Fuente Marcos}, R.:
\newblock {(309239) 2007 RW$_{10}$: a large temporary quasi-satellite of Neptune.}.
\newblock {Astron. Astrophys.} \textbf{545}, L9 (2012)



\bibitem[{de la Fuente Marcos} and {de la Fuente Marcos}, 2014]{DeDe2014}
{de la Fuente Marcos}, C., {de la Fuente Marcos}, R.:
\newblock {Asteroid 2014 OL$_{339}$: yet another Earth quasi-satellite}.
\newblock {Mon. Not. R. Astron. Soc.} \textbf{445}, 2985--2994  (2014)

\bibitem[{Deprit} et~al., 1967]{DeJaPa1967}
{Deprit}, A., {Henrard}, J., {Palmore}, J., {Price}, J.F.:
\newblock {The trojan manifold in the system Earth-Moon}.
\newblock {Mon. Not. R. Astron. Soc.} \textbf{137}, 311--335 (1967)

\bibitem[{Edelman}, 1985]{Ed1985}
{Edelman}, C.:
\newblock {Construction of periodic orbits and capture problems}.
\newblock {Astron. Astrophys.} \textbf{145}, 454--460 (1985).

\bibitem[{Gallardo}, 2006]{Ga2006}
{Gallardo}, T.:
\newblock {Atlas of the mean motion resonances in the Solar System}.
\newblock {Icarus} \textbf{184}, 29--38 (2006)

\bibitem[{Giuppone} et~al., 2010]{GiBeMi2010}
{Giuppone}, C.A., {Beaug\'{e}}, C., {Michtchenko}, T.A., and {Ferraz-Mello}, S.:
\newblock {Dynamics of two planets in co-orbital motion}.
\newblock {Mon. Not. R. Astron. Soc.} \textbf{407}, 390--398 (2010)

\bibitem[{Hadjidemetriou} et~al., 2009]{HaPsVo2009}
{Hadjidemetriou}, J.~D., {Psychoyos}, D., {Voyatzis}, G.:
\newblock {The 1/1 resonance in extrasolar planetary systems}.
\newblock {Celest. Mech. Dyn. Astron.} \textbf{104}, 23--38 (2009).

\bibitem[{Hadjidemetriou} and {Voyatzis}, 2011]{HaVo2011}
{Hadjidemetriou}, J.D.,  {Voyatzis}, G.:
\newblock {The 1/1 resonance in extrasolar systems. Migration from planetary to satellite orbits}.
\newblock {Celest. Mech. Dyn. Astron.} \textbf{111}, 179--199 (2011)

\bibitem[{Henon}, 1965a]{He1965}
{Henon}, M.:
\newblock {Exploration num\'{e}rique du probl\`{e}me restreint. I. Masses  \'{e}gales ; orbites p\'{e}riodiques}.
\newblock {Annales d'Astrophysique} \textbf{28}, 499 (1965a)

\bibitem[{Henon}, 1965b]{He1965a}
{Henon}, M.:
\newblock {Exploration num\'{e}rique du probl\`{e}me restreint. II. Masses
  \'{e}gales, stabilit\'{e} des orbites p\'{e}riodiques}.
\newblock {Annales d'Astrophysique} \textbf{28}, 992. (1965b)

\bibitem[{Henon}, 1969]{He1969}
{Henon}, M.:
\newblock {Numerical exploration of the restricted problem, V}.
\newblock {Astron. Astrophys.} \textbf{1}, 223--238 (1969)

\bibitem[{Henon} and {Guyot}, 1970]{HeGu1970}
{Henon}, M., {Guyot}, M.:
\newblock {Stability of Periodic Orbits in the Restricted Problem}.
\newblock{In: Giacaglia, G.E.O. (ed.) Periodic Orbits, Stability and Resonances, Reidel, Dordrecht-Holland}, 349--374 (1970)

\bibitem[{Jackson}, 1913]{Ja1913}
{Jackson}, J.:
\newblock {Retrograde satellite orbits}.
\newblock {Mon. Not. R. Astron. Soc.} \textbf{74}, 62--82 (1913)

\bibitem[{Kinoshita} and {Nakai}, 2007]{KiNa2007}
{Kinoshita}, H., {Nakai}, H.:
\newblock {Quasi-satellites of Jupiter}.
\newblock {Celest. Mech. Dyn. Astron.} \textbf{98}, 181--189 (2007)

\bibitem[{Kogan}, 1990]{Ko1990}
{Kogan}, A.I.:
\newblock {Quasi-satellite orbits and their applications}.
\newblock In: Jehn, R. (ed.) Proceedings of the 41st Congress of the International Astronautical Federation, 90–307. (1990)

\bibitem[{Kortenkamp}, 2005]{Ko2005}
{Kortenkamp}, S.J.:
\newblock {An efficient, low-velocity, resonant mechanism for capture of satellites by a protoplanet}.
\newblock {Icarus} \textbf{175}, 409--418 (2005)

\bibitem[{Kortenkamp}, 2013]{Ko2013}
{Kortenkamp}, S.J.:
\newblock {Trapping and dynamical evolution of interplanetary dust particles in
  Earth's quasi-satellite resonance}.
\newblock {Icarus}, \textbf{226}, 1550--1558 (2013)

\bibitem[{Lidov} and {Vashkov'yak}, 1993]{LiVa1993}
{Lidov}, M.L., {Vashkov'yak}, M.A.:
\newblock {Theory of perturbations and analysis of the evolution of
  quasi-satellite orbits in the restricted three-body problem}.
\newblock {Kosmicheskie Issledovaniia} \textbf{31}, 75--99 (1993)

\bibitem[{Lidov} and {Vashkov'yak}, 1994a]{LiVa1994}
{Lidov}, M.~L., {Vashkov'yak}, M.A.
\newblock {On quasi-satellite orbits for experiments on refinement of the gravitation constant}.
\newblock {Astron. Lett.} \textbf{20}, 188--198 (1994a)

\bibitem[{Lidov} and {Vashkov'yak}, 1994b]{LiVa1994a}
{Lidov}, M.L., {Vashkov'yak}, M.A.:
\newblock {On quasi-satellite orbits in a restricted elliptic three-body
  problem}.
\newblock {Astron. Lett.} \textbf{20}, 676--690 (1994b)

\bibitem[{Meyer} and {Hall}, 1992]{MeHa1992}
{Meyer}, K.R., {Hall}, G.R.:
\newblock {Introduction to Hamiltonian Dynamical Systems and the N-Body Problem}.
\newblock vol 90 of AMS. Springer, New York (1992)

\bibitem[{Mikkola} et~al., 2004]{MiBrWi2004}
{Mikkola}, S., {Brasser}, R., Wiegert, P.A., {Innanen}, K.:
\newblock {Asteroid 2002 VE68, a quasi-satellite of Venus}.
\newblock {Mon. Not. R. Astron. Soc.} \textbf{351}, L63--L65 (2004)

\bibitem[{Mikkola} and {Innanen}, 1997]{MiIn1997}
{Mikkola}, S., {Innanen}, K.:
\newblock {Orbital Stability of Planetary Quasi-Satellites}.
\newblock In {Dvorak}, R. \& {Henrard}, J. (eds.), {  {The Dynamical
  Behaviour of our Planetary System}}. Kluwer Academic Publishers, 345--355 (1997)

\bibitem[{Mikkola} et~al., 2006]{MiInWi2006}
{Mikkola}, S., {Innanen}, K., Wiegert, P.A., {Connors}, M., {Brasser}, R.:
\newblock {Stability limits for the quasi-satellite orbit}.
\newblock {Mon. Not. R. Astron. Soc.} \textbf{369}, 15--24  (2006)

\bibitem[{Moeller}, 1935]{Mo1935}
{Moeller}, J.P.:
\newblock {Zwei Bahnklassen im probleme restreint}.
\newblock {Publikationer og mindre Meddeler fra Kobenhavns Observatorium} \textbf{99}, 1--II (1935)

\bibitem[{Morais}, 2001]{Mo2001}
{Morais}, M. H.M.:
\newblock {Hamiltonian formulation of the secular theory for Trojan-type motion}.
\newblock {Astron. Astrophys.} \textbf{369}, 677--689 (2001)

\bibitem[Namouni, 1999]{Na1999}
Namouni, F.:
\newblock {Secular Interactions of Coorbiting Objects}.
\newblock {Icarus} \textbf{137}, 293--314 (1999)

\bibitem[Namouni et~al., 1999]{NaChMu1999}
Namouni, F., Christou, A., Murray, C.:
\newblock {New coorbital dynamics in the solar system}.
\newblock {Phys. Rev. Let.} \textbf{83}, 2506--2509 (1999)

\bibitem[{Nesvorn\'{y}} et~al., 2002]{NeThFe2002}
{Nesvorn\'{y}}, D., {Thomas}, F., {Ferraz-Mello}, S., {Morbidelli}, A.:
\newblock {A Perturbative Treatment of The Co-Orbital Motion}.
\newblock {Celest. Mech. Dyn. Astron.} \textbf{82}, 323--361 (2002)

\bibitem[{Robutel} and {Gabern}, 2006]{RoGa2006}
{Robutel}, P., {Gabern}, F.:
\newblock {The resonant structure of Jupiter's Trojan asteroids - I. Long-term
  stability and diffusion}.
\newblock {Mon. Not. R. Astron. Soc.} \textbf{372}, 1463--1482 (2006)

\bibitem[{Robutel} et~al., 2016]{RoNiPo2015}

{Robutel}, P., {Niederman}, L., {Pousse}, A.:
\newblock {Rigorous treatment of the averaging process for co-orbital motions
  in the planetary problem}.
\newblock {Comp. Appl. Math.} \textbf{35}, 675--699  (2016)

\bibitem[{Robutel} and {Pousse}, 2013]{RoPo2013}
{Robutel}, P., {Pousse}, A.:
\newblock {On the co-orbital motion of two planets in quasi-circular orbits}.
\newblock {Celest. Mech. Dyn. Astron.} \textbf{117}, 17--40 (2013)

\bibitem[{Sidorenko} et~al., 2014]{SiNeAr2014}
{Sidorenko}, V.V., {Neishtadt}, A.I., {Artemyev}, A.V., {Zelenyi}, L.~M.:
\newblock {Quasi-satellite orbits in the general context of dynamics in the 1:1
  mean motion resonance: perturbative treatment}.
\newblock {Celest. Mech. Dyn. Astron.}, \textbf{120}, 131--162 (2014)

\bibitem[{Str{\"o}mgren}, 1933]{St1933}
{Str{\"o}mgren}, E.:
\newblock {Connaisance actuelle des orbites dans le probl{\`e}me des trois corps}.
\newblock {Bulletin Astronomique} \textbf{9}, 87--130 (1933)

\bibitem[{Wajer}, 2009]{Wa2009}
{Wajer}, P.:
\newblock {2002 AA $_{29}$: Earth's recurrent quasi-satellite?}
\newblock {Icarus} \textbf{200}, 147--153 (2009)

\bibitem[{Wajer}, 2010]{Wa2010}
{Wajer}, P.:
\newblock {Dynamical evolution of Earth's quasi-satellites: 2004 GU$_{9}$ and
  2006 FV$_{35}$}.
\newblock {Icarus} \textbf{209}, 488--493 (2010)

\bibitem[{Wajer} and {Kr{\'o}likowska}, 2012]{WaKr2012}
{Wajer}, P., {Kr{\'o}likowska}, M.:
\newblock {Behavior of Jupiter Non-Trojan Co-Orbitals}.
\newblock {Acta Astronomica} \textbf{62}, 113--131 (2012)

\bibitem[{Wiegert} et~al., 2000]{WiInMi2000}
{Wiegert}, P., {Innanen}, K., {Mikkola}, S.:
\newblock {The Stability of Quasi Satellites in the Outer Solar System}.
\newblock {Astronomical Journal} \textbf{119}, 1978--1984 (2000)

\end{thebibliography}

\end{document}